\documentclass[a4paper,12pt]{article}
\usepackage{jheppub}
\usepackage[utf8]{inputenc}
\usepackage [english] {babel} 
\usepackage{amsmath,amssymb,latexsym}
\usepackage{slashed}
\usepackage{graphicx}
\usepackage{epstopdf}
\usepackage{mathtools}

\usepackage{xcolor}
\usepackage{hyperref}

\hypersetup{
  colorlinks=false
  urlbordercolor=gray,
  citebordercolor=gray,
  linkbordercolor=gray,
  pdfborderstyle={/S/U/W 1}
}

\DeclareGraphicsExtensions{.pdf}

\def \sec{\begin{section}}
\def \esec{\end{section}}

\renewcommand{\tilde}{\widetilde}

\def \al {\alpha}

\def \om {\omega}

\def \ep {\epsilon}
\def \th {\theta}

\def \Oc {\mathcal{O}}

\def \Fc {\mathcal{F}}

\def \Hc {\mathcal{H}}

\def \Lc {\mathcal{L}}

\def \Cc {\mathcal{C}}

\def \altt {\tilde{\alpha}}

\def \Jtt {\tilde{J}}

\def \pr {\partial}
\def \ra {\rightarrow}

\def \beq { \begin{equation}}
\def \eeq {\end{equation}}

\DeclareMathOperator*{\Tr}{Tr}

\DeclareMathOperator*{\Sch}{Sch}
\DeclareMathOperator*{\sgn}{sgn}

\renewcommand\Im{\operatorname{Im}}
\newcommand\const{\operatorname{const}}

\def \l {\left(}
\def \r {\right)}

\def \bra {\langle}
\def \ket {\rangle}

\def \nono {\nonumber \\}

\title{Non-local reparametrization action in coupled Sachdev--Ye--Kitaev models}
\author{Alexey Milekhin}
\affiliation{Department of Physics, University of California at Santa Barbara, Santa Barbara, CA 93106,
U.S.A.}
\emailAdd{milekhin@ucsb.edu}
\abstract{
We continue the investigation of coupled Sachdev--Ye--Kitaev(SYK) models without Schwarzian
action dominance.
Like the original SYK, at large $N$ and low energies these models have an approximate reparametrization symmetry. 
However, the dominant action for reparametrizations is non-local due to the presence of irrelevant local operator with small
conformal dimension.
We semi-analytically study different thermodynamic properties and the 4-point function and 
demonstrate that they significantly differ from the Schwarzian prediction. However, the residual entropy and 
maximal chaos exponent are the same as in Majorana SYK.
We also discuss chain models and finite $N$ corrections.
}

\begin{document}
\maketitle
\flushbottom
\section{Introduction}
Sachdev--Ye--Kitaev(SYK) model \cite{SachdevYe,KitaevTalks,Gross:2016kjj} and related
Kondo models \cite{parcollet1998, GPS} and tensor models 
\cite{Gurau:2009tw,Witten:2016iux,Klebanov:2016xxf,Klebanov:2017nlk} are remarkable quantum mechanical
models which exhibit emergent conformal symmetry, maximal chaos \cite{Maldacena:2015waa} and non-zero residual
entropy. The most salient feature shared by all these models is the emergence
of reparametrization symmetry at low energies which is 
explicitly(by kinetic term) and
spontaneously(by the form of 2-point function) broken.
It was shown by Maldacena and Stanford \cite{ms} and Kitaev and Suh \cite{SuhFirstPaper} that in the original SYK model
the corresponding (Euclidean) action for reparametrizations is governed by the Schwarzian action:
\beq
\label{k:sch}
S_{Sch} = -\frac{N \alpha^S_{Sch}}{ J} \int du \ \Sch\l \tau[u], u \r, \ 
\Sch \l \tau[u],u \r =  \frac{\tau'''}{\tau'} - \frac{3}{2} \l \frac{\tau''}{\tau'} \r^2.
\eeq
Because of this it has been conjectured that the original SYK model provides a 
UV completion
for two-dimensional Jackiw--Teitelboim (JT) gravity.

In a variety of examples\footnote{Refs. \cite{MQ,Fu:2016vas,cotler2017black,gu2017spread,Yoon:2017nig,GurAri2018Does,
wormhole_form,syk_bath, bath_old, pengfei, KitaevRecent, su2021page} and many
others.} it has been explicitly demonstrated that the Schwarzian does indeed
dominate in various physical observables at low energies in both in- and
out-of-equilibrium. 
However, it had remained an open
question whether there were models where the reparametrizations are governed
by some other action.

\textit{The purpose of this paper is to present such model and argue that the low
energy physics is dominated by a non-local action for reparametrizations:}
\beq
\label{k:s_nonloc}
S_{nonloc} = -\frac{N \alpha^S_{2h}}{J^{2h-2}} \int du_1 du_2  
\l \frac{\tau'(u_1) \tau'(u_2)}{(\tau(u_1)-\tau(u_2))^2} \r^{h}.
\eeq
This action was conjectured by Maldacena, Stanford and Yang(MSY) \cite{cft_breaking} when \footnote{Similar
action was recently studied in the context of chaotic 2d CFTs, \cite{Haehl:2018izb,Nguyen:2020jqp}.}
the spectrum of conformal dimensions contains an irrelevant local operator with the
dimension $h$ within the interval $1<h<3/2$. 
The original SYK does not have such operators.
In this study we present a microscopic model where such operators are present.
And then provide some analytic and extensive numerical evidence that the
non-local action indeed dominates. Our strategy is to study various large
$N$ exact equations numerically.
This paper is a more extensive and detailed presentation of our results reported in \cite{short}. 
In addition, at the end of this paper we present some results about finite $N$ corrections.

It is important to emphasize that the Schwarzian is still present in the model
we explored.  Numerical results clearly shows its presence. 
The main point is that at large $N$ it gives a subleading(in
$1/\beta J$) contribution.

The microscopic model we consider is simply two coupled SYK models
with twisted kinetic terms. It has $2N$ Majorana fermions $\psi^a_i,\ i=1,\dots,N, a=1,2$.
 The Lagrangian has the following form:
\begin{align}
\label{eq:L_T}
\Lc_T = \Lc_0 + \Lc_{int},
\end{align}
where
\begin{align}
\Lc_0 = \sum_i \frac{(1-\xi)}{2}  \psi^1_i \pr_u \psi^1_i + \frac{(1+\xi)}{2} \psi^2_i \pr_u 
\psi^2_i + \frac{1}{4!}\sum_{ijkl} \Bigg( J^1_{ijkl}  \psi^1_i \psi^1_j \psi^1_k \psi^1_l +  
J^2_{ijkl} \psi^2_i \psi^2_j \psi^2_k \psi^2_l \Bigg),
\end{align}
\begin{align}
\Lc_{int} = \frac{3}{2} \alpha \sum_{ijkl}
C_{ij;kl} \psi^1_i \psi^1_j \psi^2_k \psi^2_l .
\label{eq:Lint}
\end{align}
Disorder tensors $J^1,J^2,C$ are all independent and drawn from Gaussian ensemble.
We specify their variances and symmetry properties in the main text.

Let us state some elementary properties of this model:
\begin{itemize}
\item Without the two-side coupling, $\alpha=0$, it is just two decoupled Majorana SYK
models and $1 \pm \xi$ can be reabsorbed into $J^{1},J^2$. Obviously,
Schwarzian dominates in this case.
\item Without twisting, $\xi=0$, it is just coupled SYK model with a marginal
interaction which was studied in
\cite{Gu2017Local,Altland:2019lne,bath_old}. This model
has $\mathbb{Z}_2$ symmetry and is dominated by Schwarzian at 
any coupling $\alpha$.
\end{itemize}
For general $\alpha, \xi$ the discussion is very similar to standard
SYK model. However, we do not expect the Schwarzian to dominate. In the large
$N$ limit one can integrate out the disorders and write down exact
Schwinger--Dyson(SD) equations. Recall that the low energy
conformal solution in SYK is obtained by neglecting the kinetic term.
The same happens here. In fact, at low energies parameter $\xi$ drops out and
the 2-point function has the same form as in the original SYK.

Parameter $\alpha$ controls the
anomalous dimension $h$ in the non-local action. The dimension $h$ can be anywhere between $1$ and $2$. Specifically the
interesting range is $|\alpha|>1$, where $1<h<3/2$. The interaction strength $\alpha^S_{2h}$ depends
on both $\xi$ and $\alpha$. For small $\xi$ we expect it to depend quadratically
on $\xi$, however we have observed that for large $\xi$ there are deviations
from this behavior. We expect the dependence on $\alpha$ in $\alpha^S_{2h}$ to
be complicated.

From a holographic point of view the action (\ref{k:s_nonloc}) has a simple
interpretation: we have a matter field in $AdS_2$ dual to a boundary operator $\Oc_h$ of dimension
 $h$. In the large $N$ limit we expect the
matter to be non-interacting. Adding $\Oc_h$ to the boundary action 
and integrating out the matter produces a boundary-to-boundary propagator
$1/(u_1-u_2)^{2h}$ integrated over the whole boundary. Dressing it with
reparametrizations produces exactly the action (\ref{k:s_nonloc}). From
this point of view, non-quadratic $\xi$ dependence is quite puzzling.
Perhaps a simple explanation is that operator the $\Oc_h$ enters in the action with
a coefficient non-linear in $\xi$: holographic description of SYK(and our coupled model)
works in the IR only and various operators undergo a finite renormalization
between UV and IR. For this reason one should not treat the $\xi$ 
deformation in the UV
Lagrangian (\ref{eq:L_T}) as a simple addition
of $\xi \Oc_h$ in the IR. We discuss this issue more in the main text.

Unfortunately, we were not able to demonstrate analytically that the non-local
action indeed dominates in this coupled SYK model. Therefore our strategy is to obtain various
physical predictions of the non-local action analytically and check them against the numerics.

We performed an extensive
numerical analysis of large $N$ exact equations.
As we mentioned above, at infinite $N$ it is possible to write down exact SD equations 
for 2-point functions.
Our strategy was to first solve the Euclidean SD equations 
to obtain exact(valid at all times, not just in low energy)  
2-point functions. We did this using a uniform discretization in the time/frequency domain and a standard 
iteration procedure \cite{ms}. It is straightforward to extract the energy from the 2-point functions.
Also we studied the connected 4-point function. It is $1/N$ effect, but one can obtain
an exact(but somewhat formal) expression in terms of a certain functional kernel build from 2-point functions.
By numerically diagonalizing the kernel we argued that the non-local action dominates in the 4-point function too.
In fact, one can look at 4-point function computation as the derivation of the non-local action.

Also we discuss the physics of the non-local action. In general it is applicable at low temperatures: 
$T \ll J$. If temperatures are not too low, $J/N^{1/(2h-2)} \ll T$ it can be
treated classically. We mostly study this temperature range. 
We show that the residual entropy and chaos exponent are the same as in SYK.
We study elementary thermodynamic quantities and also transport coefficients in the
chain models. We demonstrate that
the diffusion constant becomes temperature dependent(in the Schwarzian-dominating case it does not depend on the
temperature). However, the thermal conductivity remains linear in the temperature.
We have summarized our findings in Table
\footnote{Residual entropy depends on the form on the conformal solution
only so the matching between the two columns in trivial. We included it for completeness.}
\ref{t:sum}. It is worth noting that the leading non-conformal correction $\delta G$ to 2-point function is always
different from SYK answer as long as $\alpha \neq 0, \xi \neq 0$. We have found that at zero temperature
\beq
\frac{\delta G}{G} \propto \frac{1}{(J u)^{h-1}},
\eeq
whereas in SYK:
\beq
\frac{\delta G_{SYK}}{G_{SYK}} \propto \frac{1}{J u}.
\eeq
This happens because the coupled model
always has the operator with dimension $1<h<2$. It is only for $|\alpha|>1$ that this operator 
dimension becomes less than $3/2$ and it starts to dominate over the Schwarzian in 4-point function and thermodynamic quantities.
\begin{center}
\begin{table}[!ht]
\begin{tabular}{|c | c | c |}
& Schwarzian & Non-local action \\
\hline
Residual entropy(Sec. \ref{sec:entropy}) & $2 S_{0,SYK}$ & $2 S_{0,SYK}$ \\
Energy vs temperature(Sec. \ref{sec:pert}),& $T^2$ & $T^{2h-1}$ \\
Late-time OTOC(Sec. \ref{sec:chaos}) & $\beta J e^{2\pi t/\beta}$ & $(\beta J)^{2h-2} e^{2 \pi t/\beta}$ \\
Diffusion constant(chain models, Sec. \ref{sec:conduct}) & $\const$ & $T^{3-2h}$ \\
Thermal conductance(chain models, Sec. \ref{sec:conduct}) & $T$ & $T$ \\
\end{tabular}
\caption{Summary of our results. We kept only the most relevant factors: 
temperature $T$, inverse temperature $\beta$,  Lorentzian time $t$.}
\label{t:sum}
\end{table}
\end{center}

The paper is organized as follows. 

Section \ref{sec:model} is devoted to the elementary properties of the coupled model.
In Section \ref{sec:micro} we discuss in more detail the microscopics of the model: 
we describe the properties of disorder couplings, derive SD equations and the spectrum of anomalous dimensions.
After this, in Section \ref{sec:pert} we review the perturbative MSY argument for the non-local action 
and study the thermodynamics numerically.
Then we continue this analysis and discuss $\xi$-dependence in Section \ref{sec:xi_dep}.

In Section \ref{sec:physics} we investigate the physics of the non-local action.
We start by discussing the residual entropy in Section \ref{sec:entropy}. Section \ref{sec:chaos}
contains the computation of the out-of-time ordered 4-point function and demonstrates the maximality
of chaos exponent. In Section \ref{sec:time_order} we examine the time ordered 4-point function and
its relation to energy-energy correlators. 
Section \ref{sec:density} computes 1-loop $N^0$ correction to the free energy. We conclude by
studying the chain models in Section \ref{sec:conduct}, where we derive the low-energy effective action and 
study transport. 

Section \ref{sec:derivation} is dedicated to a detailed discussion of the 4-point function and derivation of the
non-local action.
We start by reviewing Maldacena--Stanford \cite{ms} derivation of the Schwarzian in Section \ref{sec:review}.
After that Section \ref{sec:spectral} explores the subleading correction to the conformal 2-point functions in the 
coupled model. In Section \ref{sec:kernel} we discuss the properties of the kernel. Section
\ref{sec:shift} contains the results of the numerical diagonalization of the kernel. Kernel spectrum
is sensitive to the precise form of the non-local action. We see a good agreement with the analytical prediction, 
which we take as the most important evidence for the non-local action dominance.
In Section \ref{sec:Tdep} we continue the exploration of the kernel eigenvalues and discuss the prefactor in the non-local
action.

Section \ref{sec:ED} contains some exact diagonalization(ED) results at finite $N$. 
In Section \ref{sec:E0} we compare the ground state energy obtained two ways: by numerically solving large $N$
Schwinger--Dyson equations and performing ED. 
In Section \ref{sec:densityN} we probe the density of states near the ground state. 
This quantity is sensitive to $1/N$ corrections. 
Section \ref{sec:2ptN} contains the numerical evaluation of 2-point function at very late times, $\tau \gg N/J$.
Section \ref{sec:level} is dedicated to the study of the energy levels statistics.

In Conclusion we summarize our results and describe numerous open questions.

In Appendix \ref{sec:lorentz} we write Schwinger--Dyson equations in Lorentzian signature.
\section{The model}
\label{sec:model}
\subsection{Microscopic formulation}
\label{sec:micro}
The model we consider has two\footnote{Throughout the paper index $a$ labels the two sides. It will be 
equal either $1,2$(for individual fermions) or $11,22$(for 2-point functions).} 
independent Majorana SYK with a marginal interaction:
\beq
\label{eq:H_T}
H_T = \sum_{ijkl=1}^N \l  \frac{1}{4!} J^1_{ijkl} \psi^1_i \psi^1_j \psi^1_k \psi^1_l + \frac{1}{4!}J^2_{ijkl} 
\psi^2_i \psi^2_j \psi^2_k \psi^2_l + 
\frac{6 \alpha}{(2!)^2} C_{ij ; kl} 
\psi^1_i \psi^1_j \psi^2_k \psi^2_l     \r.
\eeq
However the anti-commutation relations are twisted because of the twisted kinetic term:
\beq
\{\psi_i^a,\psi_j^b \} = \frac{1}{1-\xi_a} \delta_{ij} \delta^{ab}, \quad \xi_1 = \xi,\ \xi_2 = -\xi.
\eeq
In principle, we can make the kinetic term standard by rescaling the fermions. However, we prefer not to do that.
Tensors $J^1, J^2$, are usual SYK disorders: totally antisymmetric and the components are independent and Gaussian.
Tensor $C_{ij;kl}$ has a Gaussian distribution too, but it has a separate skew-symmetry in $ij$ and $kl$ indices:
\beq
C_{ij;kl} = -C_{ji;kl} = -C_{ij;lk}.
\eeq
However, it does not mix $ij$ and $kl$. Because of that, 
integrating it out only produces $G_{11}^2$ and $G_{22}^2$ and
would not introduce mixed correlators $G_{12}, G_{21}$.
We adopt the following normalizing for the variances:
\beq
\bra  \l J^a_{ijkl} \r^2 \ket = \frac{3! J^2}{N^3} ,\ a=1,2;\
\bra \l  C_{ij;kl} \r^2 \ket = \frac{J^2}{6 N^3}.
\eeq
As in SYK, up to $1/N^2$ corrections there is no difference between quenched and annealed averages.
Treating $J^{1,2},C$ as annealed(i.e. normal quantum fields) and integrating them out,
we get the following Euclidean $G \Sigma$ action:
\begin{align}
S_{G \Sigma}  = \frac{1}{2} \sum_{a=11,22} \l \Tr \log \l (1-\xi_a)\pr_u - \Sigma_a \r -  
\int du_1 du_2 \Sigma_a(u_1,u_2) G_a(u_1,u_2) \r + \nonumber \\
+ \frac{1}{8} \int du_1 du_2 \l G_{11}^4 + G_{22}^4 + 6 \alpha^2 G_{11}^2 G_{22}^2  \r, \xi_{11} = \xi,\ \xi_{22} = -\xi,
\label{eq:s_gsi}
\end{align}
and Euclidean Schwinger--Dyson equations:
\begin{align}
\label{sd:eucl}
(1-\xi) \pr_u G_{11} - J^2(G_{11}^3 + 3 \alpha^2 G_{11} G_{22}^2) * G_{11} = \delta(u),  \nonumber \\
(1+\xi) \pr_u G_{22} - J^2(G_{22}^3 + 3 \alpha^2 G_{22} G_{11}^2) * G_{22} = \delta(u),
\end{align}
where $*$ denotes convolution in imaginary time $u$.

At low energies(Euclidean times $u \gg 1/J$) and low temperatures($\beta J \gg 1$) we can neglect 
the kinetic term. Notice that $\xi$ parameter drops out. Then SD equations
admit symmetric $G_{11}=G_{22}$ solution given by SYK conformal
solution:
\beq
\label{eq:Gconf}
G_{11} = G_{22} = G_{conf} = \frac{b \sgn(u)}{(1+3 \alpha^2)^{1/4}} \l \frac{\pi}{J \beta \sin\l \frac{\pi |u|}{\beta} \r} \r^{1/2},
\ 1/J \ll |u|,\ \beta J \gg 1,
\eeq
with $b = 1/(4 \pi)^{1/4}$. 
By dropping the kinetic term, we acquired time-reparametrization symmetry.
However, because of non-zero $\alpha$, $G_{11}$ and $G_{22}$ are still coupled, so there is only one reparametrization mode
which acts on $G_a$ as
\beq
\label{eq:reparam}
G_a(u_1,u_2) \ra  \l \tau(u_1)' \tau(u_2)' \r^{1/4} G_a(\tau(u_1),\tau(u_2)), \ a=11,22.
\eeq

\begin{figure}[!ht]
\centering
\includegraphics[scale=0.7]{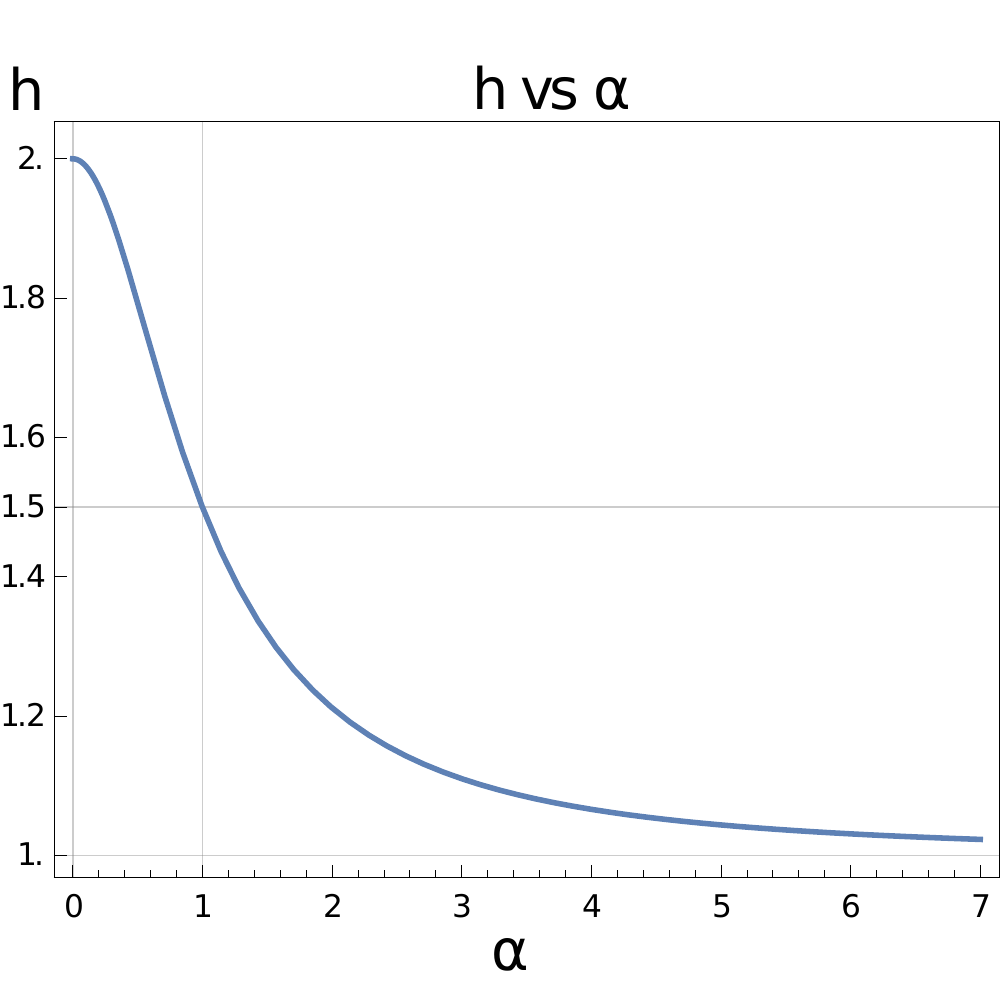}
\caption{The dimension $h$ of operator (\ref{eq:theoperator}) as a function of $\alpha$. 
$h$ approaches $1$ for $\alpha \ra \infty$.}
\label{fig:h}
\end{figure}

Above conformal solution (\ref{eq:Gconf}) tells us that elementary fermions $\psi^{1,2}_{i}$ has conformal dimension
$1/4$.
Let us discuss the spectrum of conformal dimension of bilinear operators. 
Using standard techniques, 
it can be shown \cite{Kim:2019upg} that the dimension $h$ of operator
\beq
\label{eq:theoperator}
\Oc_{2,0} = \sum_i \l \psi^1_i \pr_u \psi^1_i - \psi^2_i \pr_u \psi^2_i \r,
\eeq
is determined by the smallest
\footnote{The rest of the solutions determine the dimensions of
$
\Oc_{2,n} = \sum_i \psi^1_i \pr^{2n+1}_u \psi^1_i - \psi^2_i \pr^{2n+1}_u \psi^2_i
$.
Also, there is $\mathbb{Z}_2$-even sector
$
\Oc_{1,n} = \sum_i \psi^1_i \pr^{2n+1}_u \psi^1_i + \psi^2_i \pr^{2n+1}_u \psi^2_i
$
with the same dimensions as in SYK, which are determined by $h_A(h)=1$.
}
 $h$ solution of
\beq
\label{eq:dimension}
\frac{1-\alpha^2}{1+3 \alpha^2} g_A(h) = 1, \quad g_A(h) = -\frac{3}{2} \frac{\tan \l \pi(h-1/2)/2 \r}{h-1/2}.
\eeq
This spectrum strongly depends on $\alpha$. 
One can easily see that for $|\alpha|>1$, the dimension $h$ is in the range we are looking for: $1<h<3/2$
- Figure \ref{fig:h}. It will be important that this operator is $\mathbb{Z}_2$-odd.

Before proceeding to the detailed investigation of this operator, let us discuss the possible symmetry breaking in this model.
It is important because in the $\mathbb{Z}_2$ symmetry-broken phase the conformal solution (\ref{eq:Gconf})
does not represent the thermodynamically dominating phase and the whole argument would not work.
A closely related model, but with $\xi=0$, was studied by Kim--Klebanov--Tarnopolsky--Zhao(KKTZ) \cite{Kim:2019upg}:
\beq
H_{\mathbb{Z}_2} = \frac{1}{4!}\sum_{ijkl} J_{ijkl} \l \psi^1_i \psi^1_j \psi^1_k \psi^1_l +\psi^2_i \psi^2_j \psi^2_k \psi^2_l +  
6 \alpha \psi^1_i \psi^1_j \psi^2_k \psi^2_l     \r.
\eeq
In fact, the above models have the same spectrum of conformal dimension in the
antisymmetric\footnote{``antisymmetric'' refers to time dependence, not $\mathbb{Z}_2$ parity.
Operators $\Oc_{1,n},\ \Oc_{2,n}$ are said to be in antisymmetric sector because the 2-point function
$\bra T \psi^1_i(u) \psi^1_i(0) \ket$ is antisymmetric under $u \ra -u$. In contrast, under general assumptions the correlator 
$\bra T \psi^1_i(u) \psi^2_i(0) \ket$ is symmetric in $u$. } $\psi^1 \psi^1$, $\psi^2 \psi^2$ bilinear sector. 
However the problem is that in the original KKTZ there is $\mathbb{Z}_2$
symmetry breaking for $|\alpha|>1$. Actual ground state is separated by a gap from the rest of the spectrum.
At certain critical temperature $T_{crit}(\al) \sim N^0$ there is second-order phase transition. 
Below this temperature $\mathbb{Z}_2$ symmetry is broken and
the actual physical behavior is not described by the conformal solution. However above $T_{crit}$ the physics
is described by the conformal solution. Hence we expect that KKTZ model, once augmented with $\xi$-term, 
is also dominated by the non-local
action, but only in some window of temperatures $ T_{crit}(\alpha) < T \ll J$.
Notice that after integrating out $J_{ijkl}$ in KKTZ model,
SD equations contain mixed Green's functions $G_{12}, G_{21}$. 
The symmetry breaking is triggered by the operator 
\beq
\Oc_4 = \sum_i \psi^1_i \psi^2_i
\eeq
in the symmetric sector which acquires complex scaling dimension for $|\alpha|>1$.
In our case mixed correlators $G_{12}$ do not appear at all up to 
$1/N$ order. 
Therefore we
conjecture that the symmetry breaking does not occur in our model and the non-local action dominates
all the way to temperatures as low as $J/N^{1/(2h-2)}$. We verify this statement with
finite $N$ exact diagonalization in Section \ref{sec:ED}.

\subsection{A perturbative argument and thermodynamics}
\label{sec:pert}
As we just found out, the coupled model does contain an operator with dimension $1<h<3/2$.
Obviously, this irrelevant operator does not affect the conformal solution. How do we describe 
the influence of this operator on thermodynamics and other physical observables?

Let us review the arguments of \cite{SuhFirstPaper,cft_breaking}. 
In the standard SYK story(and in our coupled model) one obtains the conformal solution by neglecting
the kinetic term in the SD equations. One way to recover the low energy physics
is to consider conformal perturbation theory \cite{SuhFirstPaper}(see \cite{Tikhanovskaya:2020elb, 
Tikhanovskaya:2020zcw} for
a recent discussion). One starts from the artificial ``exactly conformal'' SYK without 
the kinetic term: 
\beq
\Lc_{conf} = \sum_{ijkl} J_{ijkl} \psi_i \psi_j \psi_k \psi_l.
\eeq
This theory taken literary is obviously pathological, as $\psi_i$ operators square to zero and lead to null states.
However, the exact 2-functions are given by conformal solutions proportional to the one in eq. (\ref{eq:Gconf}).
We proceed by perturbing this theory by a set of irrelevant operators which are meant to mimic the kinetic term:
\beq
\label{eq:irrels}
\Lc_{SYK} = \Lc_{conf} + \sum_h \alpha_{h} \Oc_h.
\eeq
The most important operator in this sum is $h=2$ operator:
\beq
\Oc_{h=2} = \psi_i \pr_u \psi_i.
\eeq
However, there are other terms with higher conformal dimensions. Notice that all of them come with
unknown\footnote{To the best of our knowledge, there are no recipes for computing them \textit{ab initio}.
One possibility in the original SYK is to find them in $1/q$ expansion. Unfortunately, for our coupled model
large $q$ limit is more complicated.} coefficients $\alpha_h$. 
Therefore one should be very careful in translating the operators in the UV Lagrangian to IR expansion in 
eq. (\ref{eq:irrels}).
Specifically, we expect that in our case some $\alpha_h$ are non-linear in $\xi$.

Operator with $h=2$ gives rise to Schwarzian and has to be treated separately. 
We can try to treat other, $h \neq 2$ operators $\Oc_h$ in our model
in a perturbative fashion. Naively, the leading contribution to the free energy
comes from dressing the two-point function $\bra \Oc_h \Oc_h \ket$ with reparametrizations:
\beq
\bra \Oc_h(u_1) \Oc_h(u_2) \ket \propto \frac{1}{(u_1-u_2)^{2h}} \ra
 \l \frac{\tau'(u_1) \tau'(u_2)}{(\tau(u_1)-\tau(u_2))^{2}} \r^h.
\eeq
This leads to a non-local action for reparametrizations (\ref{k:s_nonloc}) with some unknown
coefficient $\alpha_{2h}^S$.
Crucially, the above computation assumes that 1-pt function $\bra \Oc_h \ket$ vanishes.

Let us now describe elementary consequences of this.
As long as temperatures are not too low, $T \gg J/N^{1/(2h-2)}$, the action (\ref{k:s_nonloc}) can be treated classically
because of the overall factor of $N$. It is easy to check that the thermal solution is the same as in the
Schwarzian case: $\tau(u) = \tan(\pi u/\beta)$. 
Plugging this solution into the Schwarzian action trivially yields the following free energy:
\beq
\Delta F_{Sch}/N = - \frac{2\pi^2 \alpha_{Sch}^S}{J} T^{2} \ra  \Delta E_{Sch}/N = \frac{2\pi^2 \alpha_{Sch}^S}{J} T^2.
\label{energy:sch}
\eeq
The non-local action requires a bit more work. Assuming a fixed energy cutoff at $\sim J$, naive evaluation of the action yields a divergent term
\beq
\label{eq:pred}
-\beta \Delta F_{nonloc}/N = \frac{\alpha_{2h}^S}{J^{2h-2}} \beta^2 
\int_0^1 d \tilde{u} \ \l \frac{\pi}{\beta \sin^2 \l \pi \tilde{u} \r} \r^{2h} =
\frac{\#}{T} +   T^{2h-2} \frac{\alpha_{2h}^S \pi^{2h-1/2}}{J^{2h-2}} \frac{\Gamma \l 1/2-h \r}{\Gamma \l 1 - h \r }.
\eeq 
Fortunately, this divergent term is
proportional to $1/T$, hence it is simply a shift in the ground state energy \cite{cft_breaking}. 
Throughout the paper we will be using the following notation for the free energy:
\beq
\label{eq:F_T}
F/N = E_0/N -T S_0 - f_{2h} T^{2h-1} -f_{Sch} T^2 + \dots,
\eeq
and energy:
\beq
\label{eq:E_T}
E/N =E_0/N + c_{2h} T^{2h-1} +  c_{Sch} T^2 + \dots \ .
\eeq
These coefficients are related by:
\beq
c_{Sch} = f_{Sch},\ c_{2h} = (2h-2) f_{2h}.
\eeq
In this notation eq. (\ref{eq:pred}) says that
\beq
f_{2h} = \frac{\alpha_{2h}^S \pi^{2h-1/2}}{J^{2h-2}} \frac{\Gamma \l 1/2-h \r}{\Gamma \l 1 - h \r }.
\label{eq:FS}
\eeq

\begin{figure}[!ht]
\centering
\includegraphics[scale=0.75]{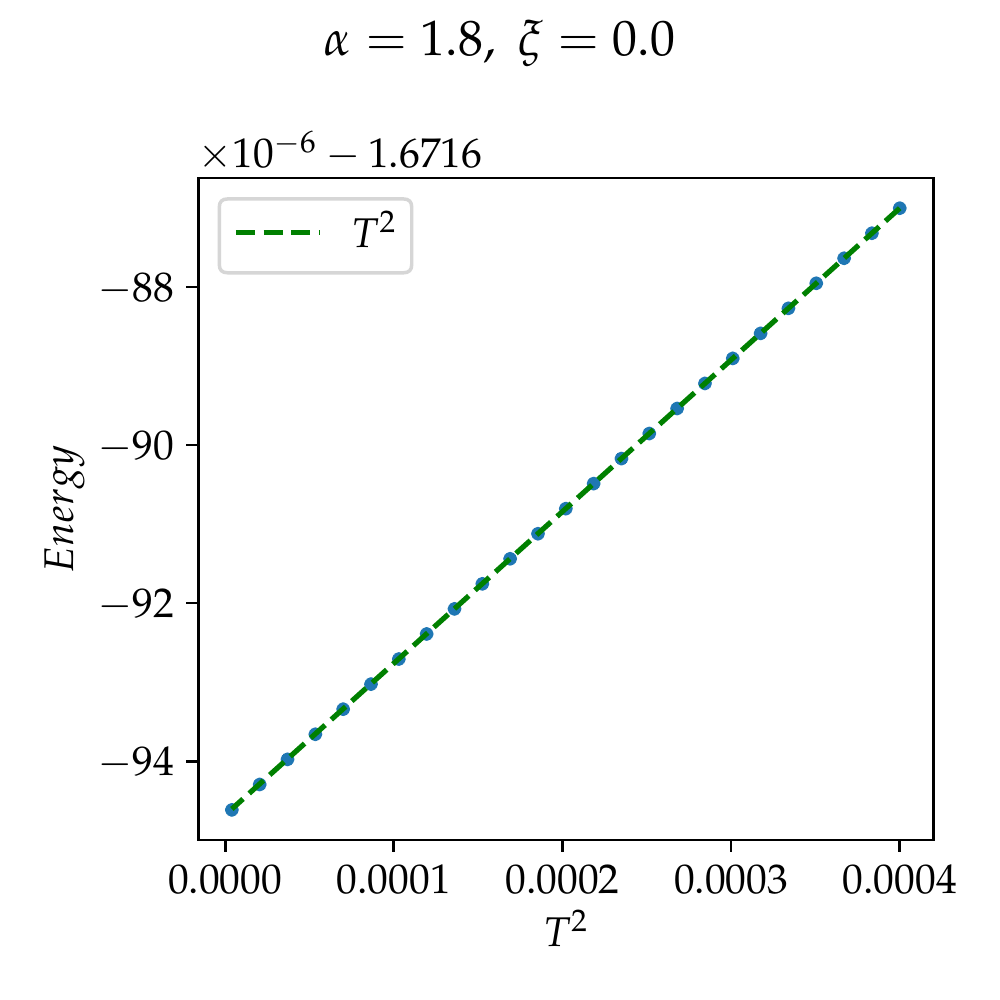}
    \caption{Energy vs $T^2$ for $J=2\pi$. Blue points are numerical data. For $\xi=0$ we expect Schwarzian answer.}
\label{fig:e_sch}
\end{figure}

\begin{figure}[!ht]
\centering
\minipage{0.47\textwidth}
\includegraphics[scale=0.73]{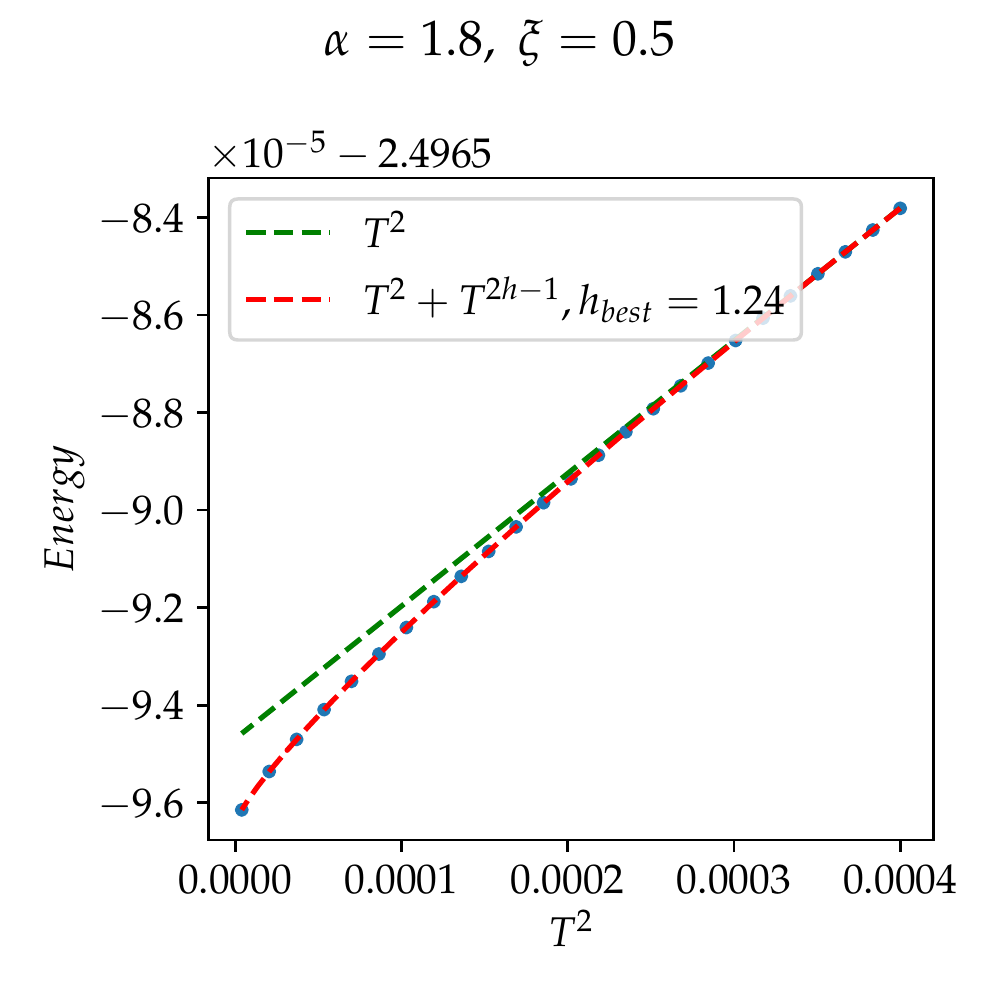}
\endminipage
\minipage{0.47\textwidth}
\includegraphics[scale=0.73]{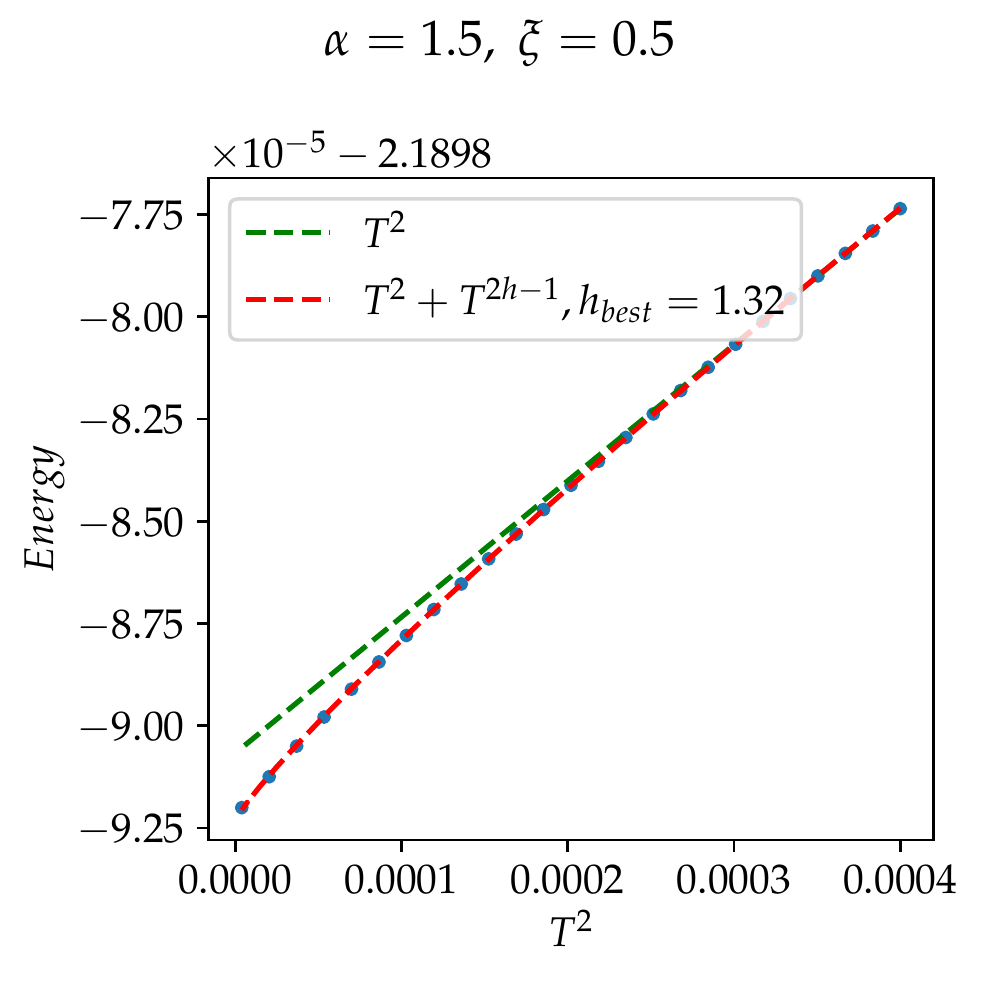}
\endminipage
\caption{Energy vs $T^2$ for $J=2\pi$. Blue points are numerical data. 
    We see a clear deviation from the Schwarzian prediction(dashed
    green is a straight line to guide the eye). For $\alpha=1.8$, $h_{theor}=1.24$ and for $\alpha=1.5$,
    $h_{theor}=1.31$. Changing the number of discretization points, temperature range and
    removing $c_{Sch} T^2$ term from the fit produces
    $h_{best} = 1.24 \pm 0.04$ for $\alpha=1.8$ and $h_{best}=1.32 \pm 0.02$ for $\alpha=1.5$.}
\label{fig:e_nonloc}
\end{figure}

We can easily check predictions from the Schwarzian and the non-local action against the numerical solution of SD equations.
Using by now standard methods of solving SD equation in Euclidean time, we plotted
energy versus temperature squared $T^2$.
First consider the benchmark case with $\xi=0$ and $\alpha=1.8$ - 
Figure \ref{fig:e_sch}. Since $\xi=0$, we expect Schwarzian answer.
We see that the energy is indeed proportional to $T^2$. We have performed this analysis for a wide
range of $\beta$ between $50$ and $500$(not shown) and verified that the energy stays proportional to $T^2$.
Now we switch to non-zero $\xi=0.5$ - Figure \ref{fig:e_nonloc}. We see a clear
deviation from $T^2$ law. To quantify it, we have fitted the data with eq. (\ref{eq:E_T})
keeping\footnote{It is computationally
costly to go to very low temperatures, therefore we have included the subleading Schwarzian $c_{Sch} T^2$ term.
By performing a fit with and without it one can estimate the uncertainty in $h$.} 
$c_{Sch}, c_{2h}$ and power $h$ unknown(i.e. they are extracted from the data). 
We see that $h_{best}$ are very close to theoretical values. Analysis at other values of $\alpha$(not shown) 
lead to similar results.

\subsection{Twist $\xi$ dependence}
\label{sec:xi_dep}
\begin{figure}[!ht]
\centering
\minipage{0.47\textwidth}
\includegraphics[scale=0.72]{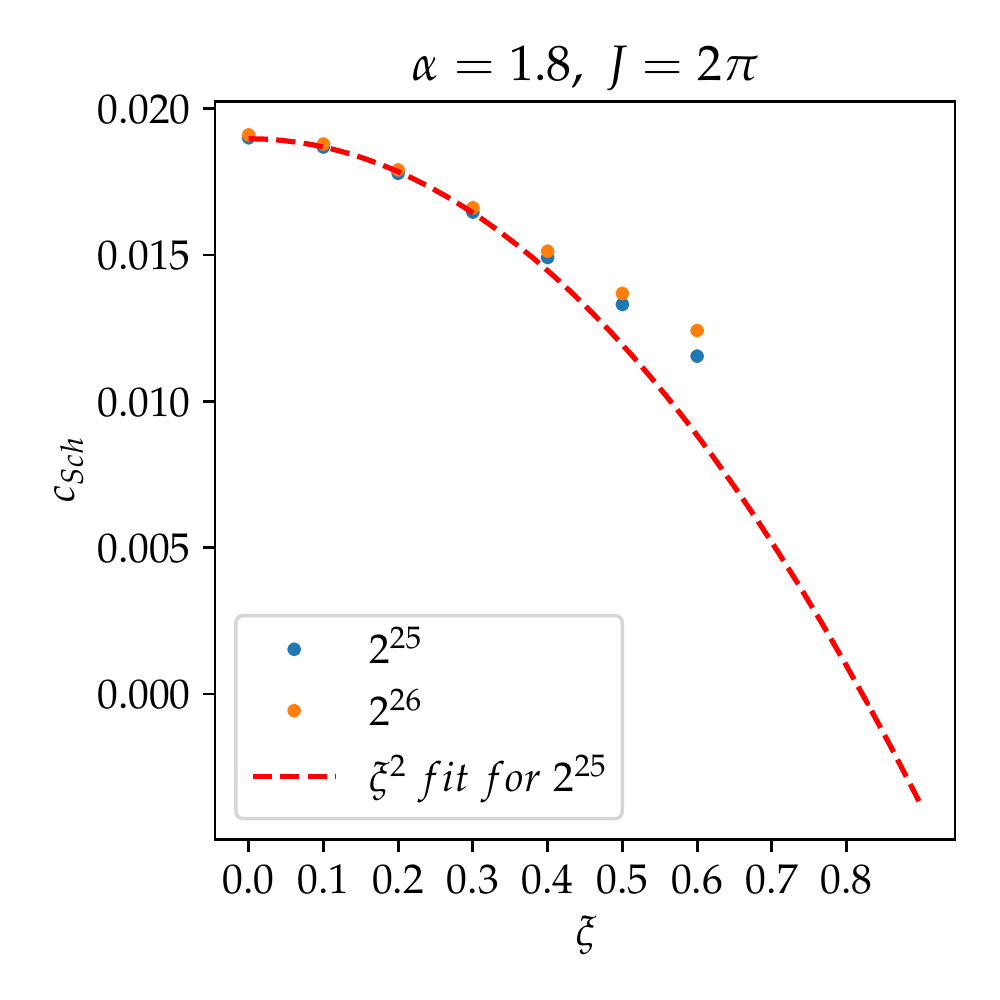}
\endminipage
\minipage{0.47\textwidth}
\includegraphics[scale=0.72]{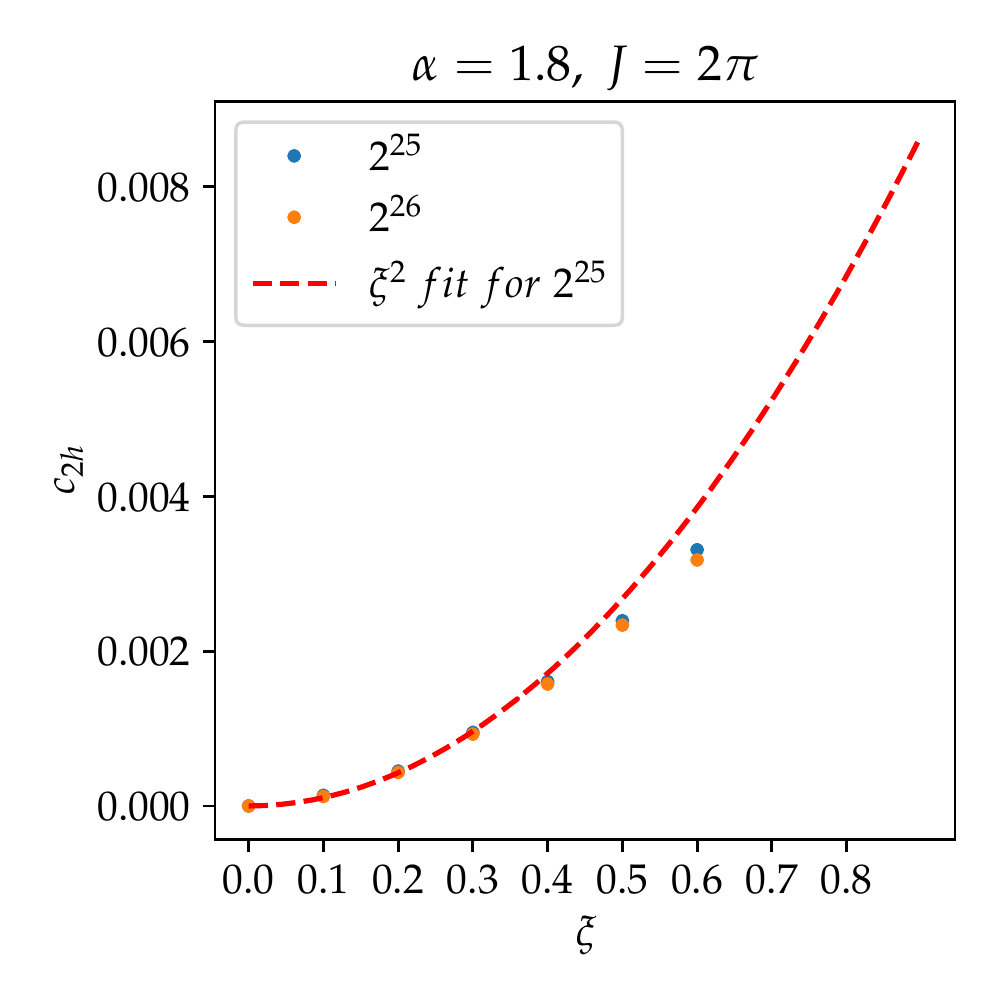}
\endminipage
\caption{Coefficients $c_{Sch}$(Left) and $c_{2h}$(Right) versus $\xi$ for different number of discretization
points $2^{25}, 2^{26}$.}
\label{fig:c_xi}
\end{figure}

As we have mentioned before, we do not really know $\xi$-dependence of coefficients $\alpha_h$ in the
expansion (\ref{eq:irrels}).
We addressed this question by numerically extracting
coefficients $c_{2h}$ and $c_{Sch}$ in the energy, eq. (\ref{eq:E_T}) for different values of $\xi$.
It is challenging to perform this computation because time discretization has to
be smaller than the inverse $J_{eff} = J/(1-\xi)^2$, which becomes very big for $\xi \ra 1$.
This is why we plotted $c_{Sch}, c_{2h}$ vs $\xi$ for different number of discretization points 
to make sure we converge. The results are presented in Figure \ref{fig:c_xi}.
We see that both $c_{Sch}$ and $c_{2h}$ start quadratic but then deviate from $\xi^2$ law. 
For large $\xi$ the dependence becomes slower than quadratic.

The Schwarzian coefficient $c_{Sch}$ decreases with $\xi$.
From JT gravity perspective, adding $\xi$ terms introduces extra light matter in the bulk.
We can try to compare this result to a similar problem: Schwarzian coupled to 2D CFT.
This problem is exactly soluble \cite{Yang:2018gdb} and CFT does lower Schwarzian coefficient.

\section{Physics of non-local action}
\label{sec:physics}
This Section is dedicated to various physical properties of the non-local action.
Everywhere, except Section \ref{sec:density}, 
we assume that $N$ is large and the temperatures are not too low($T \gg J/N^{1/(2h-2)}$)
so that the non-local action can be treated classically. In Section \ref{sec:density} we compute leading $N^0$ correction
to the free energy, which amounts to 1-loop computation.
We do not perform any numerics here.

In many places we will need the form of quadratic fluctuations around the thermal solution.
Expanding the non-local action (\ref{k:s_nonloc}) near the zero-temperature solution $\tau(u)=u+\ep(u)$ would yield
\beq
S_{nonloc,\beta=\infty} \propto \int dp \ \ep(p) |p|^{2h+1} \ep(-p).
\eeq
However, we are interested in the finite temperature case $\tau(u) = \tan\l \pi (u+\beta \ep(u))/\beta \r$.
In this case the fluctuations can be expanded in Fourier modes $\ep = \sum_n \ep_n e^{2 \pi i u n/\beta}$
giving rise to the following action
\footnote{We are grateful to D.~Stanford and Z.~Yang for the discussion about this
computation and the subsequent chaos exponent computation.}
\beq
\label{eq:snonloc_n}
S_{nonloc, \beta} = \frac{N \alpha_{2h}^S m_h}{\l \beta J\r^{2h-2} } \sum_n \ep_{n} g_h(n) \ep_{-n},
\eeq
with
\beq
\label{eq:gh}
g_h(n) = n^2 \l \frac{\Gamma(n+h)}{\Gamma(1+n-h)} -
\frac{\Gamma(h-1)}{\Gamma(-h)} \r,
\eeq
and numerical coefficient $m_h$ is
\beq
\label{eq:mh}
m_h = - (2 \pi)^{2h+1} \frac{\pi (h-1)^2 }{\cos(\pi h) \Gamma(2 h)}.
\eeq
For large $n$ we expect zero-temperature answer $g_h(n) \propto n^{2h+1}$. However even at small $n$ this is 
a good approximation. For bookkeeping, $\ep(u)$ and $\ep_n$ will always be dimensionless.
\subsection{Residual entropy}
\label{sec:entropy}
For a warm-up, let us start from the zero-temperature entropy.
Recall that the residual entropy at $T=0$ can be computed
\footnote{Modulo some UV subtleties} from
evaluating $\Tr \log G$ on the conformal solution \cite{parcollet1998}. 
In the model we are considering the conformal solution is exactly the 
same as in SYK model. Therefore the residual entropy is just twice Majorana SYK
residual entropy:
\beq
S_0 = 2 S_{0,SYK},
\eeq
\beq
S_{0,SYK} = \int^{1/4}_{0} dx \ \frac{\pi x}{\tan(\pi x)} = 0.2324\dots \ .
\eeq
The actual residual entropy is $N$ times this, eq. (\ref{eq:F_T}).
Our numerical results are consistent with this prediction.

\subsection{Chaos exponent}
\label{sec:chaos}
In this Section we show that the non-local action (\ref{k:s_nonloc}) leads to
a maximal chaos exponent \cite{Maldacena:2015waa} in the out-of-time ordered correlation(OTOC) function.

The OTOC can be computed as follows.
Since the reparametrizations is the only dominant physical mode at low energies, we need to dress 
the product of two 2-point functions with reparametrizations and
average over them. In SYK one has to use the Schwarzian action (\ref{k:sch}), 
however in our case it is the non-local action (\ref{eq:snonloc_n}). 
Leading $1/N$ contribution comes from using
the linearized action (\ref{eq:snonloc_n}):
\begin{align}
& \frac{\Fc}{G_{conf}(x) G_{conf}(x')} = \frac{\bra \psi^1_i(\th_1) \psi^1_i(\th_2) \psi^1_j(\th_3) \psi^1_j(\th_4) \ket_{conn}}{G_{conf}(x) G_{conf}(x')}
=  \nono
& = \frac{(\beta J)^{2h-2}}{N}  \frac{\pi^{2-2h}}{2 m_h \alpha_{2h}^S} \sum_{|n| \ge 2} \frac{e^{i n (y'-y)}}{g_h(n)}
\left[ \frac{\sin \frac{n x}{2}}{\tan \frac{x}{2}} -
n \cos \frac{n x}{2}   \right]
\left[ \frac{\sin \frac{n x'}{2}}{\tan \frac{x'}{2}} -
n \cos \frac{n x'}{2}   \right],
\label{eq:4pt_raw}
\end{align}
where $\theta_i$ are angle variables on the thermal circle $\theta= 2\pi u/\beta$ and $y,y',x,x'$ are certain 
combinations of angles:
\beq
x= \th_1 -\th_2, \ x'=\th_3 - \th_4,\ y=\frac{\th_1+\th_2}{2},\ y'=\frac{\th_3+\th_4}{2}.
\eeq
This piece dominates over
contributions from other conformal fields due to $(\beta J)^{2h-2}$ enhancement.

In general this expression is complicated
for ordering $\theta_1 < \theta_3 < \theta_2 < \theta_4$
which is relevant for OTOC.
Fortunately, it simplifies a lot when the points are antipodal on the circle.
Specifically, we put
\beq
\th_1 = -\frac{\pi}{2} - \th, \th_3 = 0,\ \theta_2 = \frac{\pi}{2}-\th, \th_4 = \pi,
\eeq
which corresponds to
\beq
x=x'= -\pi,\ y = -\th, y' = \frac{\pi}{2}.
\eeq
So we have
\beq
\sum_{|n| \ge 2} \frac{e^{i n\l \pi/2  + \th \r} n^2 \cos^2 \frac{\pi n}{2}}{g_h(n)}.
\eeq
We see that the sum goes over even $n$ only. We can convert the sum into the
integral by introducing a factor $1/(e^{i \pi n}-1)$ and integrating over 
the contour $\Cc$ enclosing $\pm 2, \pm 4, \dots$:
\beq
\frac{1}{2}\oint_\Cc dn \ \frac{n^2}{e^{i \pi n}-1} \frac{e^{i n\l \pi/2  + \th \r}}{g_h(n)}.
\eeq
Now we can move the contour to infinity, since the integrand decays along the imaginary
axis. It will pick up the pole at $n=0$ where $e^{i \pi n}=1$ and at locations where  $g_h(n)$ has zeroes.
The zeroes are located at $n=1,0,-1$ and at other negative $n$. Poles at negative $n$ are not relevant for us, 
because after analytically continuing to OTOC, namely
$\th \ra -2\pi i t/\beta$, they will produce exponentially decaying contributions(or a constant for $n=0$).
The pole at $n=1$ yields\footnote{The coefficient $p_h$ here is
$
\frac{\pi^{3-2h} \Gamma(2-h)}{4\Gamma(1+h) ( \psi(1+h) - \psi(2-h) )} 
$
where $\psi$ is Digamma function, $\psi(x)=\Gamma'(x)/\Gamma(x)$.
} maximal chaos exponent
\beq
\frac{\Fc(t)}{G_{conf}(\pi) G_{conf}(\pi)} =
-p_h \frac{(\beta J)^{2h-2}}{N \alpha^S_{2h} m_h} 
 \exp \l \frac{2 \pi t}{\beta} \r + [\text{non-increasing}].
\eeq
The fact that the chaos exponent is still maximal should not be too surprising: late time asymptotic
growth $\sim e^{2\pi t/\beta}$ at late times $t \gg \beta$ can be found directly from the conformal solution
\cite{KitaevTalks}(and Section 3.6.1 of \cite{ms}) by computing the OTOC in the real time domain. However, the prefactor
is parametrically smaller: for Schwarzian it is $(\beta J)^1$.
It would be interesting to compute finite $\beta J$ corrections to the Lyapunov exponent and see that they
satisfy the bound proposed in \cite{Zhang:2020jhn}.

\subsection{Time ordered 4-point function and energy-energy correlator}
\label{sec:time_order}
It is also interesting to consider time-ordered 
\beq
\th_1 < \th_2 < \th_3 < \th_4,
\eeq
4-point function. This computation will highlight a certain difference with SYK: in SYK $h=2$ mode(Schwarzian) is 
the energy operator, whereas in our case it is not.

We take the general expression (\ref{eq:4pt_raw}) for the 4-point function and convert it into a contour integral:
\begin{align}
\frac{(\beta J)^{2h-2}}{N \alpha^S_{2h}} \oint_\Cc dn \frac{1}{e^{2\pi i n}-1} \frac{e^{i n (y'-y)}}{g_h(n)}
\left[ \frac{\sin \frac{n x}{2}}{\tan \frac{x}{2}} -
n \cos \frac{n x}{2}   \right]
\left[ \frac{\sin \frac{n x'}{2}}{\tan \frac{x'}{2}} -
n \cos \frac{n x'}{2}   \right].
\end{align}
where the contour $\Cc$ encloses $\pm 2, \pm 3,\dots$.
In the time-ordered case one can actually close the contour at infinity and pick up poles of 
$e^{2\pi i n}-1$ and $g_h(n)$. 
In the usual SYK story, $g_h(n) \propto n^2(n^2-1)$ and the only contributing pole is at $n=0$.
Because of this, the leading contribution to the 4-point function depends only on two variables $\th_1-\th_2$ and
$\th_3-\th_4$($y, y'$ drop out). 
Further taking the OPE limit $\th_1 \ra \th_3,\ \th_3 \ra \th_4$
will produce the expression which is independent of $\th_i$ at all. This is usually interpreted as follows: the 
OPE limit has produced
the operator $\psi \pr_\tau \psi$ which is just the stress-energy operator $T$. Obviously, the correlator
\beq
\bra T(\th_1) T(\th_3) \ket,
\eeq
should not depend on $\th_1,\th_3$ from the energy conservation. It turns out to be indeed the case. This
allows one to compute the energy-energy correlators in SYK-chain models at any frequency $|\om| \ll J$.

Let us return to our expression with complicated $g_h(n)$. Now we have to take into account poles
of $g_h(n)$ at negative $n$. Because of that, the OPE limit $\th_1 \ra \th_2,\ \th_3 \ra \th_4$
will produce an expression which is $\th_1-\th_3$ \textit{dependent}. We are forced 
to conclude that $\psi \pr_\tau \psi$ is no
longer proportional to energy. So we cannot extract energy-energy correlators easily.

However, we can still do it, but only in the hydrodynamic regime $|\om| \ll 1/\beta \ll J$ \cite{Song_2017}. 
We consider a general time reparametrization (\ref{eq:reparam}) of the 2-point function.
In the hydrodynamic regime we keep only the leading derivative of $\tau(u)$:
\beq
\tau(u) = u + \ep' \beta u + \dots \ .
\eeq
By looking at the form of conformal 2-point function, eq. (\ref{eq:Gconf}),
we see that $\ep'$ simply rescales $\beta$:
\beq
\delta \beta = - \beta^2 \ep' \ .
\eeq
Now we propagate this change into energy:
\beq
\label{eq:ide}
\delta E = - \beta^2 \frac{dE}{d\beta} \ep' \ .
\eeq
This way we identify $\ep'$ and $\delta E$. Hence, knowing the correlators of $\ep'$ we can 
obtain the correlators of energy. Obviously, correlators of $\ep'$ are governed by the non-local action.

One can cross-check this relation. First, from explicitly differentiating the partition function one has
\beq
\bra (\delta E)^2 \ket = - \frac{dE}{d \beta}.
\eeq
This implies
\beq
\bra \ep' \ep' \ket =  \frac{4 \pi^2 n^2}{\beta^2} \bra \ep_n \ep_{-n} \ket = 
- \frac{1}{\beta^4 \frac{dE}{d \beta}} .
\eeq

Now, using the explicit $\ep$ propagator (\ref{eq:snonloc_n}) and the expression for the energy (\ref{eq:pred})
one can indeed verify the above relation. 
The identification between $\ep'$ and $\delta E$ will be very useful when we discuss the chain model.

\subsection{Diffusion in chain models}
\label{sec:conduct}

We can also study a chain build from our model and study transport properties. The model we will consider
is very similar to the ones discussed in the literature \cite{Gu2017Local, Song_2017, Khveshchenko:2017mvj, Khveshchenko:2020rai}. It was previously shown that
in SYK chain models thermal conductivity and electrical resistivity are linear in the temperature, similar to strange metals.
Also, when the Schwarzian dominates, the diffusion constant is temperature-independent.
In this Section we will show that once the non-local action becomes dominant, the thermal conductivity is 
still linear in the temperature. However, the diffusion constant becomes temperature dependent.

\begin{figure}[!ht]
\centering
\includegraphics[scale=1.6]{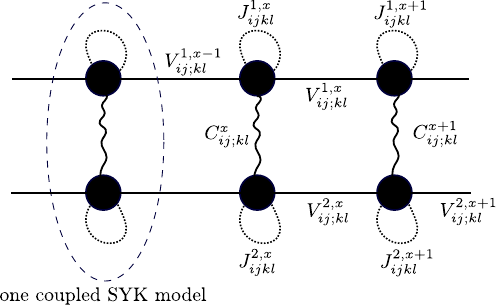}
\caption{Illustration of couplings in the chain model. }
\label{fig:chain}
\end{figure}

The model we consider is simply 1D array of independent dots:
\beq
\mathcal{L}_{T, chain} = \sum_x \Lc_{T,x},
\eeq
where for each $x$ Lagrangian $\Lc_{T,x}$ is given by eq. (\ref{eq:L_T}).
After integrating out the disorder we get a bunch of non-interacting models (\ref{eq:s_gsi}).
To make them interact, we add a tight-binding(in $x$) 
random interaction:
\beq
\label{eq:chain_int}
\Lc_{int, chain} = \frac{1}{2!^2} \sum_{x,ijkl} \l V^{1,x}_{ij;kl} 
 \psi^1_{i,x} \psi^1_{j,x} \psi^1_{k,x+1} \psi^1_{l,x+1} +
V^{2,x}_{ij;kl} \psi^2_{i,x} \psi^2_{j,x} \psi^2_{k,x+1} \psi^2_{l,x+1} \r,
\eeq
where each $V^{1/2,x}_{ij;kl}$ is skew symmetric in $ij$ and $kl$:
\beq
V^{1/2,x}_{ij;kl} = - V^{1/2,x}_{ji;kl} = -V^{1/2,x}_{ij;lk},
\eeq
but do not mix the two pairs $ij$ and $kl$. The full configuration is illustrated in Figure \ref{fig:chain}.
Assuming $x-$independent variance
\beq
\bra \l V^{1/2,x}_{ij;kl} \r^2 \ket = \frac{V^2}{N^3},
\eeq
we get the following extra terms in the $G \Sigma$-action:
\beq
\Delta S_{G \Sigma} = \frac{V^2}{8} \int du_1 du_2 \sum_x \l G_{11,x}(u_1,u_2)^2 G_{11,x+1}(u_1,u_2)^2 + 
G_{22,x}(u_1,u_2)^2 G_{22,x+1}(u_1,u_2)^2 \r.
\eeq
We have chosen the interaction term such that it does not lead to mixed $G_{12}$ correlators, which might cause 
instability. As usual, choosing $x-$independent ansatz for SD equations
\beq
G_{x,11} = G_{11}, \quad G_{x,22} = G_{22},
\eeq
we arrive at single-site SD equations (\ref{sd:eucl}) with effective $\Jtt$ and $\altt$:
\beq
\Jtt^2 = J^2 + V^2,\quad
\altt = \al \frac{J^2}{J^2+V^2}.
\eeq

Now can discuss the kernel and the effective action for reparametrizations.
We will assume that the reader is familiar with SYK 4-point function computation via ladder diagrams.
This computation is pedagogically reviewed in Section \ref{sec:review} of this paper.
We have two types of interactions in the chain model: on-site and between-site next-neighbor 
interaction, eq. (\ref{eq:chain_int}). 
Performing a Fourier transform in the $x$ space makes the ladder diagrams(and the kernel) depend on momentum $p$.
On-site interaction would produce $p$-independent part. Next-neighbor interaction will yield $\cos(p)$ dependence. 
So that the total kernel is
\beq
K_{chain} = K_{ren} + K_p,
\eeq
$K_{ren}$ is simply single-site kernel (\ref{eq:2Kernel}), but with renormalized $J,\alpha$: 
\begin{align}
    K_{ren} v =  \begin{pmatrix}
3 \Jtt^2 G_{11} * \l \l G_{11}^2 + \altt^2 G_{22}^2 \r v_{1} + \
 2 \altt^2 G_{11} G_{22} v_{2} \r * G_{11}  \\
3 \Jtt^2 G_{22} * \l \l G_{22}^2 + \altt^2 G_{11}^2 \r v_{2} + \
 2 \altt^2 G_{11} G_{22} v_{1} \r * G_{22}  \\
    \end{pmatrix}
\end{align}
The remaining part is proportional to $\cos(p)-1$:
\beq
K_p v  = 2 V^2 (\cos(p)-1) 
\begin{pmatrix}
G_{11} * (G_{11}^2 v_{1}) * G_{11} & 0 \\
0 & G_{22} * (G_{22}^2 v_{2}) * G_{22} \\
\end{pmatrix}
\eeq
Note that $K_p$ is not proportional to $K_{ren}$. So in general it would be hard to find eigenvalues
of $K_{chain}$. Fortunately, $K_p$ has the form of SYK kernel and in the 
leading conformal approximation $G_{11}, G_{22}$ are proportional to the 
standard SYK conformal solution. 
So reparametrizations of $G$ again produce the kernel eigenvector with the eigenvalue close to $1$.
In the small $p$ limit this is enough for us, since this term is proportional to $1-\cos(p) \approx p^2/2$. 

Kernel $K_{ren}$ is analysed in Section \ref{sec:kernel} in detail. 
The upshot is that at $|\altt|>1$ its eigenvalue
shift is controlled by the non-local action instead of the Schwarzian. Note that the conformal dimensions now
are controlled by $\altt$, not $\al$.
Putting this together we learn that the leading(in $1/\beta \Jtt$ and $p$)
eigenvalue shift for reparametrizations is
\beq
1-k(2,n,p) = \frac{\al^K_{2h}}{(\beta \Jtt)^{2h-2}} \frac{g_h(n)}{|n|(n^2-1)} + 
p^2 \frac{V^2}{3 \Jtt^2(1+3 \altt^2)}.
\eeq
So that the action for infinitesimal reparametrizations is given by \footnote{The overall factor can be
determined from requiring that $p=0$ reparametrizations reproduce the answer for a single copy of our model, see 
Sections \ref{sec:review}, \ref{sec:shift}
of this paper for a detailed discussion.}
\beq
S =  \frac{\pi^4 b^4 N}{4}\sum_{n,p} \ep_{n,p} \l \frac{\al^K_{2h}}{(\beta \Jtt)^{2h-2}} g_h(n) + 
p^2 |n| (n^2-1) \frac{V^2}{3 \Jtt^2(1+3 \altt^2)} \r \ep_{-n,-p}.
\eeq
One final step is to switch to Minkowski space and consider the limit of small time frequencies. Recall the analytic
expression  eq. (\ref{eq:gh}) for $g_h(n)$. We need to continue $i n \ra \frac{\beta \om }{2 \pi}$ and consider the 
limit $\om \ra 0$ with fixed $\beta$. This way only the factor $n^2$ in $g_h(n)$ gives a finite contribution.
Putting everything together we get:
\beq
S_{hydro} =  \frac{\pi^4 b^4 N}{4} \frac{\beta}{2 \pi} \int d\om dp \ \ep_{\om,p} \l \frac{\al^K_{2h}}{(\beta \Jtt)^{2h-2}} 
\frac{(2h-1) \Gamma(h)}{\Gamma(2-h)} \frac{\om^2 \beta^2}{4 \pi^2}  +
i p^2 \frac{\om \beta}{2 \pi} \frac{V^2}{3 \Jtt^2(1+3 \altt^2)} \r \ep_{-\om,-p}.
\eeq
Similarly to a single-site case, we identify $\pr_t \ep_x$ with energy at site $x$, eq. (\ref{eq:ide}).
We see a typical diffusion pole $\om + i D p^2$ in the energy-energy correlator\footnote{Note that $p$ is
dimensionless in our conventions.}. The diffusion constant is given by
\beq
\label{eq:D}
D =  T^{3-2h} \frac{2 \pi \Gamma(2-h)}{3(2h-1) \Gamma(h)} \frac{V^2 \tilde{J}^{2h-2}}{\alpha^K_{2h} 
\tilde{J}^2(1+3\tilde{\alpha}^2)}.
\eeq
A few comments are in order.

The important part is the temperature dependence $T^{3-2h}$. Schwarzian yields temperature-independent
diffusion constant \cite{Gu2017Local,Song_2017}. Here, the dimension $h$ of the irrelevant operator controls
the temperature power. 

Using the identification (\ref{eq:ide}) between the energy and reparametrizations, one can 
also compute the energy-energy correlator and extract the thermal conductivity. It is proportional
to the specific heat, eq. (\ref{eq:E_T}): 
\beq
c_{v} = N T^{2h-2}(2h-2)(2h-1) \frac{\alpha_{2h}^S \pi^{2h-1/2}}{\tilde{J}^{2h-2}} 
\frac{\Gamma \l 1/2-h \r}{\Gamma \l 1 - h \r },
\eeq
times the diffusion constant, eq. (\ref{eq:D}).
Thus it is linear in the temperature as in
the chains where Schwarzian dominate:
\beq
\kappa = c_v D \propto N T.
\eeq

In the conventional SYK chain \cite{Gu2017Local} the butterfly velocity is 
\beq
v_B^2 = 2 \pi D T,
\label{eq:vb}
\eeq
which 
agrees with the holographic expectations \cite{Blake:2016wvh}. However, careful computation of the butterfly
velocity requires the knowledge of the subleading correction to the 4-point function \cite{Gu2017Local}.
In our case this seems complicated because the two kernels $K_{ren}$ and $K_p$ are not the same.
Nonetheless, we conjecture that in our model the relation (\ref{eq:vb}) still holds. 
Physically it is motivated by the fact that the same mode(reparametrizations) governs both OTOC
chaos exponent and energy diffusion. This happens in the conventional SYK chains too.
On the computational level, ignoring the subleading correction to 4-point function results in picking the pole
at $\om + i D p^2 = 0$, leading to eq. (\ref{eq:vb}).

Unfortunately, we cannot study electric conductivity in our model because we do not have $U(1)$ symmetry.
There are two ways to introduce $U(1)$ symmetry. We can simply promote the Majorana fermions to complex fermions.
In this case one has to study operators in the symmetric sector which is not related to the
Schwarzian. For example, in complex SYK \cite{KitaevRecent} fluctuations in $U(1)$ phase are governed by simple
$U(1)$-sigma model. However, there are also t-J models where resistivity is related to time 
reparametrization mode \cite{Guo_2020}. It would be interesting to see how the change from Schwarzian to non-local
action affects the transport in t-J models. 
We leave this question for future work. 

\subsection{$N^0$ correction}
\label{sec:density}
In this Section we will compute $N^0$ correction to the free energy and try to infer the density of states $\rho(E)$
near the ground state. We would like to emphasize that in this Section we will be interested in the energies 
close to ground state: $|E-E_0| \sim N^0$. Whereas the thermodynamic results in eq. (\ref{eq:pred})
corresponded to $|E-E_0| \sim N^1$. Also we will discuss below the validity of our $N^0$ computation.

We can easily compute $N^0$ correction to the free energy.
Again, we will need some knowledge about the kernel eigenvalues so we refer to Sections \ref{sec:review} and \ref{sec:shift} 
for the detailed discussion.
The $N^0$ correction is given by the determinant of fluctuations around the
thermal solution. Since reparametrizations are enhanced we expect that they will dominate in the determinant too.
It can be argued diagramatically \cite{PS} and by path-integral techniques \cite{ms} that $N^0$ correction to  
$\log Z$ is given the sum of kernel eigenvalues
\beq
- \frac{1}{2} \sum_{h,n} \log \l 1-k(h,n) \r.
\eeq
Obviously, the leading contribution will come from $k$ close to $1$, which are exactly reparametrizations.

As usual, let us start from the standard SYK case, where the Schwarzian dominates. 
Then the eigenvalue shift is proportional to, eq. (\ref{review:k}):
\beq
\l 1 - k(2,n) \r_{Sch} \propto \frac{|n|}{\beta J}, \quad |n| \ge 2.
\eeq
The sum over $n$ has to be cut at $n \sim \beta J$. This produces the following answer:
\beq
\label{z:sch}
\l \log Z \r_{Sch,1-loop} \propto \# \beta  -\frac{3}{2} \log \l \beta J \r.
\eeq
The term proportional to $\beta$ has unknown coefficient, but it simply gives the shift to the ground state energy.

Let us discuss the non-local action now.
From the linearized action (\ref{eq:snonloc_n}), 
we expect the eigenvalue shift to be, eq. (\ref{xi:k}):
\beq
\l 1 - k(2,n) \r_{nonloc} \propto \frac{1}{(\beta J)^{2h-2}}\frac{g_h(n)}{|n|(n^2-1)}.
\eeq
Recall that $g_h(n)$ is given by eq. (\ref{eq:gh}). This sum is harder to evaluate. It can be simplified by
noticing that one can separate the first term in the parenthesis:
\beq
\label{eq:gh2}
g_h(n) = n^2 \frac{\Gamma(n+h)}{\Gamma(1+n-h)} \eta_h(n),
\eeq
with
\beq
\eta_h(n) = \l  1 - \frac{\Gamma(h-1) \Gamma(1+n-h)}{\Gamma(-h)\Gamma(n+h)} \r.
\eeq
We see that at large $n$, $\eta_h \propto 1-1/n^{2h-1}$ hence the sum $\sum_{|n| \ge 2} \log(\eta_h)$ actually converges and gives 
something of order $(\beta J)^0$. We are not interested in this contribution.
Now it is possible to evaluate the sum $\sum_{2 \le |n| \le \beta J} \log \l g_h(n)/(|n|(n^2-1)\eta_h(n)) \r$. The answer is
\beq
\label{z:nonloc}
\l \log Z \r_{nonloc,1-loop} \propto \# \beta -\frac{3}{2} (2h-2) \log \l \beta J \r.
\eeq

We can try to convert this into the energy density by doing the inverse Laplace transform:
\beq
\rho(E) = \int d\beta \ Z(\beta) e^{-\beta E}.
\eeq
In this equation the energy $E$ is measured from the ground state and it includes a factor of $N$.
The most interesting regime, which actually can be probed with exact diagonalization, is the
vicinity of the ground state,
$E \sim N^0 J$. However, one has to be extremely careful with the range of validity of (\ref{z:sch}) and (\ref{z:nonloc}).
In this regime the above integral is dominated by $\beta \sim N/J$(in the Schwarzian case) and by 
$\beta \sim  N^{1/(2h-2)}/J$(in the non-local case)  and we cannot trust the above 1-loop computation anymore.

Fortunately for the Schwarzian, it is 1-loop exact \cite{Stanford:2017thb}, so we can actually trust eq. (\ref{z:sch}) and
obtain square-root edge:
\beq
\label{e:sch}
\rho(E)_{Sch,1-loop} = \rho(E)_{Sch,exact} \propto \sqrt{E}.
\eeq

Unfortunately, we do not know if the non-local action has the same property. Naively using 1-loop result (\ref{z:nonloc})
we get
\beq
\label{e:nonloc}
\rho(E)_{nonloc,1-loop} \propto E^{3h-4}.
\eeq
\textbf{In fact, our exact diagonalization results at finite $N$ do not do not support this result.}
This suggests that the non-local action partition function is not 1-loop exact. We will discuss this more
in Section \ref{sec:ED}.

\section{Non-local action from 4-point function}
\label{sec:derivation}
This Section is entirely devoted to numerical, but \textit{ab initio} 4-point function computation in 
our coupled model.
One can look at this computation as the derivation of the non-local action.
We will start by pedagogically reviewing the same computation in SYK.
Then in Section \ref{sec:spectral} we discuss the leading non-conformal correction to 2-point functions.
This Section can be read independently and the results are interesting on their own.

Schematically, the plan in this. In order to compute the 4-point function and identify the corresponding
dominant low energy mode we need to:
\begin{itemize}
\item Write down ladder diagrams and find the corresponding kernel.
\item Find the leading non-conformal correction to the 2-point function.
\item Compute the kernel eigenvalue shift coming from this correction.
\item Interpret the answer as an integral over reparametrizations with some action.
\end{itemize}
\subsection{Review of Schwarzian derivation}
\label{sec:review}
Let us recall how Maldacena--Stanford(MS) \cite{ms} argued that in SYK the low-energy physics
 is dominated by the Schwarzian action.
Standard SYK has the following Hamiltonian:
\beq
H_{SYK} = \frac{1}{4!} \sum_{ijkl} J_{ijkl} \psi_i \psi_j \psi_k \psi_l, \ \bra J_{ijkl}^2 \ket = 
\frac{3! J^2}{N^3}.
\eeq
Summation of melonic diagrams lead to (Euclidean) Schwinger--Dyson equations for the two-point function
 $G(u_1 - u_2) \equiv G(12) = \bra T \psi_i(u_1) \psi_i(u_2) \ket$:
\beq
(-i \om_n - \Sigma(\om_n)) G(\om_n) = 1,\ \om_n = \frac{2\pi}{\beta} \l n + \frac{1}{2} \r,
\eeq
\beq
\Sigma(u) = J^2 G(u)^3.
\eeq
In the strict large $N$ these equations are exact.
Compared to the rest of the paper, here the conformal solution differs by a factor of $\sqrt{1+3\alpha^2}$:
\beq
G_{conf} = b \sgn(u) \frac{\pi}{\sqrt{J \beta \sin\l \frac{\pi |u|}{\beta} \r }} ,
\ 1/J \ll |u|,\ \beta J \gg 1.
\eeq

The (connected) 4-point function is given by the sum of ladder diagrams - Figure \ref{fig:ladder}:
\begin{figure}[!ht]
\centering
\includegraphics{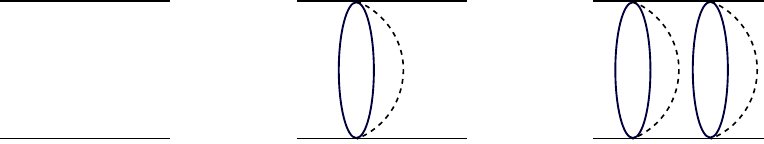}
\caption{First few ladder diagrams contributing to the (connected) 4-point function. Solid lines are fermionic
propagators and dashed lines indicate disorder contractions.}
\label{fig:ladder}
\end{figure}
\beq
\label{eq:4pt_full}
\bra \psi_i(\th_1) \psi_i(\th_2) \psi_j(\th_3) \psi_j(\th_4) \ket = G(12) G(34) + \frac{1}{N} \Fc,
\eeq
\beq
\bra \psi_i(\th_1) \psi_i(\th_2) \psi_j(\th_3) \psi_j(\th_4)  \ket_{conn} = 
\frac{1}{N} \Fc(\th_1,\dots,\th_4) = \frac{1}{N} \frac{1}{1-K} \Fc_0,
\eeq
where $\Fc_0$ is the leading (connected) 4-point function:
\beq
\Fc_0(12;34) = -G(13) G(24) + G(14) G(23) .
\eeq
Explicitly the kernel $K=K(12;34)$ is given by
\beq
K(12;34) = - 3J^2 G(13) G(24) G(34)^2 .
\label{eq:syk_kernel}
\eeq
The kernel acts by convolution with the last two ($34$) variables. 
For example, the next-to-leading correction $\Fc_1$ is
\beq
\Fc_1(12;34) = K \Fc_0 = \int d3' d4' K(12;3'4') \left[ - G(3'3) G(4'4) + G(3' 4) G(4' 3) \right].
\eeq
A natural thing to do is to expand $\Fc_0$ in the basis of eigenfunctions of $K$. Schematically:
\beq
\label{eq:decomposition}
\Fc = \sum_k \frac{1}{1-k} \frac{\bra \Psi_k | \Fc_0 \ket}{\bra \Psi_k | \Psi_k \ket} \Psi_k.
\eeq 
It turns out, in the conformal limit $K$ has a set of eigenvalues equal to one: $k(2,n)=1$. 
The corresponding eigenfunctions
are proportional to reparametrizations of $G$. The convenient basis of reparametrizations is
\beq
u \ra u +  \beta \sum_n \ep_n e^{- 2\pi i n u/\beta}.
\eeq
They give enhanced contribution to the 4-point function
\beq
\label{eq:4pt}
\text{(4-pt)} \supset  \frac{1}{N} \frac{2}{\pi^4 b^4} \sum_n 
\frac{1}{|n|(n^2-1)} \frac{k(2,n)}{1-k(2,n)} \delta_{\ep_n} G \ \delta_{\ep_{-n}} G,
\eeq
with
\beq
\delta_{\ep_{n}} G(u_1-u_2) = \pi G(u_1-u_2) \left[ \frac{\sin \frac{n \pi u}{\beta}}{\tan \frac{\pi u}{\beta}} -
n \cos \frac{n \pi u}{\beta}  \right] e^{i \pi n(u_1+u_2)/\beta}.
\eeq
The factor $2 k(2,n)/(b^4)$ came from the overlap between the $\Fc_0$ and the kernel eigenvector
and $\pi^4 |n|(n^2-1)$ came from the normalization of $\delta_{\ep_n} G$.
The leading order answer can be obtained by taking the conformal answer everywhere except in $1-k(2,n)$. 
The difference $1-k(2,n)$ is determined by the leading non-conformal correction $\delta G$ to $G$. 
By analysing the large $q$ limit MS argued that the leading
correction goes as $1/\beta J$:
\beq
\label{eq:leading}
\frac{\delta G}{G_{conf}} = - \frac{\alpha^G_{Sch}}{\beta J} f_0.
\eeq
Function $f_0$ is given by
\beq
f_0 = 2 + \frac{\pi - |\th|}{\tan \frac{|\th|}{2}}, \quad \theta=2\pi u/\beta.
\eeq
In the next sub-Section we will describe how to find these corrections in a systematic way.

From now on it is straightforward(but tedious) to find $1-k(2,n)$. MS did it analytically and found: 
\beq
\label{review:k}
k(2,n) = 1 - \frac{\alpha^K_{Sch} |n|}{\beta J}.
\eeq
Hence,
\beq
\label{eq:4pt_final}
\text{(4-pt)} \supset  \frac{1}{N} \frac{2\beta J}{\alpha^K_{Sch} \pi^4 b^4} \
 \sum_n \frac{1}{n^2(n^2-1)}\delta_{\ep_n} G \ \delta_{\ep_{-n}} G.
\eeq
This answer can be understood as follows. We start from the leading (disconnected) contribution to the 4-point function,
eq. (\ref{eq:4pt_full}):
\beq
G(u_1-u_2) G(u_3 - u_4),
\eeq
where both Green functions are taken at finite (inverse) temperature $\beta$.
Then we dress them with an infinitesimal reparametrization $u \ra u + \beta \ep(u)$ and then 
average over $\ep$ with the action
\beq
\label{eq:sch_lin}
S_{Sch} =  \frac{N}{\beta J} 8 \pi^4 \alpha^S_{Sch} \sum_n \ep_n n^2(n^2-1) \ep_{-n} ,\quad
\alpha^S_{Sch} = \frac{\alpha^K_{Sch} b^4 \pi^4}{32}.
\eeq
Now, if we take the Schwarzian action
\beq
S = -\frac{N \alpha^S_{Sch}}{J} \int du \Sch(\tau[u],u),
\eeq
and expand it near the thermal solution:
\beq
\tau(u) = \tan \l \frac{\pi}{\beta} \l u + \beta \ep(u) \r \r,
\eeq
we get exactly the action (\ref{eq:sch_lin}).
Hence the Schwarzian reproduces the correct answer for the 4-point function. 
Notice that it has the correct $n$ dependence and 
correct $\beta J$ dependence.

\subsection{Correction to conformal solution}
\label{sec:spectral}
In the previous Section we promised to present a general approach for computing corrections to the conformal 2-point 
function. This approach is nothing more than a simple conformal perturbation theory associated with the 
deformation (\ref{eq:irrels}). For example \cite{Tikhanovskaya:2020elb, Tikhanovskaya:2020zcw}, the
leading correction from an operator $\Oc_h$ of dimension $h$ is given by the 3-point function: 
\beq
\label{eq:Gleading}
(\delta G)_h(u) = \alpha_h \int du' \ \bra \Oc_h(u') \psi_i(u) \psi_i(0) \ket \propto \frac{1}{(J u)^{h-1/2}}.
\eeq
Answer $f_h = u^{1/2-h}$ is valid either at zero temperature or in the regime $1/J \ll |u| \ll \beta$. 
More generally,
this correction is given by a hypergeometric function.
In principle, one can compute even the second-order correction \cite{Tikhanovskaya:2020elb, Tikhanovskaya:2020zcw}:
\beq
\label{eq:Gsubleading}
(\delta^2 G)_{h_1 h_2}(u) = \alpha_{h_1} \alpha_{h_2} \int du_1 du_2 \ 
\bra \Oc_{h_1}(u_1) \Oc_{h_2}(u_2) \psi_i(u) \psi_i(0) \ket \propto \frac{1}{(J u)^{h_1 + h_2 -3/2}}.
\eeq
Again, the final answer $f_{h_1+h_2} \propto 1/u^{h_1+h_3-3/2}$ is valid for $1/J \ll |u| \ll \beta$. 

It can be shown that the leading correction in SYK(eq. (\ref{eq:leading})) comes from $h=2$ operator.
However, we see right away that any operator with $h<2$(not necessarily $h<3/2$) will dominate over this
$h=2$ correction. We would like to verify this statement in our coupled model.

Given the simplicity of $\beta = \infty$ answers, we will find the exact 2-point function numerically
\textit{at zero temperature in Lorentzian time}. The procedure is described in Appendix \ref{sec:lorentz}.
Specifically, we will examine the spectral density:
\beq
\rho_{11/22} = \Im G_{R,11/22}(\omega),
\eeq
where $G_R$ is retarded 2-point function and $\om$ is real frequency.
From eqns. (\ref{eq:Gleading}) and (\ref{eq:Gsubleading}) $\rho$ has the following expansion:
\beq
\rho_{11/22} \times \sqrt{\omega} = \const +  \sum_h A^{11/22}_h \omega^{h-1} + 
\sum_{h_1,h_2} B^{11/22}_{h_1 h_2} \omega^{h_1 + h_2 -2} + \dots .
\eeq
We have multiplied $\rho$ by $\sqrt{\om}$ because the leading conformal answer goes as $1/\sqrt{\om}$.
We will concentrate on the leading $\om^{h-1}$ correction. We expect it to be $\xi$-odd. 
So it should have a
different sign for $G_{11}$ and $G_{22}$.
Our strategy is the following: find $\rho_{11/22} \sqrt{\om}$ at a given $J$ and $\alpha$ and perform the fit with
\beq
b_1+ b_2 \om^{h-1},
\eeq
with unknown $b_1,b_2$ in two ways: put $h=h_{theor}(\al)$ obtained from eq. (\ref{eq:dimension}) or allow $h$ to be inferred from the data.
In other words, perform the fit with unknown $h$ and obtain $h_{best}$.
The uncertainty in $h_{best}$ arises from changing the fitting interval in $\om$.
One important thing to notice is that $h_{best}$ tends to overestimate $h$ by about $0.05$ 
because we have omitted the subleading corrections.
\begin{itemize}
\item Schwarzian benchmark: $\alpha=0$

Here for $\alpha=0$ the two systems decouple and for any $\xi$ we have two independent SYK.
The result for $\alpha=\xi=0$ is presented in Figure \ref{fig:alpha0}. We indeed see that for small $\omega$ there is a linear
term coming from $h=2$ mode:
\beq
\rho_{11/22} \sqrt{\omega} = b_1+ b_2 \omega.
\eeq
\begin{figure}[!ht]
\centering
\includegraphics[scale=0.5]{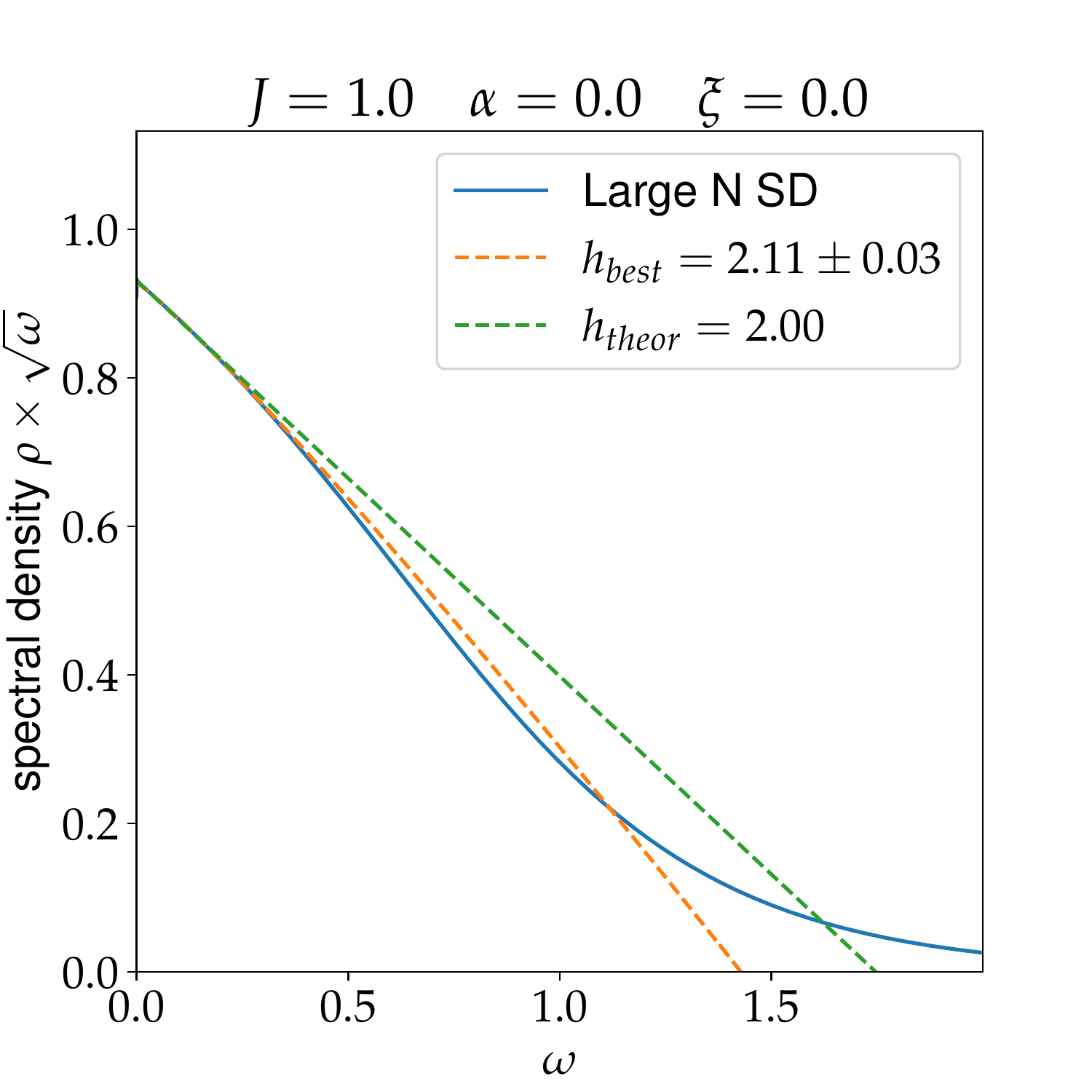}
\caption{Spectral density for original SYK. For comparison, we fitted using the theoretical value $h_{\rm theor}=2$ and
arbitrary $h$. The fit was performed with
$b_1+b_2 \omega^{h-1} $.} 
\label{fig:alpha0}
\end{figure}
\item $|\alpha|>1$: Here operator $\Oc_{2,0}$, eq. (\ref{eq:theoperator}), 
has the dimension in the interval $1<h<3/2$, as can be
easily seen from eq. (\ref{eq:dimension}).
The results for $G_{22}$ are presented in Figure \ref{fig:alpha_b1}. To make the graphs more 
expressive we have taken rather large $\xi=0.9$.
\begin{figure}[!ht]
\centering
\minipage{0.47\textwidth}
\includegraphics[scale=0.5]{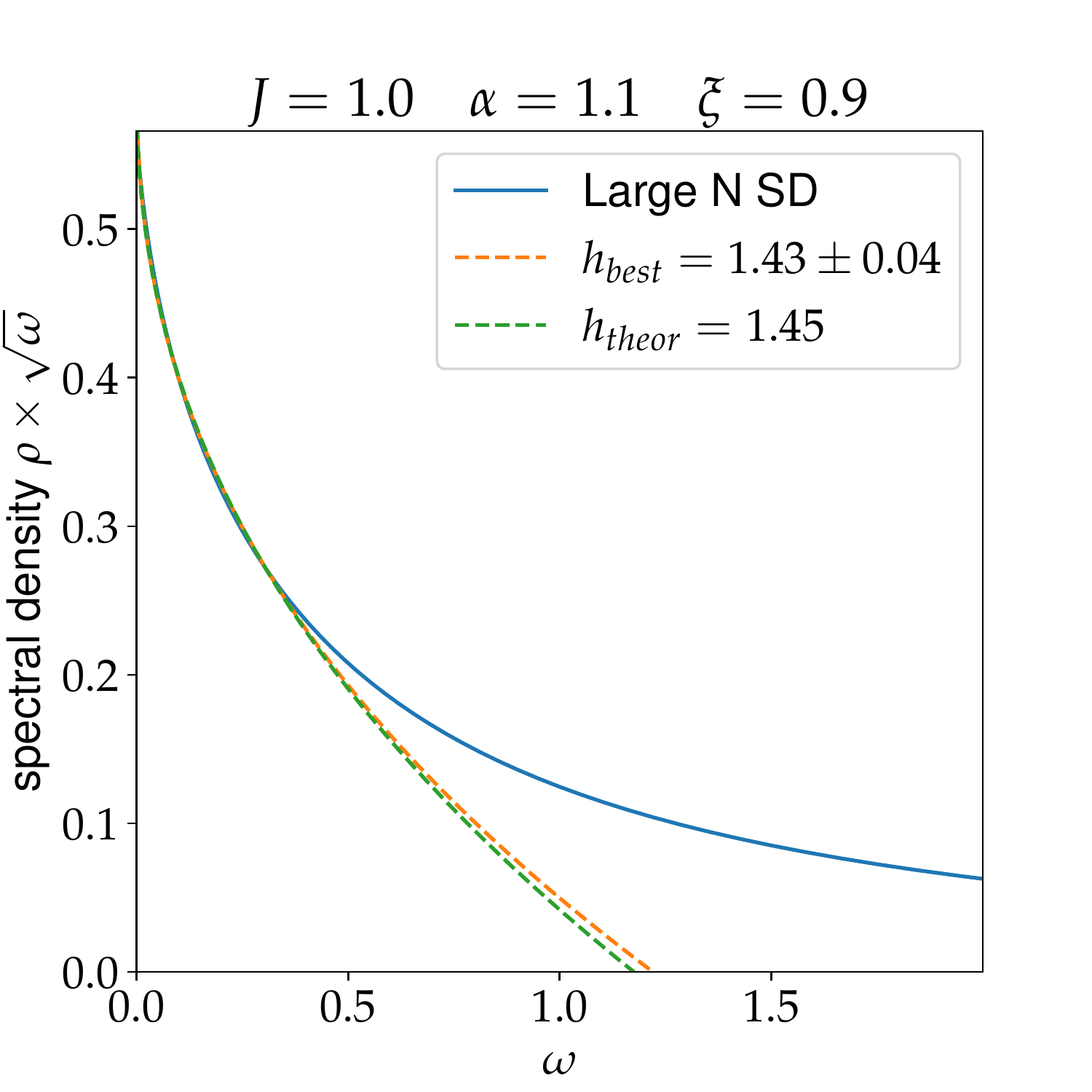}
\endminipage
\minipage{0.47\textwidth}
\includegraphics[scale=0.5]{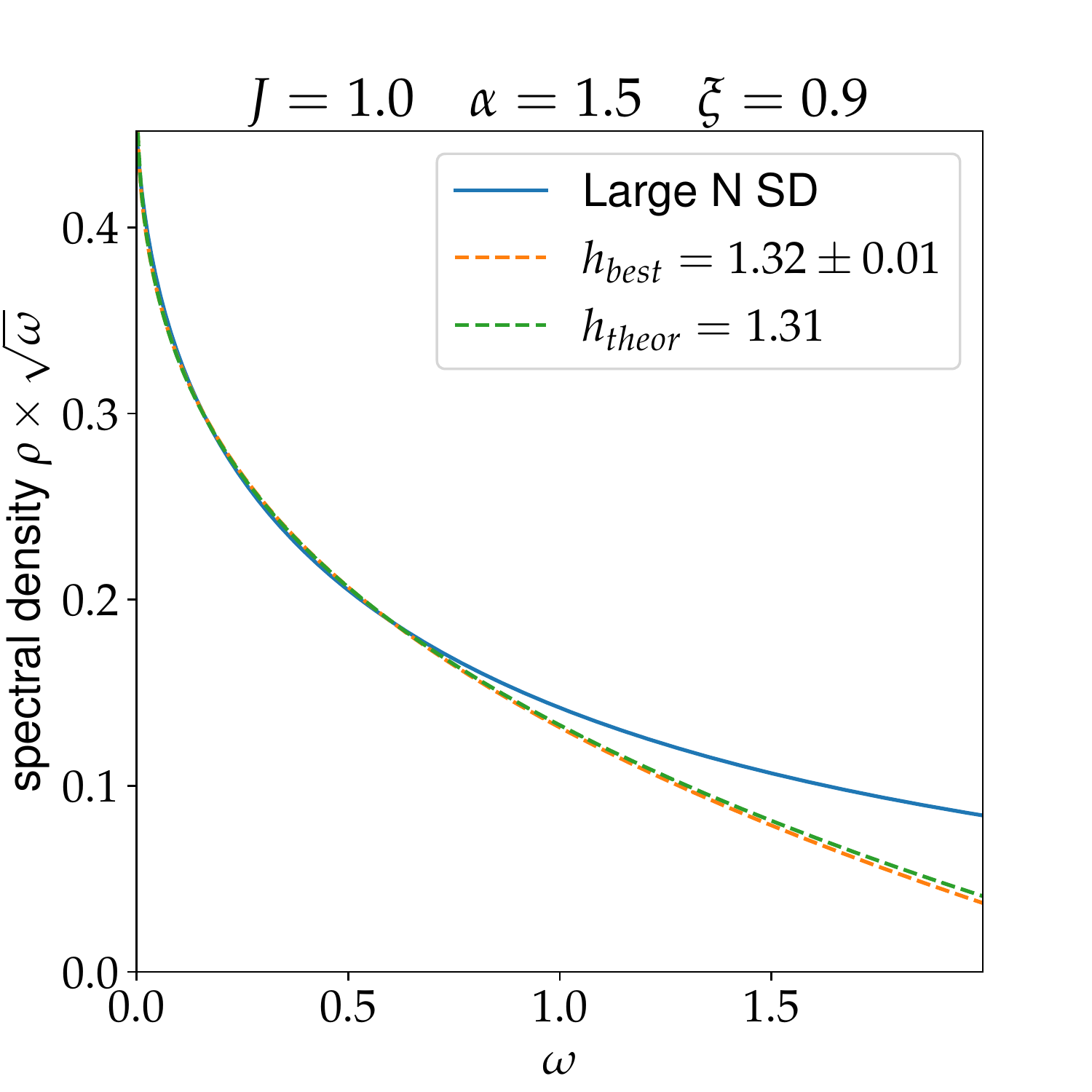}
\endminipage
\caption{Results for  $G_{22}$, $\alpha>1$. The fit was performed with $b_1+ b_2 \omega^{h-1}$.}
\label{fig:alpha_b1}
\end{figure}
Note that the leading correction has to be $\xi$-odd. Therefore we expect $\rho$ to curve in different directions 
for $G_{11}$. This is indeed the case as can be seen from Figure \ref{fig:alpha_b1_11}.
\begin{figure}[!ht]
\centering
\minipage{0.47\textwidth}
\includegraphics[scale=0.5]{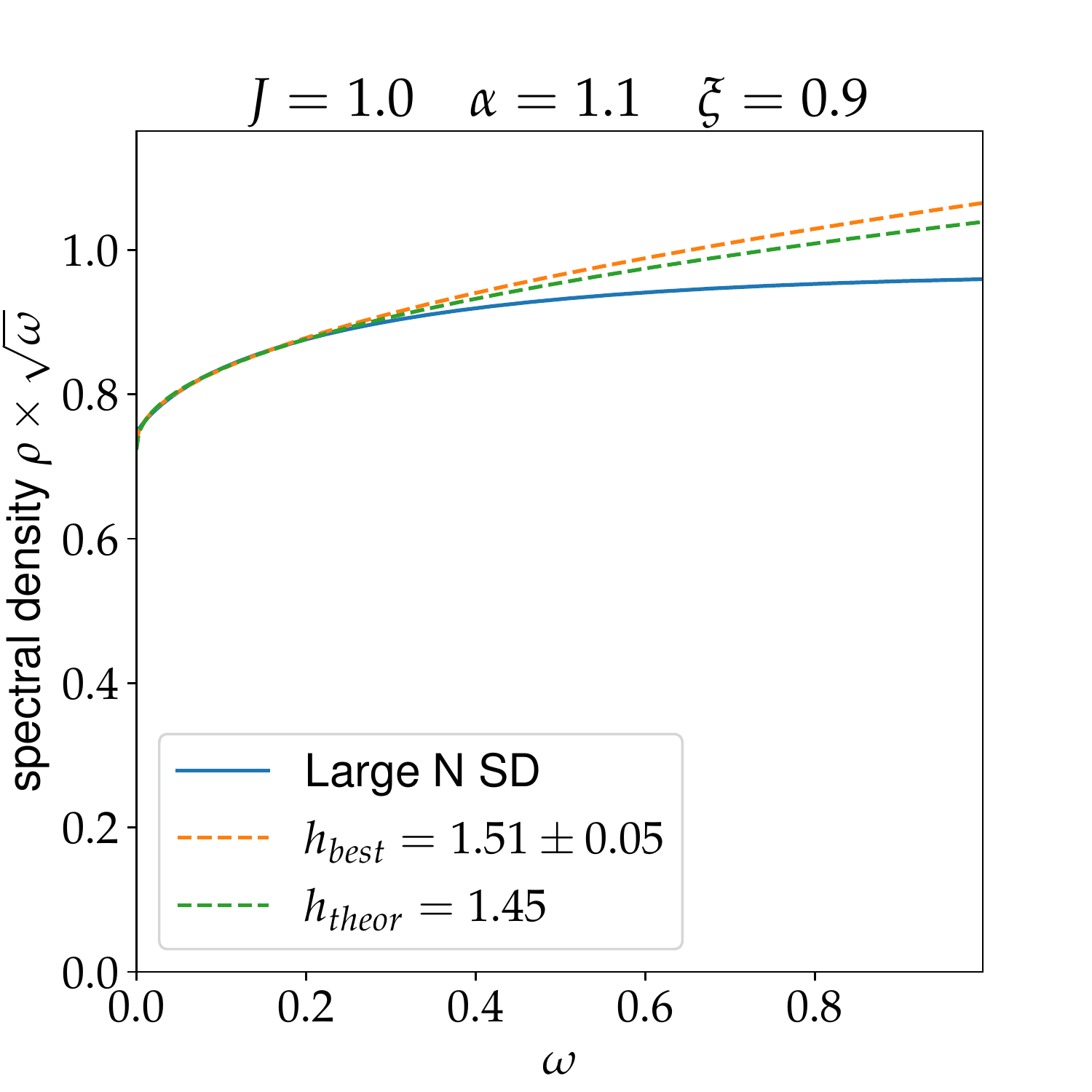}
\endminipage
\minipage{0.47\textwidth}
\includegraphics[scale=0.5]{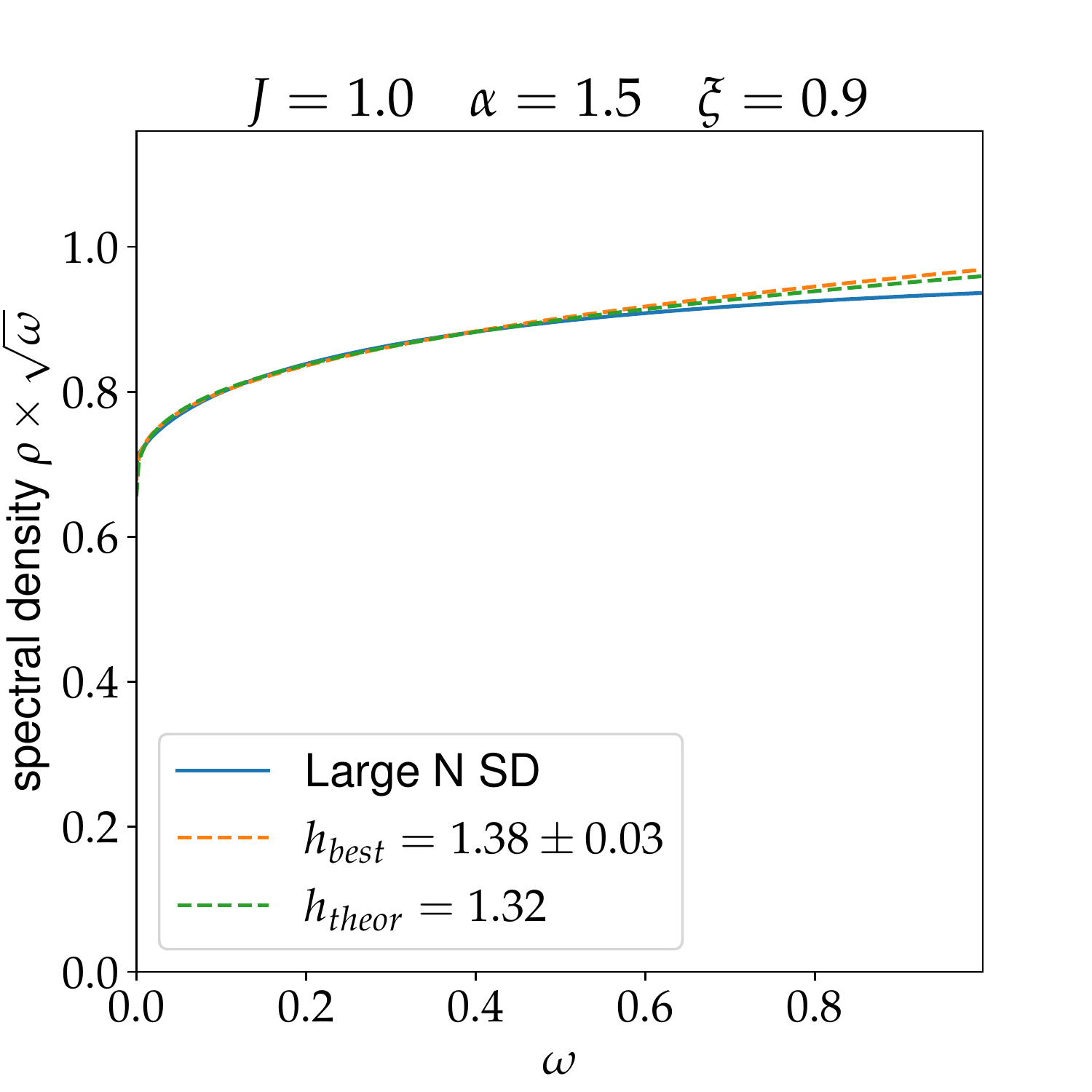}
\endminipage
\caption{Results for  $G_{11}$, $|\alpha|>1$. The fit was performed with $b_1+b_2 \omega^{h-1}$.}
\label{fig:alpha_b1_11}
\end{figure}
\item $|\alpha|<1$: Now $\Oc_{2,0}$ has the dimension $3/2<h<2$. 
In this interval we do not expect the non-local action to dominate in the 4-point function or free energy.
However, it still should dominate in the non-conformal correction.
The results are presented in Figure \ref{fig:alpha_m1}. 
Again, to make the graphs more expressive we took rather large $\xi=0.9$.
\begin{figure}[!ht]
\centering
\minipage{0.47\textwidth}
\includegraphics[scale=0.5]{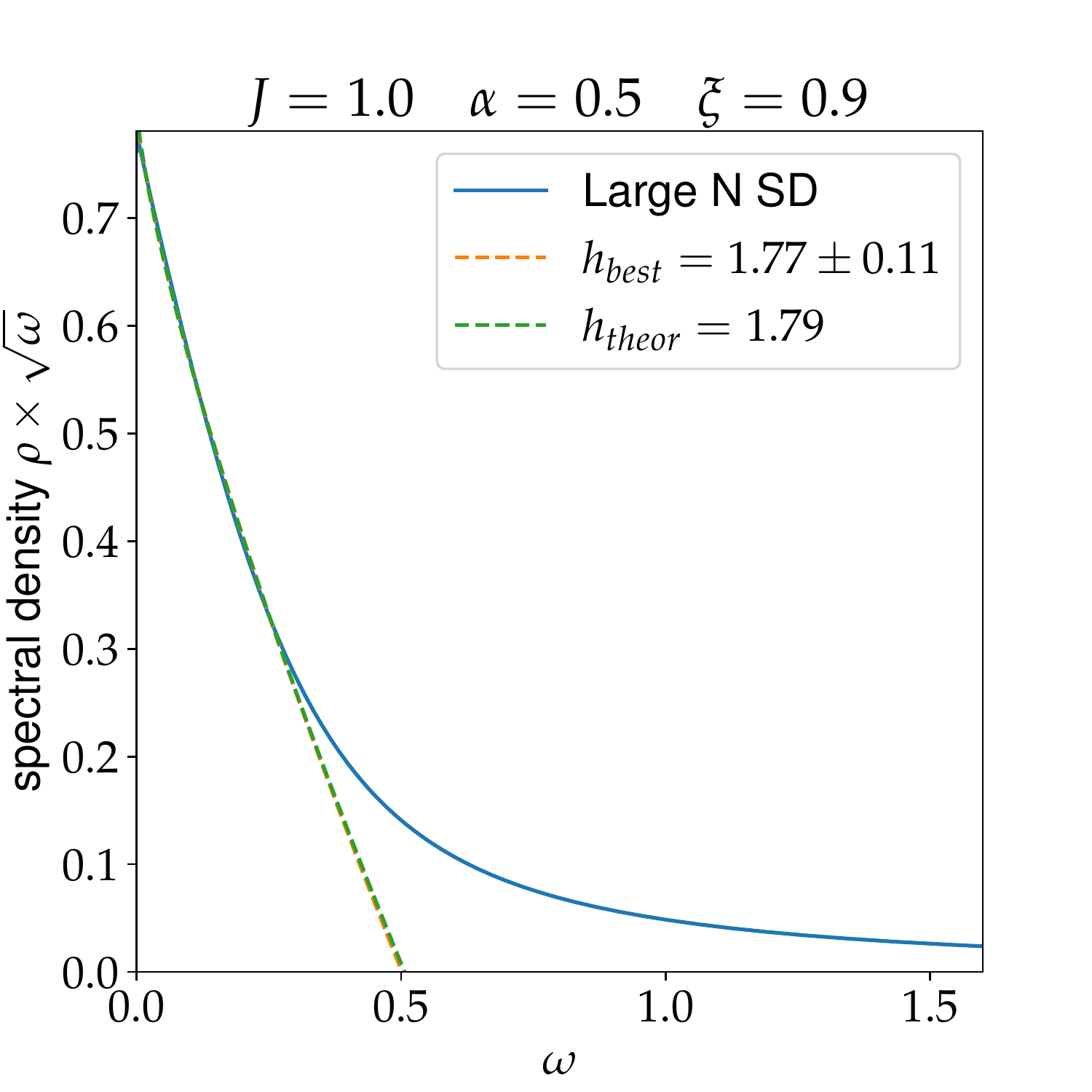}
\endminipage
\minipage{0.47\textwidth}
\includegraphics[scale=0.5]{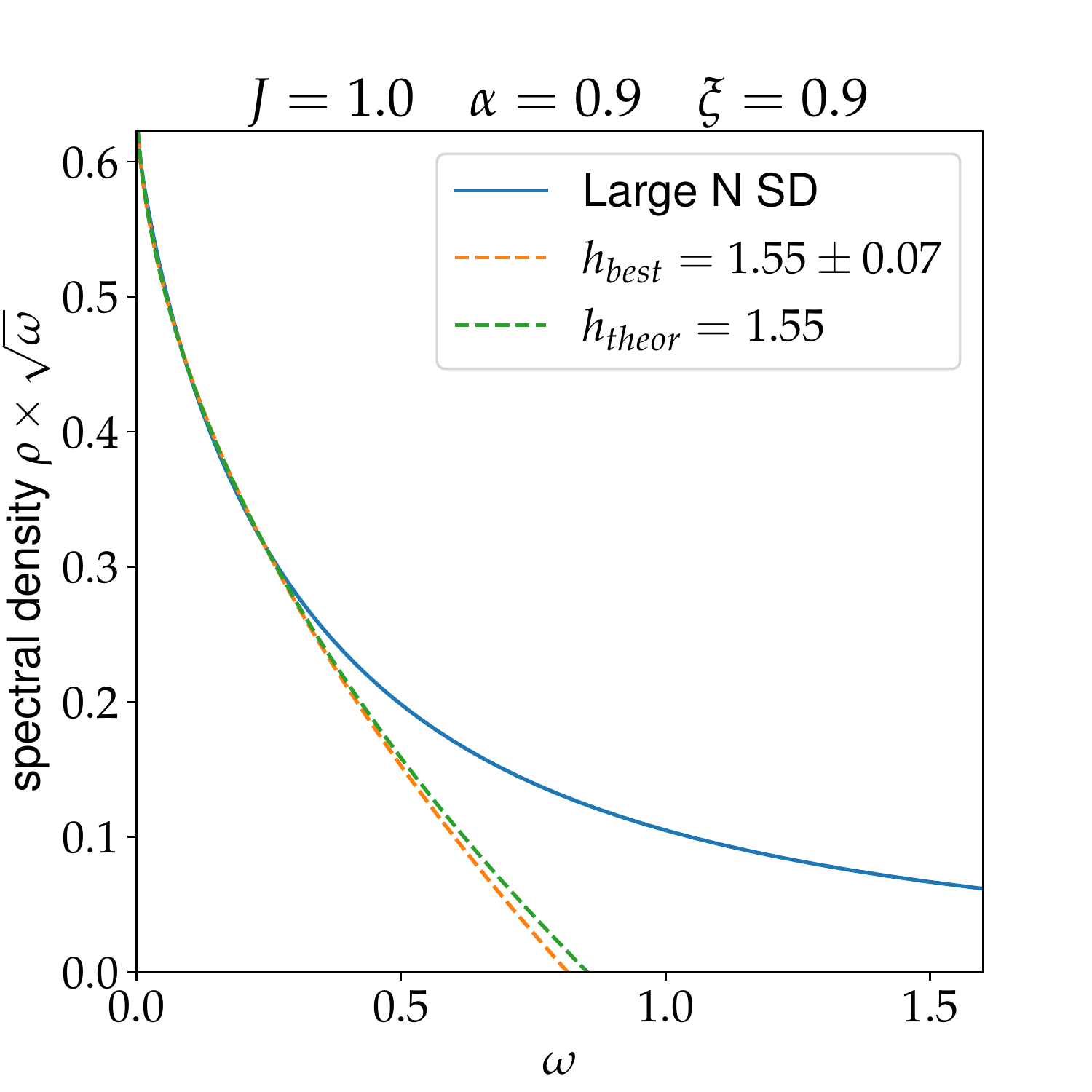}
\endminipage
\caption{Results for $G_{22}$,  $|\alpha|<1$. The fit was performed with $b_1+b_2 \omega^{h-1}$.}
\label{fig:alpha_m1}
\end{figure}

\end{itemize}

\subsection{The kernel}
\label{sec:kernel}
Let us now discuss the kernel for our coupled model. In this Section we put $\beta =2 \pi$.
It is straightforward to draw ladder diagrams.
However, the most convenient way to derive the kernel is to start from  the conformal SD equations:
\begin{align}
 -J^2(G_{11}^3 + 3 \alpha^2 G_{11} G_{22}^2) * G_{11} = \delta(u),  \nonumber \\
 -J^2(G_{22}^3 + 3 \alpha^2 G_{22} G_{11}^2) * G_{22} = \delta(u),
\end{align}
where $*$ means convolution in the Matsubara time domain,
and perturb them by $G \ra G + v$. The equations we obtain this way are
\beq
K v = v,
\eeq
where $K$ is the kernel and 
$v=(v_{1},v_{2})$ is the vector the kernel acts on. 
Explicitly we have\footnote{Possible sign difference(overall plus instead of minus) is related the last $G$ - 
it has two time-arguments exchanged compared to eq. (\ref{eq:syk_kernel}).}
\begin{align}
    K v =  \begin{pmatrix}
3 J^2 G_{11} * \l \l G_{11}^2 + \alpha^2 G_{22}^2 \r v_{1} + \
 2 \alpha^2 G_{11} G_{22} v_{2} \r * G_{11}  \\
3 J^2 G_{22} * \l \l G_{22}^2 + \alpha^2 G_{11}^2 \r v_{2} + \
 2 \alpha^2 G_{11} G_{22} v_{1} \r * G_{22}  \\
    \end{pmatrix}.
\label{eq:2Kernel}
\end{align}
Correspondingly, the kernel is a $2 \times 2$ matrix with four time variables:
\beq
K = K_{ab} \l u_1,u_2 ;u_3,u_4 \r.
\eeq
Four-point function $\Fc_{ab}$ is also a $2 \times 2$ matrix corresponding to 4 different correlators:
\beq
\Fc_{ab} = \begin{pmatrix}
\bra \psi^1 \psi^1 \psi^1 \psi^1 \ket & \bra \psi^1 \psi^1 \psi^2 \psi^2 \ket \\
\bra \psi^2 \psi^2 \psi^1 \psi^1 \ket & 
\bra \psi^2 \psi^2 \psi^2 \psi^2 \ket
\end{pmatrix}.
\label{eq:4pt_matrix}
\eeq

We can write down the expression identical to eq. (\ref{eq:decomposition}), which says
that $\Fc \propto (1-K)^{-1} \Fc_0$. We will use the conformal
solution everywhere except in $1-k$. This way the conformal kernel is proportional to SYK kernel:
\beq
K_{conf,ab}(12;34) = - 3 J^2 G_{conf}(13) G_{conf}(24) G_{conf}(34)^2
\begin{pmatrix}
1 + \alpha^2 & 2 \alpha^2 \\
2 \alpha^2 & 1+ \alpha^2 
\end{pmatrix}.
\label{eq:conf_kernel}
\eeq
Eigenvalue $1$ eigenvector $\Psi_{k=1}$ corresponds to reparametrizations. Because of non-zero $\alpha$,
$G_{11}$ and $G_{22}$ have to be reparametrized the same way, so $\Psi_{k=1}$ has equal components:
\beq
\Psi_{k=1} \propto
\begin{pmatrix}
    1 \\
    1 \\
\end{pmatrix}.
\eeq
It is indeed easy to see that this vector is an  eigenvector\footnote{The eigenvalue $1+3\alpha^2$ will
conveniently cancel with the extra $1/(1+3\alpha^2)$ in the conformal solution
(\ref{eq:Gconf}).} of the conformal kernel matrix in eq.
(\ref{eq:conf_kernel}).  Because of this, the leading answers for all four
4-point functions in eq. (\ref{eq:4pt_matrix}) are going to be the same.
Notice that the kernel does not contain $\xi$ explicitly, as it is determined by the ladder diagrams. 
The $\xi$ is actually present in the non-conformal correction to $G$ and hence in 
the eigenvalue shift $1-k(2,n)$.
In the coupled model this correction is: 
\beq
G_{exact,11/22} = G_{conf} + \delta G_{11/22}.
\eeq
Let us discuss which of the terms in $\delta G$ contribute to the eigenvalue shift $k(2,n)_{conf}-k(2,n)_{exact}=
1-k(2,n)_{exact}$.
Since the conformal $G_{11/22}$ do not 
depend on $\xi$, eq. (\ref{eq:Gconf}), they and the conformal kernel have $\mathbb{Z}_2$ 
symmetry $11 \leftrightarrow 22$.
In order to obtain the leading eigenvalue shift we need to compute $\bra \Psi_{k=1} | \delta K | \Psi_{k=1} \ket $.
The leading $(\beta J)^{h-1}$ correction to $G$ comes from computing the 3-pt function,
eq. (\ref{eq:Gleading}), where in our case the lightest operator is
\beq
\Oc_h = -\xi \sum_i \l \psi^1_i \pr_u \psi^1_i - \psi^2_i \pr_u \psi^2_i \r.
\eeq
Crucially, it is linear in $\xi$. In other words, it contributes with different signs to $\delta G_{11}$
and $\delta G_{22}$:
\beq
\frac{\delta G_{11}}{G_{conf}} = -\frac{\alpha^G_h}{\l \beta J \r^{h-1}} f_h ,
\eeq
\beq
\frac{\delta G_{22}}{G_{conf}} = +\frac{\alpha^G_h}{\l \beta J \r^{h-1}} f_h .
\eeq
Because of this asymmetry the leading correction from $f_h$,
$\bra \Psi_{k=1} | \delta K | \Psi_{k=1} \ket $ will vanish.
Including the subleading correction (\ref{eq:Gsubleading}) we have
\beq
\label{eq:delta_G}
\frac{\delta G_{11/22}}{G_{conf}} = \mp \frac{\alpha^G_h}{\l \beta J \r^{h-1}} f_h - 
\frac{\alpha^G_{2h}}{\l  \beta J \r^{2h-2}} f_{h+h} - \frac{\alpha^G_{Sch}}{\beta J} f_0 + \dots \ .
\eeq
Therefore, the leading $\delta G$ correction $f_h$ can only contribute to $\delta k$ starting at quadratic
order. However, it can mix with $f_{h+h}$: it is even in $\xi$, and 
hence can contribute to the eigenvalue shift
in the leading order. This is why the analytic computation of the eigenvalue shift seems very difficult and we resort
to numerics.

Note that because of the original $\mathbb{Z}_2$ symmetry at $\xi=0$, the one point function $\bra \Oc_h \ket$
vanishes(in the first order of perturbation theory). So the absence of $(\beta J)^{h-1}$ term in the MSY computation and in the kernel computation has the same origin.

\subsection{Eigenvalue shift}
\label{sec:shift}
We will diagonalize the kernel and find the eigenvalues closest to $1$ following ref. \cite{KitaevRecent}.
It can be done numerically by introducing a 2D grid. We will fix $\beta = 2 \pi$ and study
different $J$. Index $n$ arises by noticing that the kernel is invariant under translations 
and so $n$ is the momentum:
\beq
K_{n,ab}(u,u') = \int_0^{2\pi} ds \ K_{ab}\l s  + \frac{u}{2}, s - \frac{u}{2}; \frac{u'}{2}, -
 \frac{u'}{2} \r e^{-i n s}.
\eeq
Since we are interested in the asymmetric kernel, it would be convenient to anti-symmetrize
$u,u'$ explicitly:
\beq
K^A_{n,ab}(u,u') = \frac{1}{2} \l K_{n,ab}(u,u') - K_{n,ab}(u',u)   \r,
\eeq
it improves the numerical results.

\begin{figure}[!ht]
\centering
\minipage{0.47\textwidth}
\includegraphics[scale=0.6]{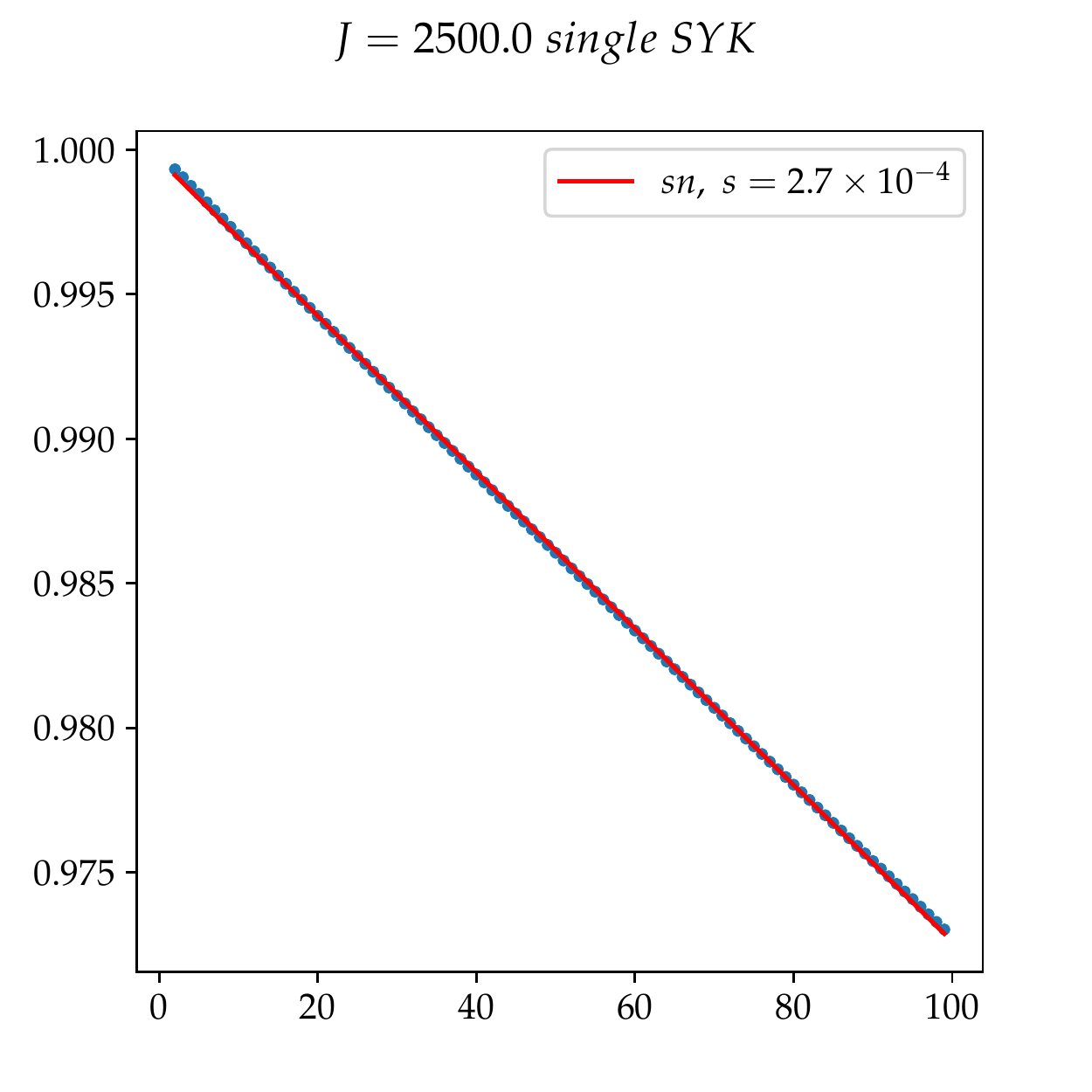}
\endminipage
\minipage{0.47\textwidth}
\includegraphics[scale=0.6]{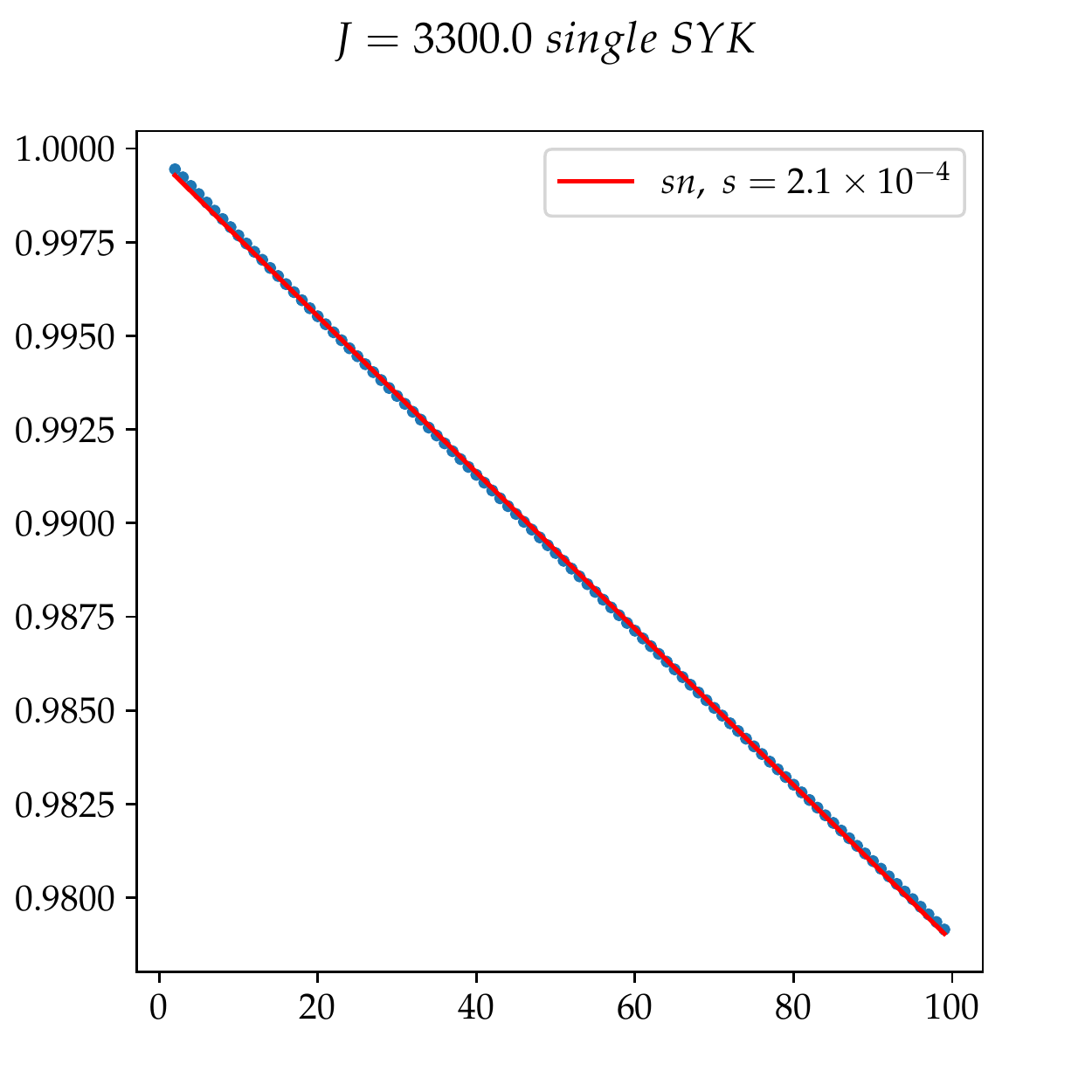}
\endminipage
\caption{$k(2,n)$ versus $n$ plot for $J=2500$ and $J=3300$ single SYK. Red line is a linear fit to guide the eye.}
\label{kn:single}
\end{figure}

Let us start from a single SYK as a benchmark - Figure \ref{kn:single}.
We see a perfect agreement with the theoretical prediction $1-k(2,n) \propto n$.

Now we need to understand what kind of eigenvalue shift we expect from the non-local action.
From the SYK discussion in Section \ref{sec:review} and eq. (\ref{eq:snonloc_n}) it follows that
the eigenvalue shift is determined by $g_h(n)/(|n|(n^2-1))$:
\beq
\label{xi:k}
(1-k(2,n))_{nonloc} = \frac{\alpha^K_{2h}}{(\beta J)^{2h-2}} 
\frac{g_h(n)}{|n|(n^2-1)}.
\eeq
Coefficients $\alpha^K_{2h}$ and $\alpha^S_{2h}$ are related by
\beq
\label{eq:KS}
\alpha^K_{2h} = \frac{4 m_h}{\pi^4 b^4} \alpha^S_{2h},
\eeq
where $m_h$ is given by eq. (\ref{eq:mh}).
Therefore, at large $n$ we expect the following behavior:
\beq
\label{k:nonloc}
\l 1-k(2,n) \r_{nonloc} =  \alpha_{2h}^K \frac{n^{2h-2}}{(\beta J)^{2h-2}}.
\eeq
In fact, for our range of $h$, $g_h(n)$ is almost indistinguishable from a power-law except for the first few $n$. 
This motivates us to try to
fit our results with a combination of a linear piece $n$(Schwarzian) and $n^{2h-2}$(non-local).
For $h$ in the range $1<h<3/2$ the power $n^{2h-2}$ is less than $1$. 
It means that for large $n$ and $h$ not too close to $3/2$, the Schwarzian will dominate at large $n$.
In order to check this we have plotted $k(2,n)$ for $n=2,\dots,100$ for various values
of $J$ and $\alpha$.

\begin{figure}[!ht]
\centering
\includegraphics[scale=0.7]{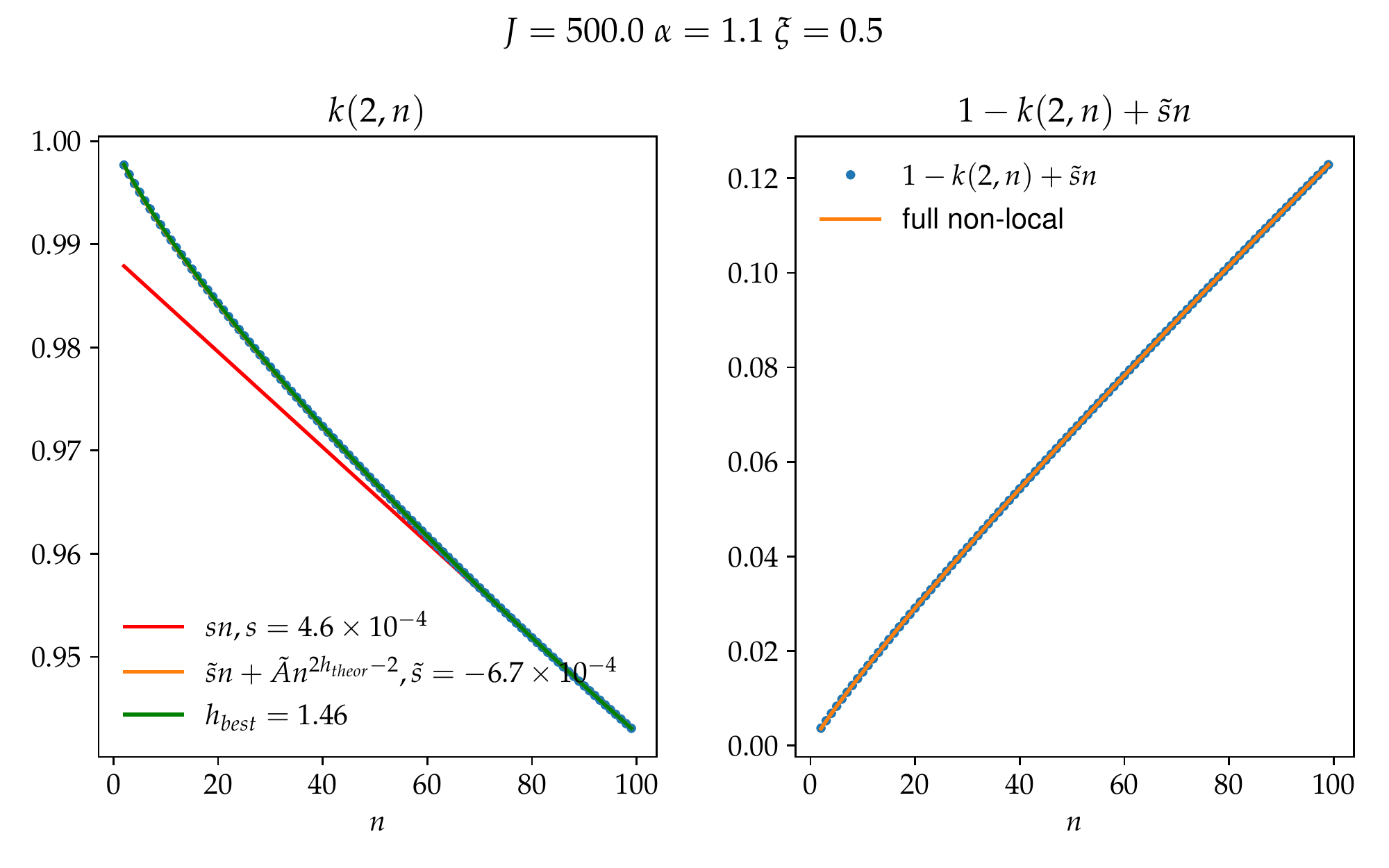}
    \caption{Results for $k(2,n)$. Details can be found in the main text.
    For $\alpha=1.1$, $h_{theor}=1.45$, whereas $h_{best}=1.46 \pm 0.01$. 
}
\label{fig:kns11}
\end{figure}

\begin{figure}[!ht]
\centering
\includegraphics[scale=0.7]{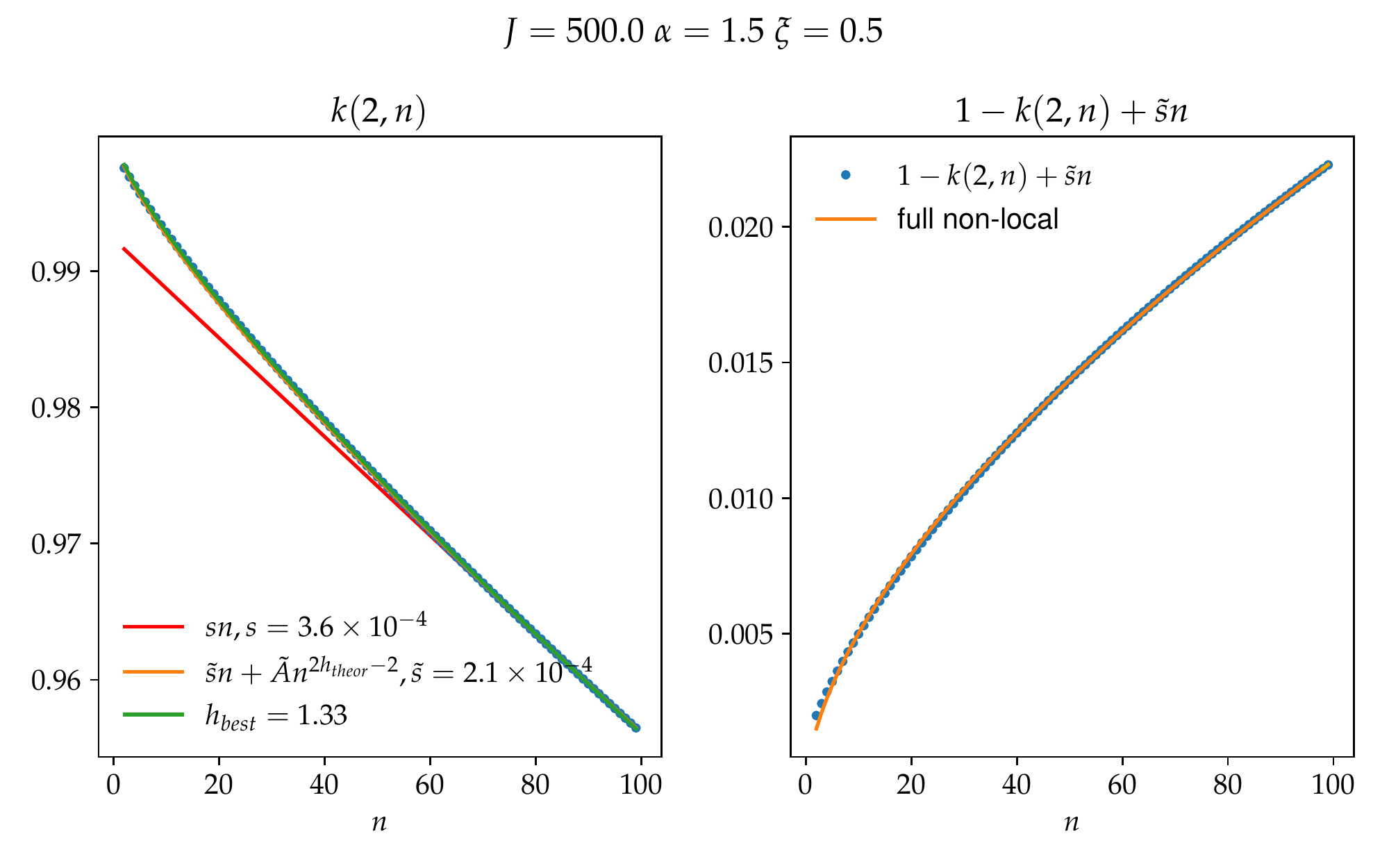}
    \caption{Results for $k(2,n)$. Details can be found in the main text.
    For $\alpha=1.5$, $h_{theor}=1.31$, whereas $h_{best}=1.35 \pm 0.02$.}
\label{fig:kns15}
\end{figure}

\begin{figure}[!ht]
\centering
\includegraphics[scale=0.7]{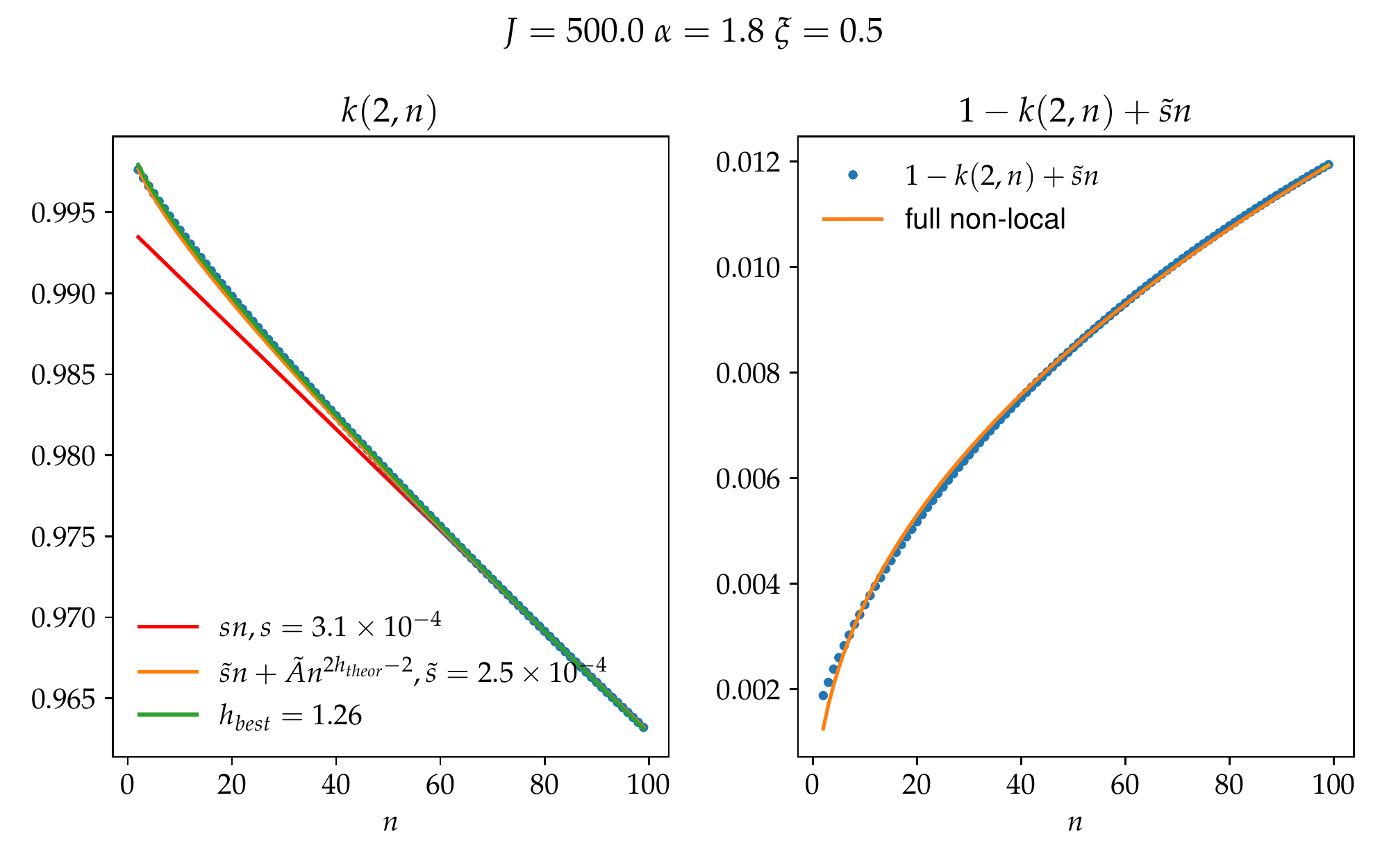}
    \caption{Results for $k(2,n)$. Details can be found in the main text.
    For $\alpha=1.8$, $h_{theor}=1.24$, whereas $h_{best}=1.28 \pm 0.04$.}
\label{fig:kns18}
\end{figure}

Let us consider $\alpha=1.1$, $\alpha=1.5$ and 
$\alpha=1.8$ - Figures \ref{fig:kns11}, \ref{fig:kns15}, \ref{fig:kns18}.
All of them show non-linear behavior which turns into a linear one for large $n$ (left side in the plots). 
Presumably for large $n$ the Schwarzian piece starts to dominate. 
Naive log-log plot is not very instructive for two reasons: non-local contribution is not exactly a
power-law and also we have a mixed expression with the linear Schwarzian contribution and the
non-linear piece (\ref{k:nonloc}).
Let us describe in detail Figures \ref{fig:kns11}, \ref{fig:kns15}, \ref{fig:kns18}. 
We start from the left part, which is $k(2,n)$:
\begin{itemize}
    \item Blue dots are numerically obtained $k(2,n)$.
    \item We could try to extract the linear term at large $n$ by fitting $k(2,n)$ 
        with a line $s n$(red line), keeping $s$ unknown.
        However, it turned out that for our range of $\beta J$ the non-linear piece is still
        not negligible even at large $n \sim 100$.
    \item So in order to extract the linear piece properly we perform the fit with the non-linear piece
       (\ref{k:nonloc}) as well:
        \beq
        1-\tilde{s} n - \tilde{A} n^{2h_{theor}-2},
        \eeq
	with unknown $\tilde{s}, \tilde{A}$ and
        where $h_{theor}$ is the theoretical value of the scaling dimension. This is the orange curve.
	As we see from the plots, slope $\tilde{s}$ is not close to naive slope $s$. This means that 
        the nonlinear piece is indeed not negligible. We will use this slope $\tilde{s}$ to subtract
        it from $k(2,n)$ and compare the result with the full non-local prediction $g_h(n)/(|n|(n^2-1))$.

    \item To double-check that we are not over-fitting by introducing to many parameters we 
        perform a fit with unknown $\hat{s}, \hat{A}$ and unknown power in the non-linear part:
        \beq
        1-\hat{s} n - \hat{A} n^{2h-2}.
        \eeq
        This is green curve. In most cases it is indistinguishable from the orange one.
        The best value of $h=h_{best}$ is close to the theoretical $h_{theor}$. From using 
	$2^{25}-2^{26}$ points for solving SD equation 
	and $60,000^2 - 140,000^2$ 2D grid points for finding the kernel eigenvalues, 
	we can estimate the uncertainty in $h_{best}$.
	We see that $h_{theor}$ and $h_{best}$ are within the uncertainty.
        As one can observe
        from the zero-temperature plots of the spectral function, Figures \ref{fig:alpha_b1}, \ref{fig:alpha_m1}, 
	the numerics seem to overestimate $h$. We attribute
        the difference $h_{best}-h_{theor}$ to this systematic overestimation.
\end{itemize}
The right side is less intricate: we subtract $\tilde{s} n$ from $k(2,n)$ and compare the result
with the full non-local shift $g_h(n)/(|n|(n^2-1))$.


\begin{figure}[!ht]
\centering
\includegraphics[scale=0.7]{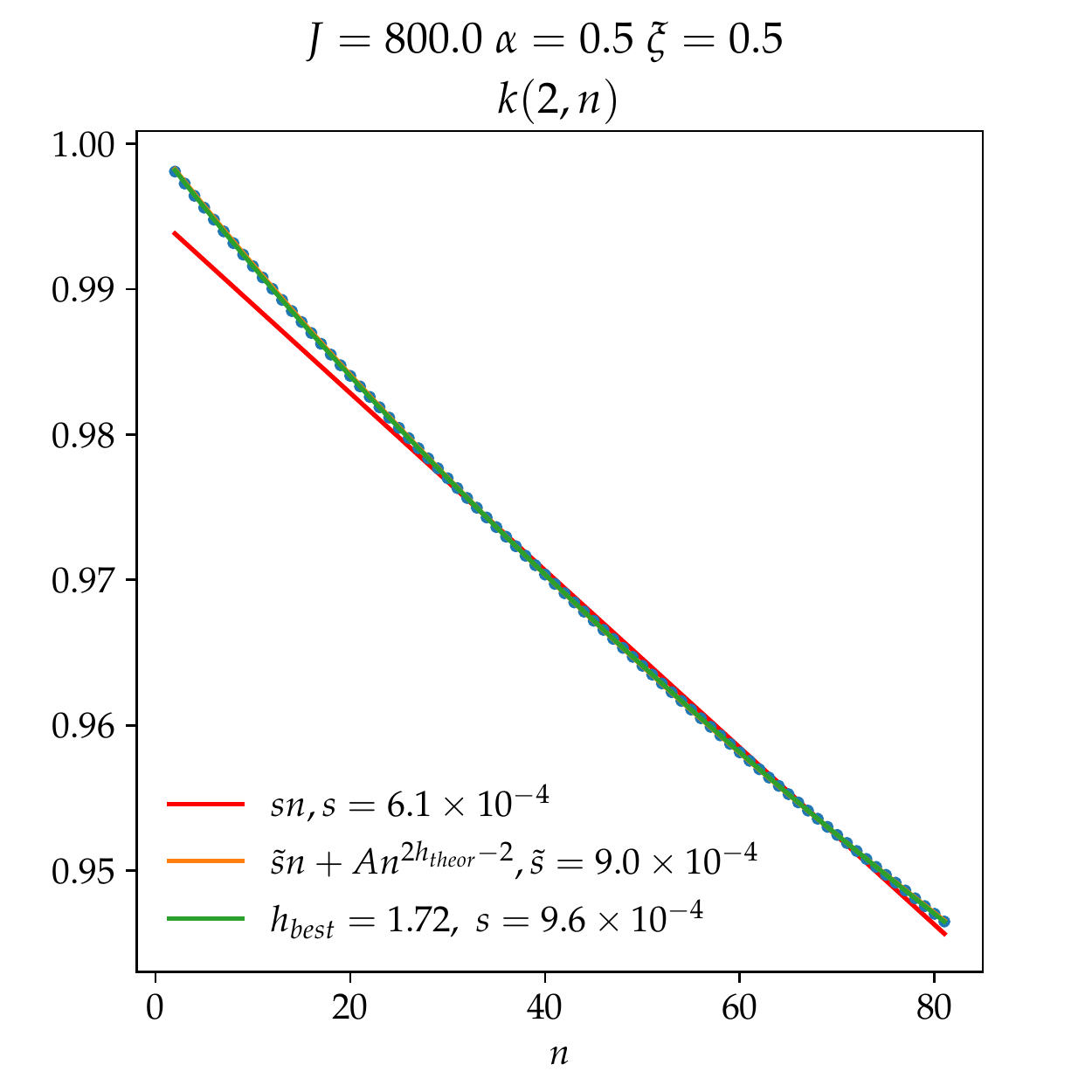}
    \caption{Results for $k(2,n)$. Details can be found in the main text.
    For $\alpha=0.5$, $h_{theor}=1.79$, whereas $h_{best}=1.72$.}
\label{fig:kns05}
\end{figure}

Finally, we can also check our results by considering $|\alpha|<1$. 
In this case we expect that the Schwarzian does dominate for large $\beta J$:
\beq
\label{eq:11fit}
1-k(2,n) = \frac{\alpha_{Sch}^K |n|}{\beta J} + \frac{\alpha^K_{2h}}{\l \beta J\r^{2h-2}} |n|^{2h-2}, n \gg 1,
\eeq
where now $h>3/2$.
It means that presumably at small $n$ the Schwarzian dominates and then for large $n$ the non-local piece starts to win.
Moreover, from the analytic expression (\ref{eq:gh})
we expect that $\alpha_{2h}^K$ is now \textit{negative}, so $1-k(2,n)$ will still curve downwards. 
We considered $\alpha=0.5$ and again fitted $k(2,n)$ with eq. (\ref{eq:11fit}), keeping $h$ unknown - Figure
\ref{fig:kns05}. We again see a very good agreement with theoretical results.

\subsection{Temperature dependence}
\label{sec:Tdep}
\begin{figure}[!ht]
\centering
\minipage{0.47\textwidth}
\includegraphics[scale=0.6]{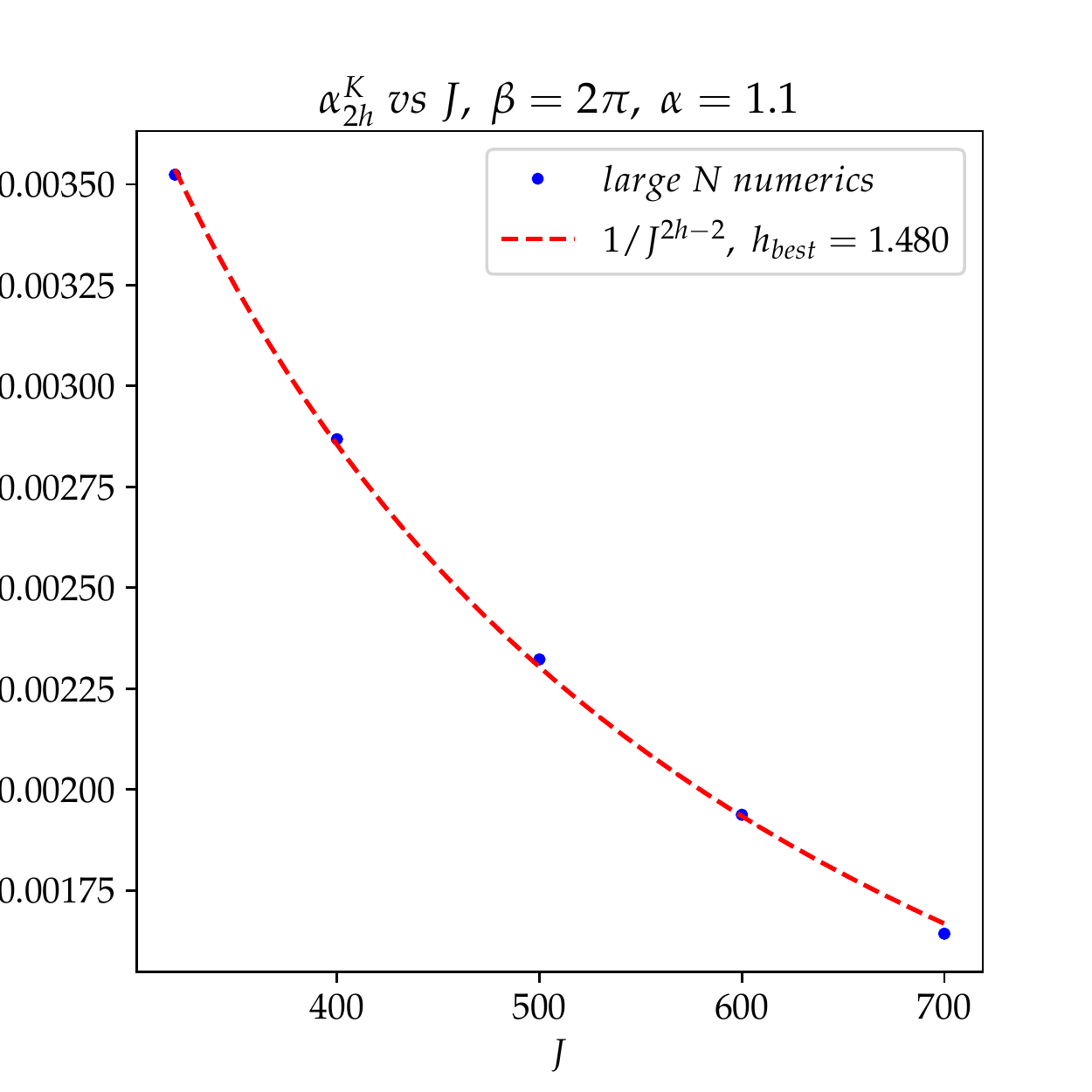}
\endminipage
\minipage{0.47\textwidth}
\includegraphics[scale=0.6]{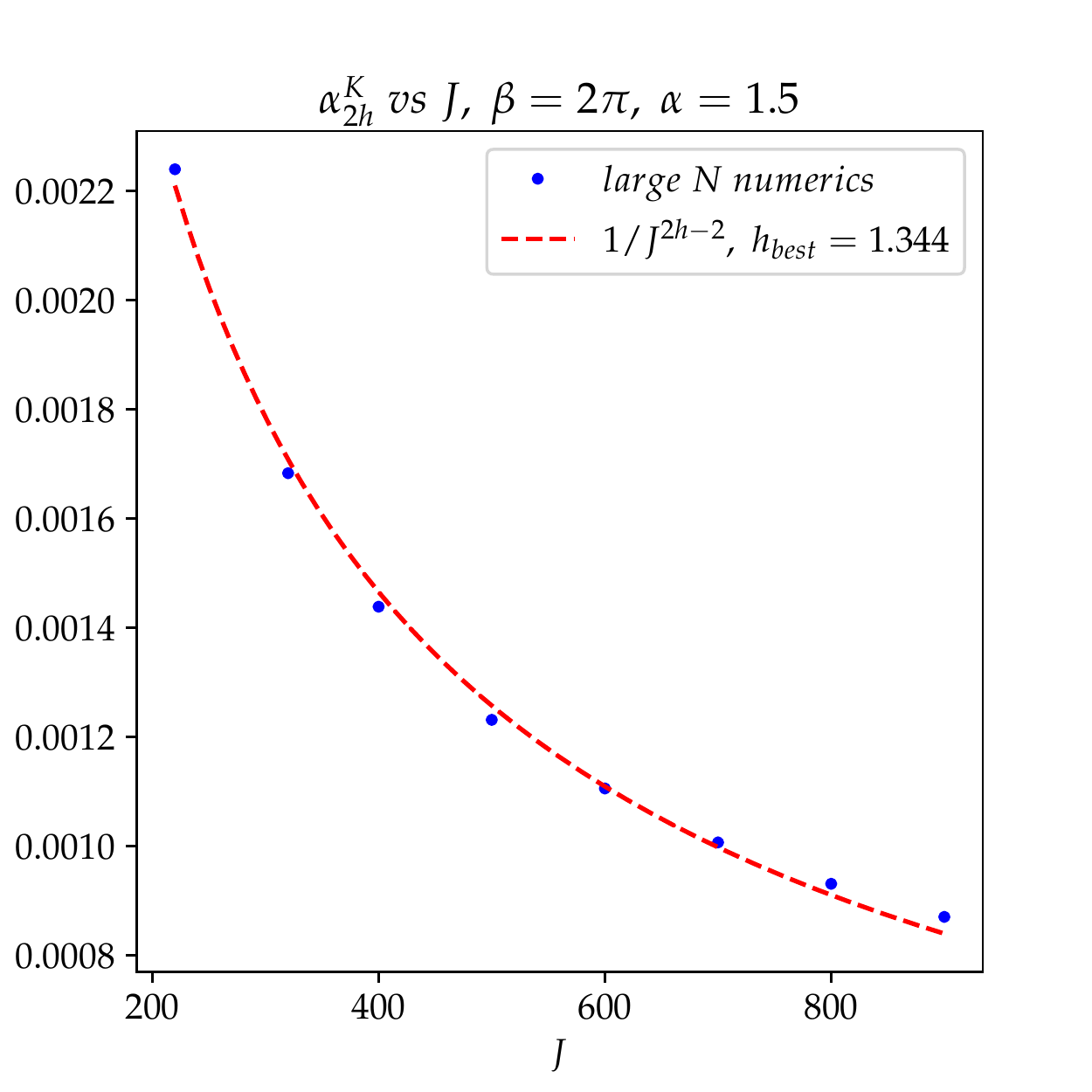}
\endminipage
\caption{Coefficient $\alpha_{2h}^K$(here for convenience 
we included $\beta J$ inside $\alpha_{2h}^K$ compared to eq. (\ref{k:nonloc})) 
as a function of $J$. Left: $\alpha=1.1$ for which $h_{theor} = 1.45$. Right: $\alpha=1.5$ for which $h_{theor}=1.31$.}
\label{fig:temp}
\end{figure}

Finally, the non-local action predicts that the non-linear term $n^{2h-2}$ in the eigenvalue shift behaves as 
$1/(\beta J)^{2h-2}$, eq. (\ref{k:nonloc}). The fitting strategy outlined in the previous Section allowed us to
extract this coefficient. We considered $\alpha=1.1$ and $\alpha=1.5$ and plotted this coefficient for different $J$.
After that, we fitted the result with
\beq
\frac{c}{J^{2h-2}},
\eeq
keeping $c$ and $h$ unknown. The results are presented in Figure \ref{fig:temp}. We see that $h_{best}$ is again very close
to the theoretical value.
One can also check that the resulting $\alpha_{2h}^K$ agrees well with $c_{2h}$ in Figure \ref{fig:c_xi}.
This computation requires using the conversions (\ref{eq:KS}) and (\ref{eq:FS}).
\label{sec:twistdep}

\section{Exact diagonalization at finite $N$}
\label{sec:ED}
One of the nice feature of SYK-like models in the opportunity to study finite-$N$ effects using
exact diagonalization of the Hamiltonian. In our case the dimension of the Hilbert space is
\beq
\dim \Hc = 2^N,
\eeq
so we can easily consider $N$ up to 16 without using any special techniques.
Similar computations for the case of SYK has been done in the literature before \cite{ms}, \cite{Bagrets2016Sachdev},
\cite{GurAri2018Does}, \cite{Kim:2019upg}.
We have performed finite $N$ exact diagonalization for four reasons:
\begin{itemize}
\item Cross-check our infinite $N$ solutions of SD equations.
\item Probe the density of states near the ground state and see if it differs from the 1-loop result (\ref{e:nonloc}).
\item See how the 2-point function behaves at very late times $\tau \gg N/J$.
\item Check if the spectral correlators obey random matrix theory predictions. A deviation from them would
indicate possible spin--glass phase at low temperatures \cite{GurAri2018Does}.
\end{itemize}

\begin{figure}[!ht]
\centering
\includegraphics[scale=0.7]{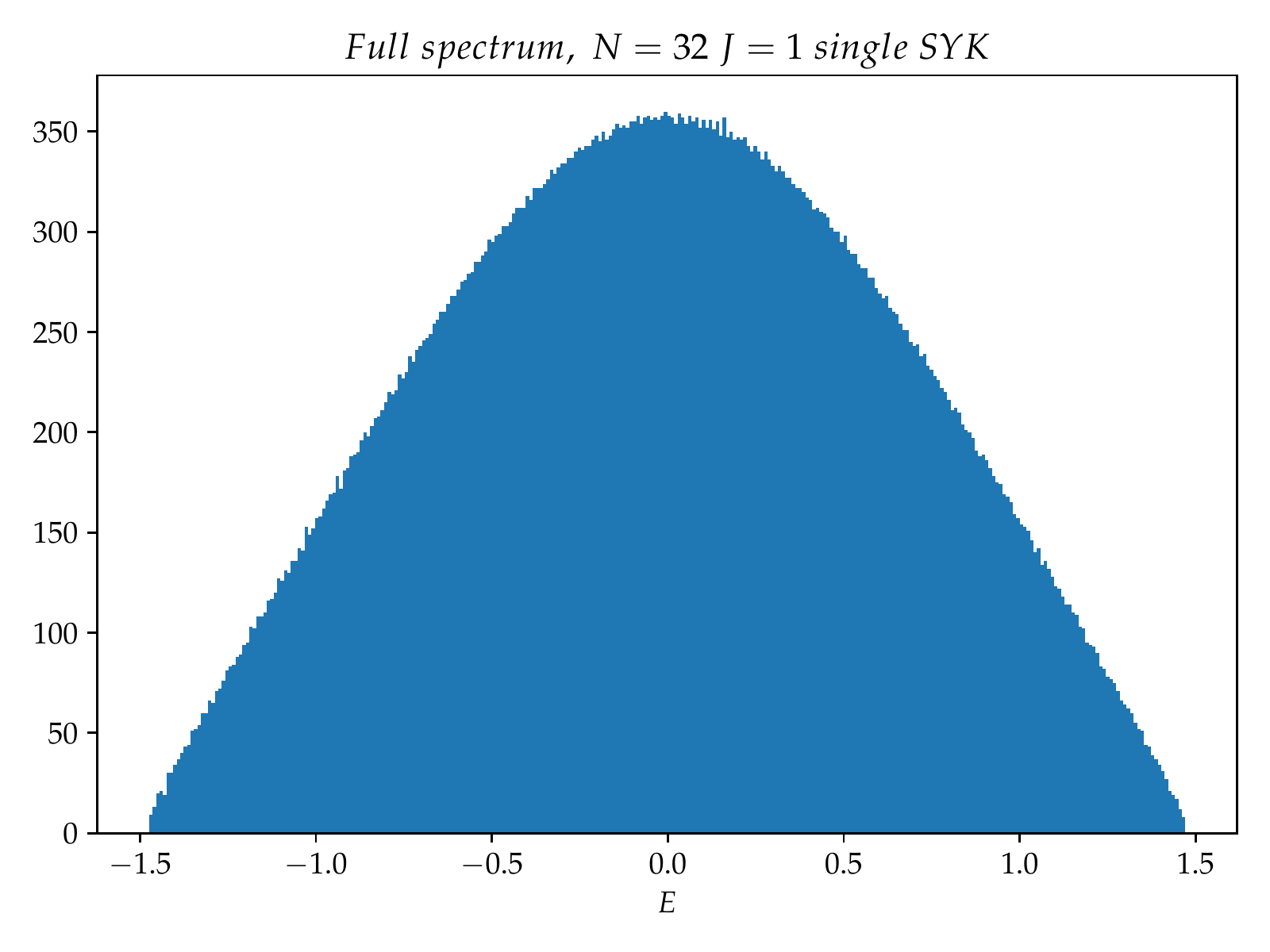}
\caption{Full spectrum of $J=1,\ N=32$ single SYK for a single disorder realization binned with 300 bins. }
\label{fig:full_N32}
\end{figure}

\begin{figure}[!ht]
\centering
\includegraphics[scale=0.7]{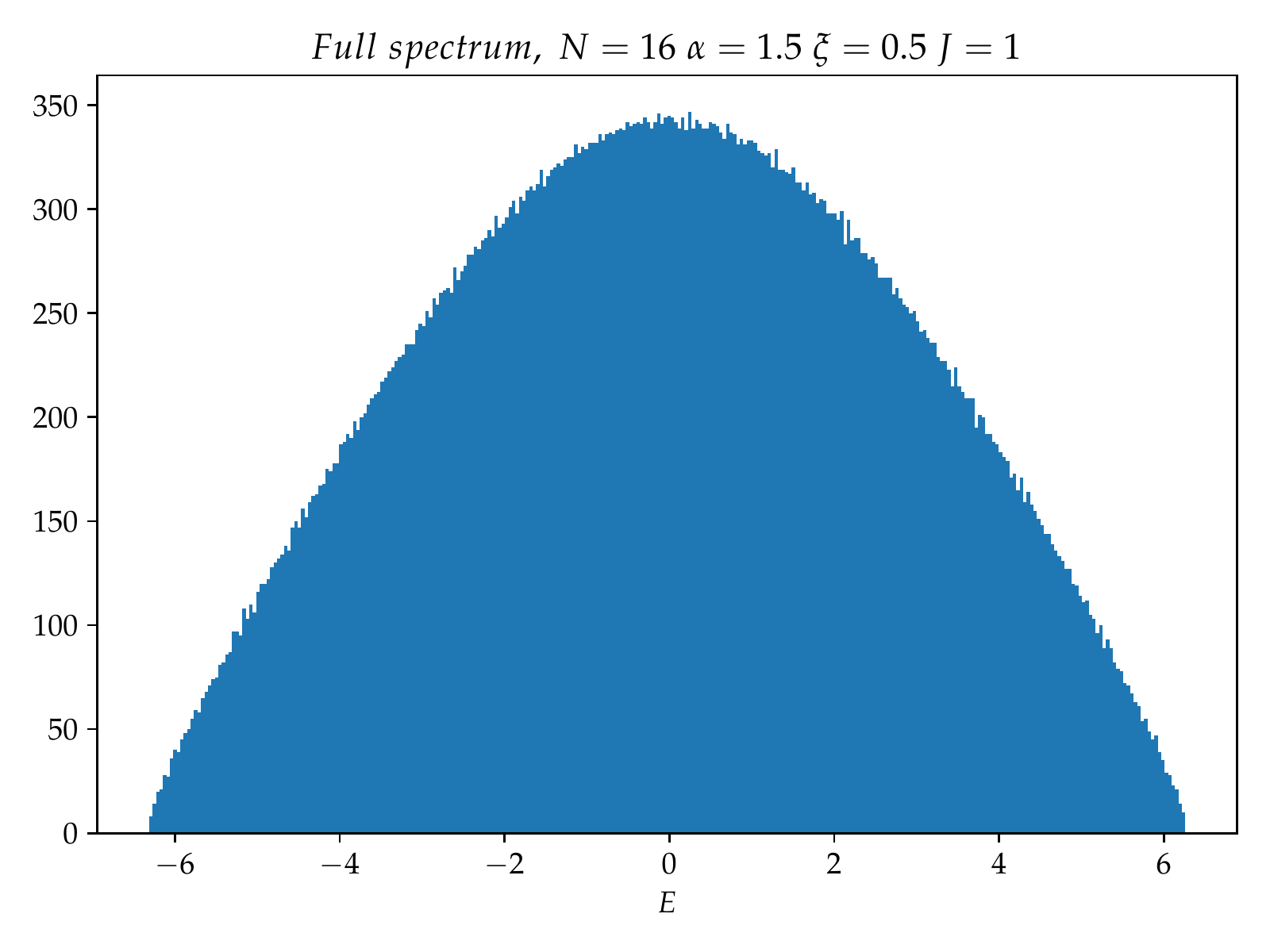}
\caption{Full spectrum of $J=1,\ N=16,\ \alpha=1.5, \xi=0.5$ model for a single 
    disorder realization binned with 300 bins. }
\label{fig:full_N16_xi}
\end{figure}

\begin{figure}[!ht]
\centering
\includegraphics[scale=0.7]{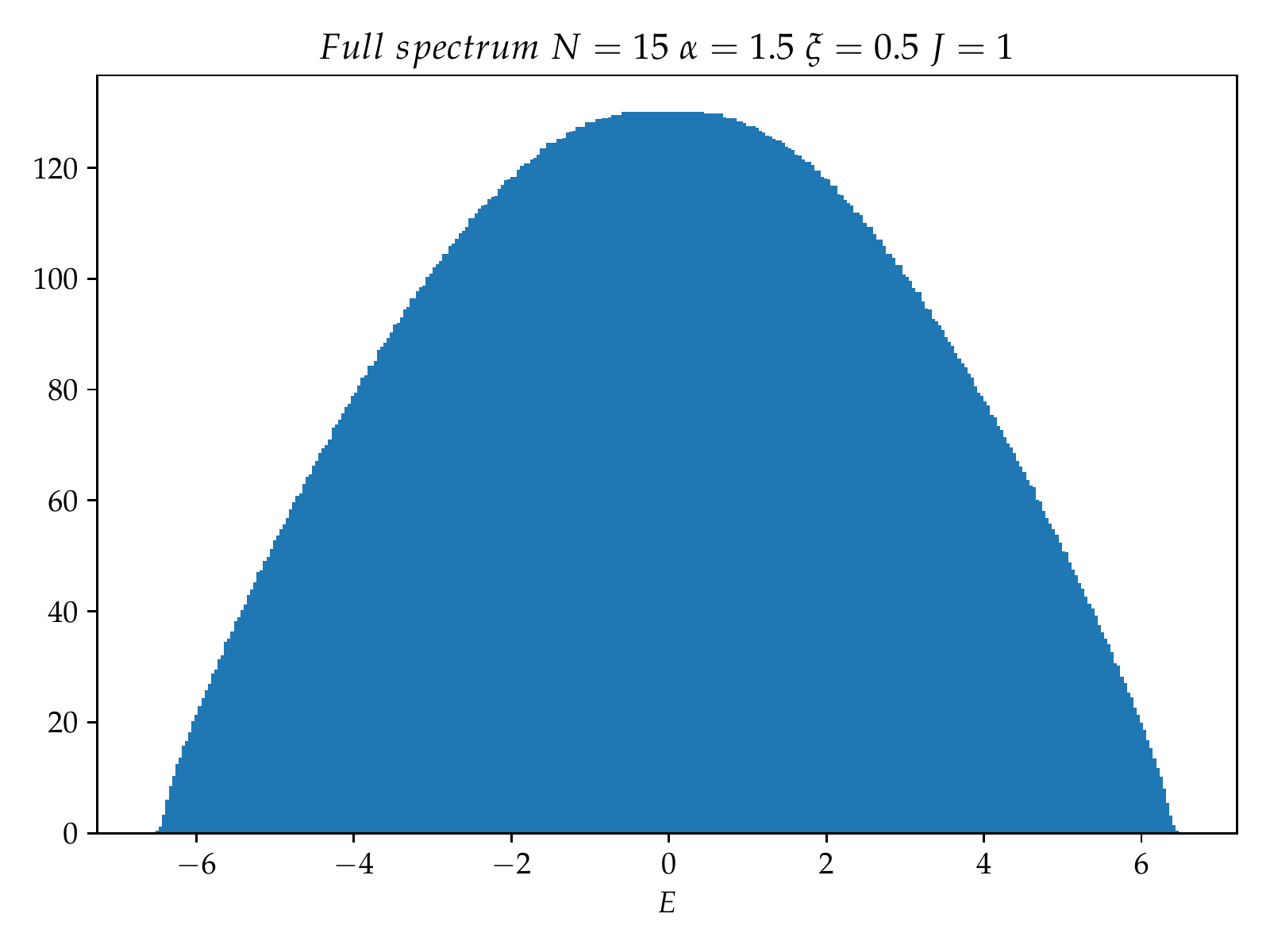}
\caption{Full spectrum of $J=1,\ N=15,\ \alpha=1.5,\ \xi=0.5$ model averaged over 30 realization and binned with 300 bins. }
\label{fig:full_N15_averaged}
\end{figure}

As a starter we present the full spectrum binned with 300 bins for a single realization of disorder.
Figure \ref{fig:full_N32} shows the full spectrum of $N=32$ original Majorana SYK.
Figure \ref{fig:full_N16_xi} shows the same quantity but for our coupled model with $\alpha=1.5,\ \xi=0.5$.
We can also average over several samples to produce a more smooth density - Figure \ref{fig:full_N15_averaged}

The main takeaway from these plots is that the coupled model does not have a gap between the ground state and the rest of the
spectrum. The presence of such gap would immediately imply that the conformal solution (\ref{eq:Gconf}) does not
represent the dominant thermodynamic solution.

\subsection{Ground state energy}
\label{sec:E0}
As we have mentioned in the Introduction, we are solving (Euclidean) SD equations (\ref{sd:eucl}) by the standard
iteration procedure \cite{ms}, when we start from a free solution $G_{11}, G_{22} \propto \sgn(u)$
and iterate the equations (\ref{sd:eucl}) until we converge(the norm between successive solutions becomes small).
A natural question is: how do we know that we converge to an actual physical solution? 

One way to check this is to compare the resulting ground state energy
to the actual ground state energy computed from finite $N$ exact diagonalization. This requires 
two extrapolations. In SD we have to extrapolate finite-temperature energy all the way to $T=0$. 
This can be done using the prediction (\ref{eq:E_T}).
In ED we have to extrapolate finite $N$ results to $N=\infty$. We can do this by assuming the following 
\footnote{Our $N$ is not very large, this is why included the subsub-leading $c_2/N$ term.
In fact, we have performed the fit with and without it and this way estimate the uncertainty in $E_0/N$.
Uncertainty in each individual point can be made very small by averaging over many samples.}
$N$ dependence in the ground state energy $E_0(N)$ at finite $N$:
\beq
E_0(N) = \l E_0/N \r N + c_1 + c_2/N,
\eeq
and extract $E_0/N, c_1, c_2$ from the fit.
The quantity $E_0/N \sim \Oc(N^0)$ is supposed to match the result from SD. 

\begin{figure}[!ht]
\centering
\includegraphics[scale=0.7]{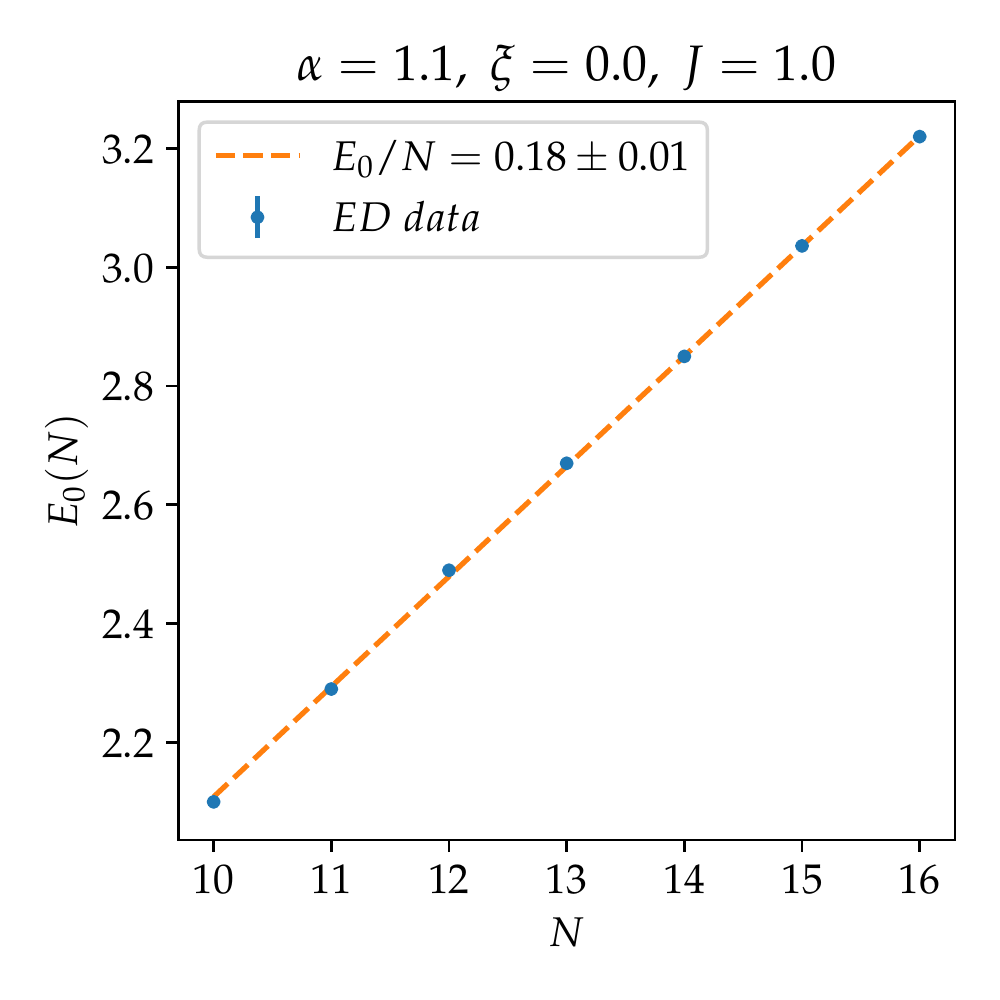}
\caption{Finite $N$ exact diagonalization results for $\alpha=1.1, \xi=0.0, J=1.0$. 
The uncertainty comes from including a subleading term $c_2/N$ in the fit.
Ground state energy from numerically
solving large $N$ 
SD equations is $E_0/N=0.175(1)$. }
\label{fig:ed_bench}
\end{figure}

\begin{figure}[!ht]
\centering
\minipage{0.47\textwidth}
\includegraphics[scale=0.7]{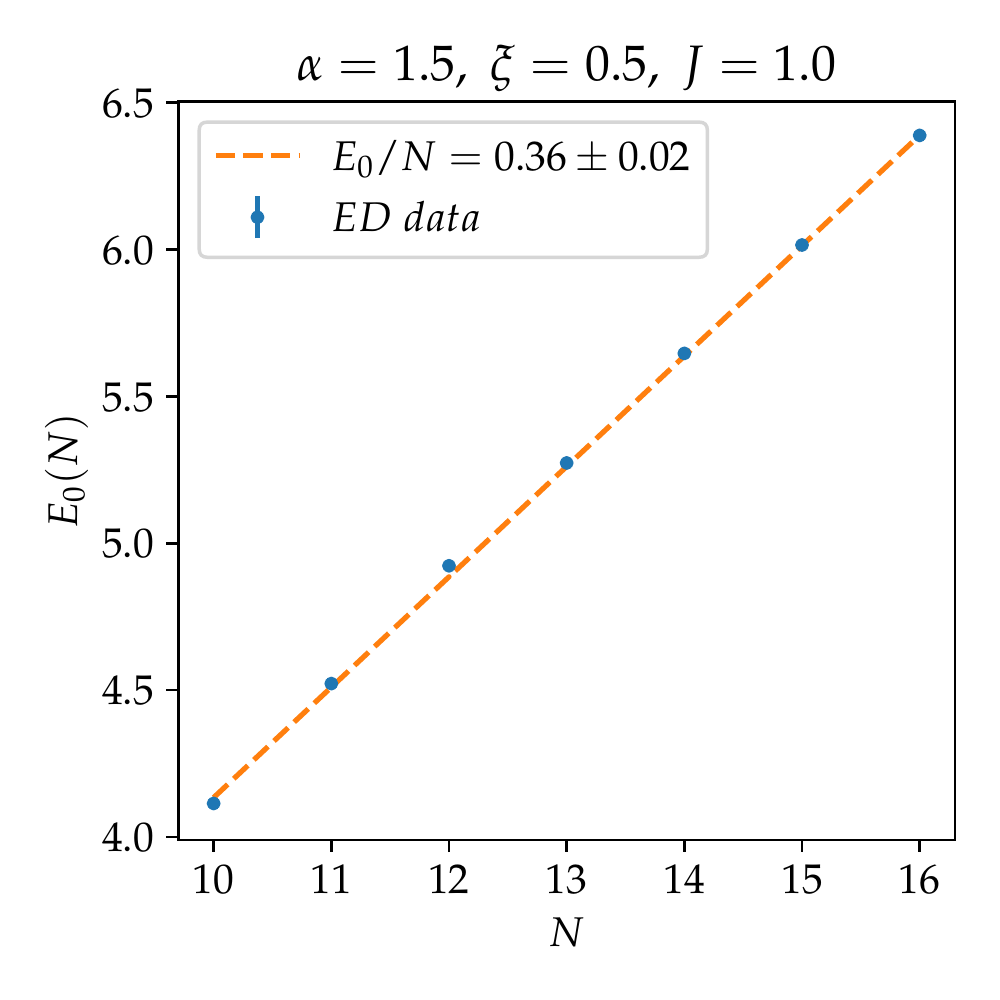}
\endminipage
\minipage{0.47\textwidth}
\includegraphics[scale=0.7]{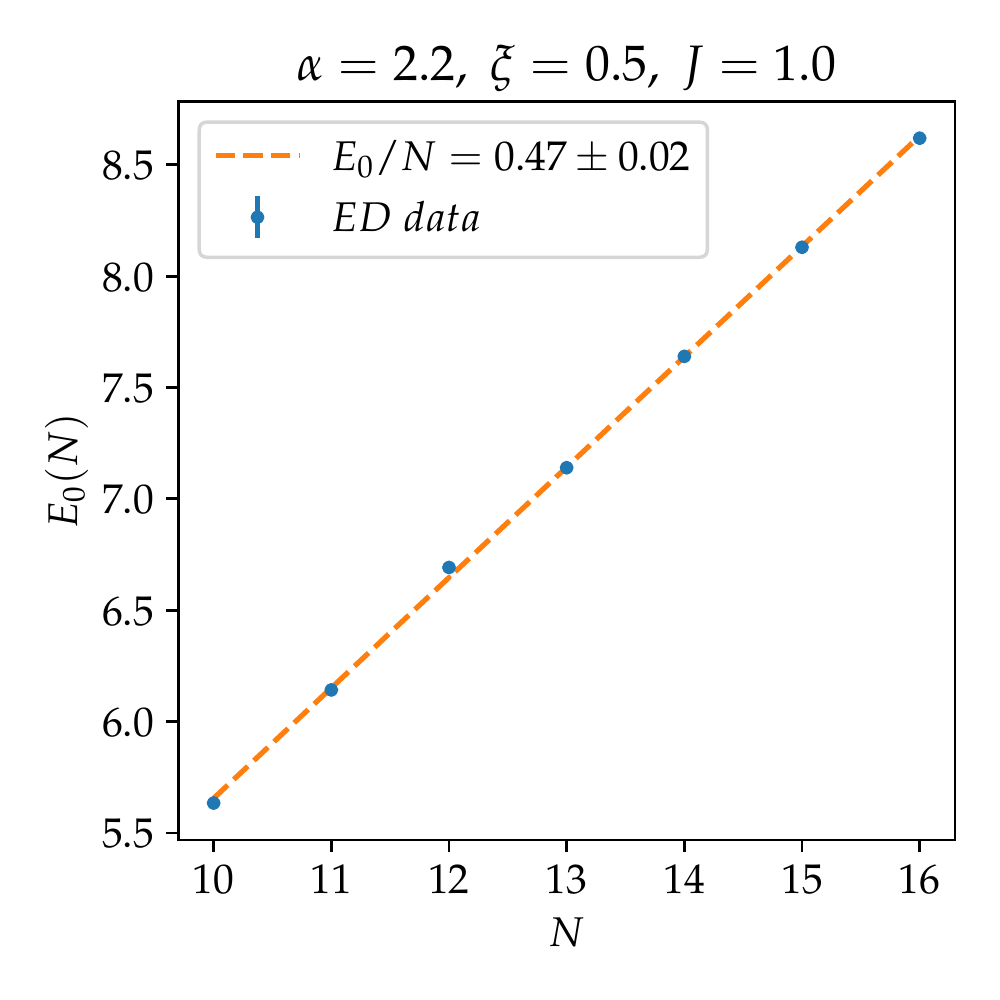}
\endminipage
\caption{Finite $N$ exact diagonalization results for $\alpha=1.5$(Left) and $\alpha=2.2$(Right).
For both cases $J=1.0, \xi=0.5$.
The uncertainty comes from adding/removing a sub-subleading term $c_2/N$ in the fit.
Ground state energy from numerically
solving large $N$ 
SD equations is $E_0/N=0.351(3)$ for $\alpha=1.5$ and $E_0/N=0.466(1)$ for $\alpha=2.2$.}
\label{fig:ed_xi}
\end{figure}

As usual, we first present the result for $\xi=0$, which is supposed to have ``conventional'' Schwarzian physics -
Figure \ref{fig:ed_bench}. In Figure \ref{fig:ed_xi} we present the results for $\xi=0.5$ and different values of $\alpha$.
In all cases we see a good agreement between SD and ED. 
\subsection{Density of states}
\label{sec:densityN}
It is very interesting to check the prediction (\ref{e:nonloc}) for the density of states:
\beq
\label{e:nonloc2}
\rho(E)_{non-loc,1-loop} \sim E^{3h-4}.
\eeq
Famous Schwarzian result predicts \cite{Stanford:2017thb} 
square-root edge $\sqrt{E}$ density of states near the ground state:
\beq
\rho(E)_{Sch,exact} \sim \sqrt{E}. 
\eeq

\begin{figure}[!ht]
\centering
\minipage{0.47\textwidth}
\includegraphics[scale=0.72]{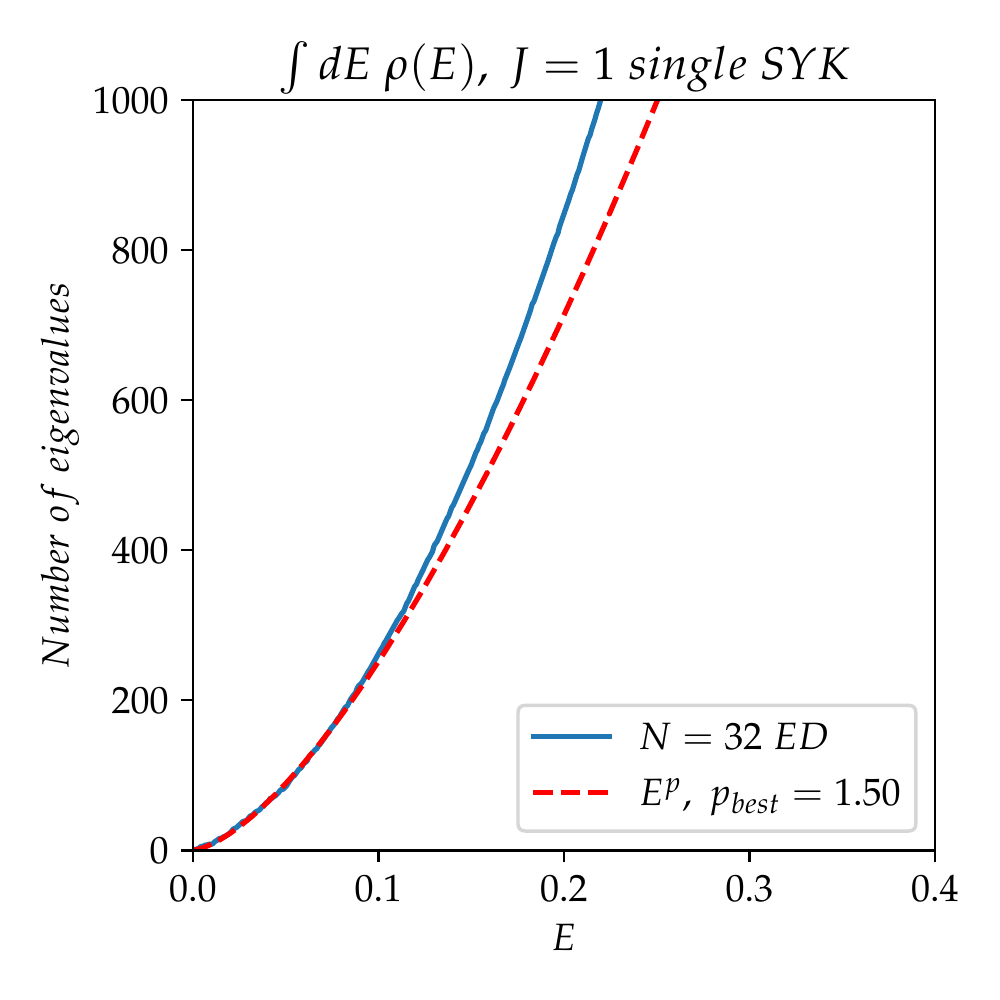}
\endminipage
\minipage{0.47\textwidth}
\includegraphics[scale=0.72]{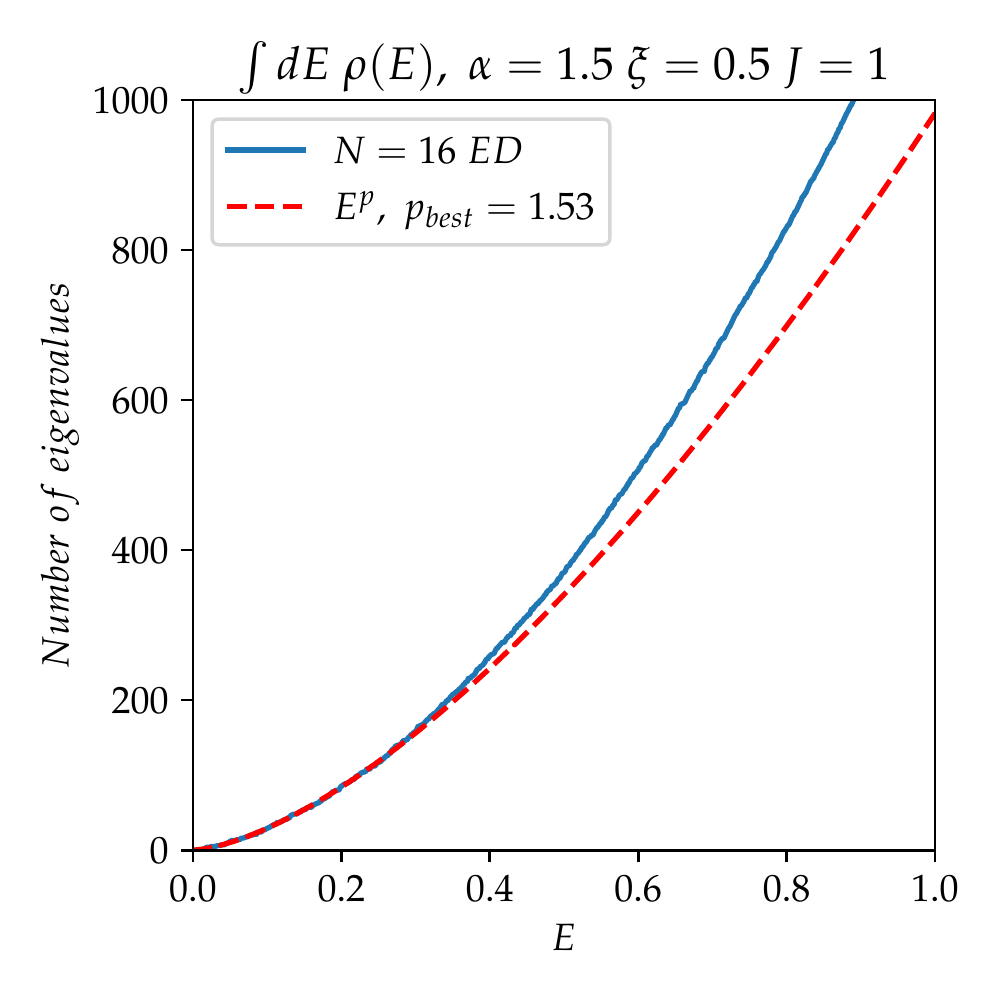}
\endminipage
    \caption{CDF for original $N=32$ SYK(Left) and coupled model with $N=16,\ \alpha=1.5, \xi=0.5$(Right). 
	We have used a single realization of disorder. In
    both cases $J=1$. The power $p_{best}$ was determined from a fit with $A E^p$. $p_{best}$ obviously
    depends on the energy interval where the fit is performed. Changing this interval
    introduces $0.1$ uncertainty for SYK case and $0.07$ uncertainty for the coupled model. }
\label{fig:CDF}
\end{figure}

On ED side, working with the density of states directly is not good, because it depends on bin size. In order
to eliminate this dependency we can plot "cumulative distribution function"(CDF) which is just the
number of states in a given energy interval from the ground state:
\beq
\text{CDF}(E) = \int_{E_0}^E \ dE' \rho(E').
\eeq
The results for the original SYK and the coupled model are shown in Figure \ref{fig:CDF}.
We see a very good agreement with $\sqrt{E}$ for the case of original SYK.
For $\alpha=1.5$, $h_{theor}=1.31$ so for the right part of Figure \ref{fig:CDF} the 1-loop result
(\ref{e:nonloc2}) predicts\footnote{The negative power should not be a concern as the density
$\rho(E)$ is still normalizable. For example, for $\mathcal{N}=1$ SUSY Schwarzian $\rho(E)_{SUSY,Sch} \propto 1/\sqrt{E}$
\cite{Fu:2016vas}. } $E^{-0.07}$ which is definitely not the case. This indicates that the 
non-local action is not 1-loop exact. This numerical analysis suggests that the density of states keeps 
the square-root edge even when the Schwarzian is not dominant.

\subsection{2-point function at very late times}
\label{sec:2ptN}
\begin{figure}[!ht]
\centering
\minipage{0.47\textwidth}
\includegraphics[scale=0.72]{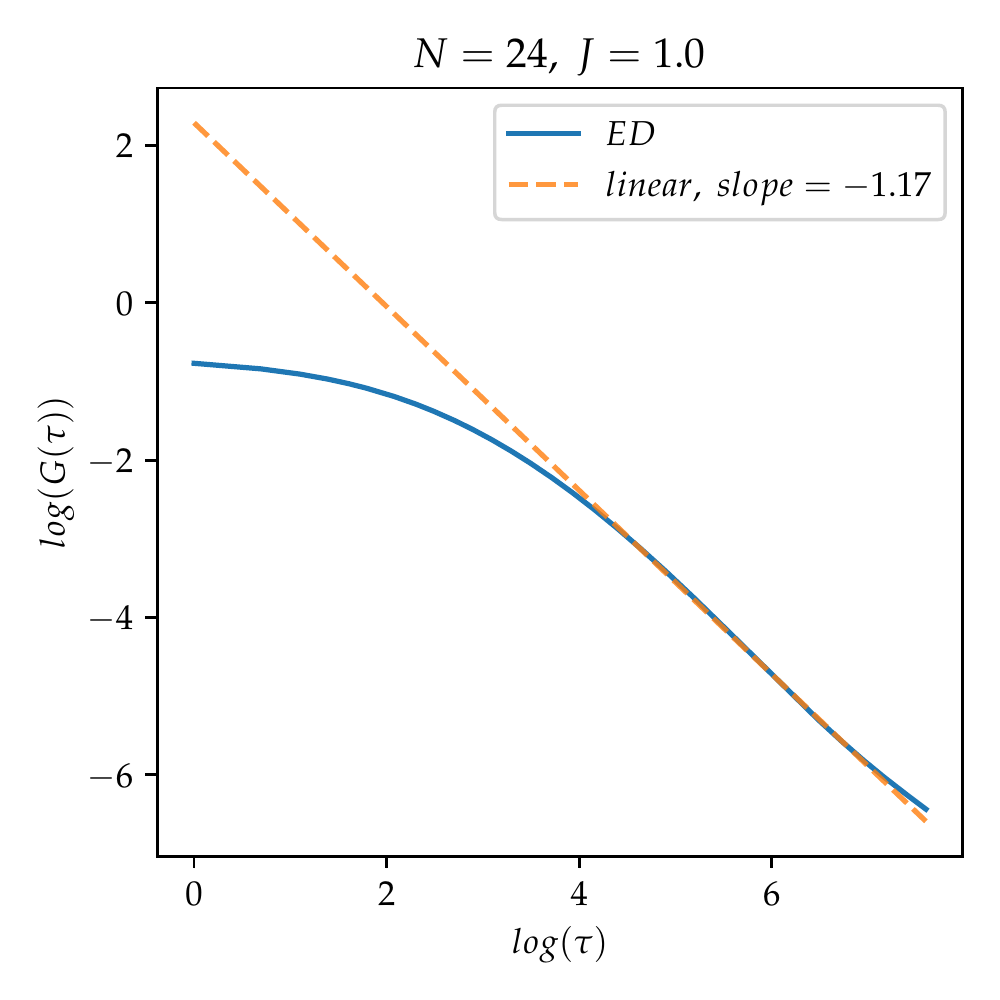}
\endminipage
\minipage{0.47\textwidth}
\includegraphics[scale=0.72]{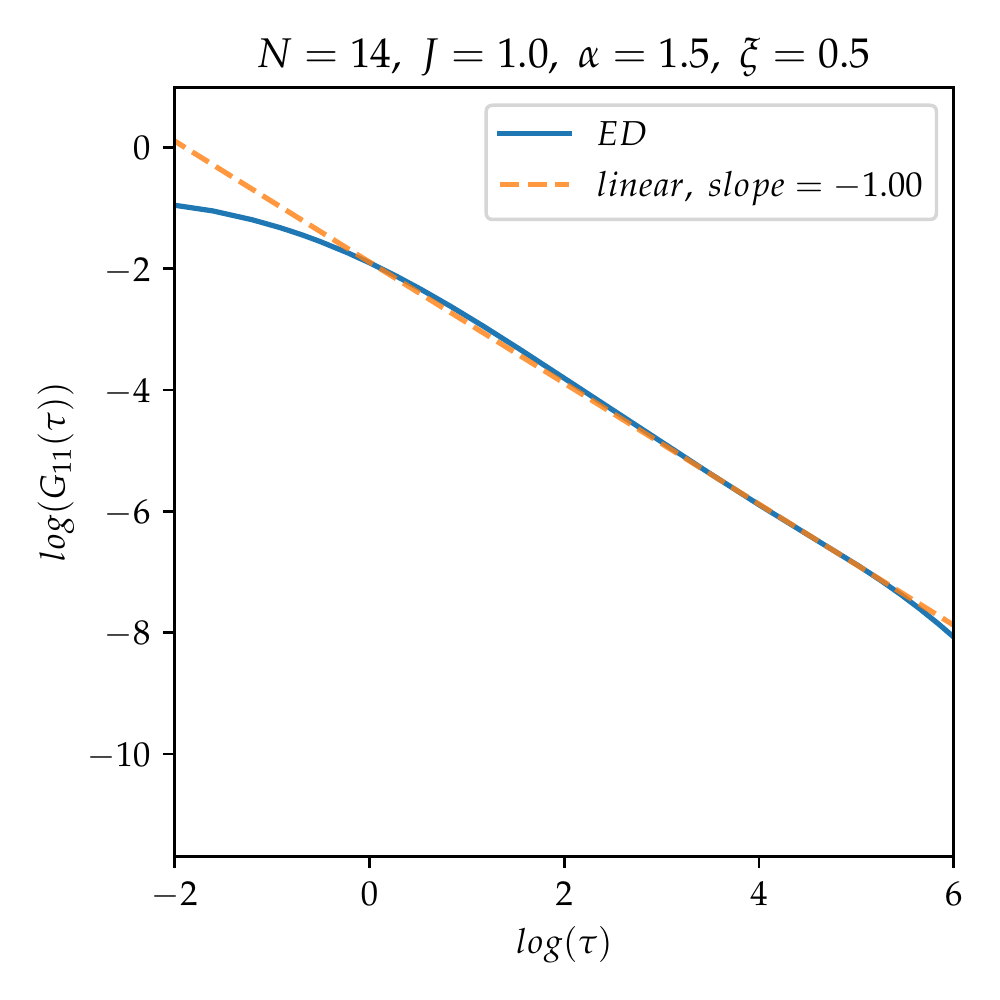}
\endminipage
\caption{2-point Green function at finite $N$ and large times. Left: original SYK. Right: the coupled model and
$G_{11}$. Almost exactly the same results hold for $G_{22}$.
In both cases we see a power-law behavior.}
\label{fig:2pt}
\end{figure}

Quantization of the Schwarzian action can be reduced to Liouville quantum 
mechanics \cite{Bagrets2016Sachdev, mertens}. At very late Euclidean times $\tau \gg N/J$ it results in a universal
behavior $N/\tau^{3/2}$ in the 2-point function.
In a single SYK, it is possible to see a power-law decay in ED even at moderate $N$. However 
one has to use large values of $N$ to
see anything close to the power $3/2$. 
We would like to see what happens in the coupled model.
Unfortunately, in the coupled model we are limited to $N=15$. 
Our results for a single SYK(for comparison) and the coupled model for $\alpha=1.5, \xi=0.5$ are presented in Figure
\ref{fig:2pt}. For this computation we did not use any approximations: we computed the full spectrum and
wavefunctions by ED and then used them to determine the 2-point function at zero temperature by the spectral decomposition:
\beq
\bra 0 |\psi_i(\tau) \psi_i(0) | 0 \ket = \sum_{E_n} |\bra n | \psi_i| 0 \ket|^2 e^{-(E_n-E_0) \tau}.
\eeq
Finally, we averaged over 100 samples.
We can confirm qualitative $1/\tau^p$ behavior, but we cannot reliable determine the 
power $p$. It seem to slowly increase with $N$. Our modest results suggest $p>1$. Presumably these results can be
easily improved by studying larger $N$, but using low-lying states only.

\subsection{Level spacing}
\label{sec:level}
Another interesting quantity is the energy level statistics. 
A general expectation for chaotic models is that after making the energy density uniform, the  energy gaps 
are distributed the same way as in a random matrix
ensemble. A deviation from this indicate possible spin-glass phase
\cite{GurAri2018Does}. In this Section we are going to show that in the 
coupled model the level statistics obey random matrix theory predictions, 
suggesting no spin-glass phase. Compared to the rest of the paper, in this Section parameter $\xi$ is absorbed into
$J^1,J^2,C$ couplings, making the fermionic operators square to one.

First of all, instead of unfolding the spectrum we consider another quantity:
the ratio $r_n$ between the adjacent energy gaps:
\beq
r_n = \frac{E_{n+1}-E_n}{E_n - E_{n-1}}.
\eeq
This quantity does not require unfolding. ``Wigner-surmise''-like computation
\cite{Atas_2013} predicts the following $r$ distribution\footnote{
Normalization factors are $Z_1 = 8/27,\ Z_2 = 4 \pi/(81 \sqrt{3}),\ Z_4 = 
4 \pi/(729 \sqrt{3})$.}:
\beq
P_\beta(r) = \frac{1}{Z_\beta} \frac{(r+r^2)^\beta}{(1+r+r^2)^{1+3\beta/2}},
\label{eq:sur}
\eeq
where as usual $\beta=1$ correspond to Gaussian Orthogonal Ensemble(GOE),
$\beta=2$ to Gaussian Unitary Ensemble(GUE) and $\beta=4$ to Gaussian 
Symplectic Ensemble(GSE). For comparison, for Poisson distributed levels
the distribution is
\beq
\label{eq:Poisson}
P_{Poisson}(r) = \frac{1}{(1+r)^2}.
\eeq

Now we need to understand what ensemble the coupled SYK 
Hamiltonian (\ref{eq:H_T}) corresponds to. 
Also we need to project out all global symmetries.
The symmetry $\psi_i^1 \leftrightarrow \psi_i^2$ is broken by $\xi$ term, so
we should not worry about it. For even $N$ we have two independent 
and commuting symmetries: $\psi^1_i \ra  -\psi_i^1$, $\psi^2_i \ra - \psi^2_i$.
The corresponding operators are
\beq
\Gamma^1 = i^{N/2} \prod_{i=1}^N \psi^1_i,
\eeq
\beq
\Gamma^2 = i^{N/2} \prod_{i=1}^N \psi^2_i.
\eeq
For odd $N$ only $\Gamma = \Gamma^1 \Gamma^2$ is a symmetry. 
Having projected on eigenvalue subspace of these operators, we need to ask
if we have any anti-linear symmetries. 
It is always possible to represent 
$\psi^1_i$ as real matrices and $\psi^2_i$ as purely imaginary matrices.
Then there are three anti-linear symmetries:
\beq
K_s = \Cc,
\eeq
\beq
K_1 =  \l \prod_{i=1}^N \psi_i^1 \r \Cc, \quad K_2 = \l \prod_{i=1}^N \psi^2_i \r \Cc,
\eeq
where $\Cc$ is complex-conjugation operator\footnote{For example, $\Cc i = 
- i \Cc$.}. 
They obey the following commutation relations for odd $N$:
\beq
K_s \Gamma = \Gamma K_s, \quad K_{1,2} \Gamma = - \Gamma K_{1,2}.
\eeq
Hence, for odd $N$ we have two sectors, $\Gamma = \pm 1$ and $K_s$ acts
within them. Since $K_s^2 =1 $ we have GOE. 
Whereas for even $N$ the commutation relations are:
\beq
K_s \Gamma^{1,2} = (-1)^{N/2} \Gamma^{1,2} K_s,
\eeq
\beq
K_1 \Gamma^{1,2} = (-1)^{N/2}  \Gamma^{1,2} K_1, \quad
K_2 \Gamma^{1,2} = (-1)^{N/2}  \Gamma^{1,2} K_2,
\eeq
\beq
K_s^2 = 1, \quad \l K_{1,2} \r^2 = (-1)^{N/2},
\eeq
and there are four sectors: $\Gamma^{1,2}=\pm 1$. 
For even $N/2$, operators $K_{s,1,2}$ act within the sectors 
and we have GOE. For odd $N/2$
individual sectors do not have any anti-linear symmetries and the ensemble
is GUE.

\begin{figure}[!ht]
\centering
\minipage{0.47\textwidth}
\includegraphics[scale=0.62]{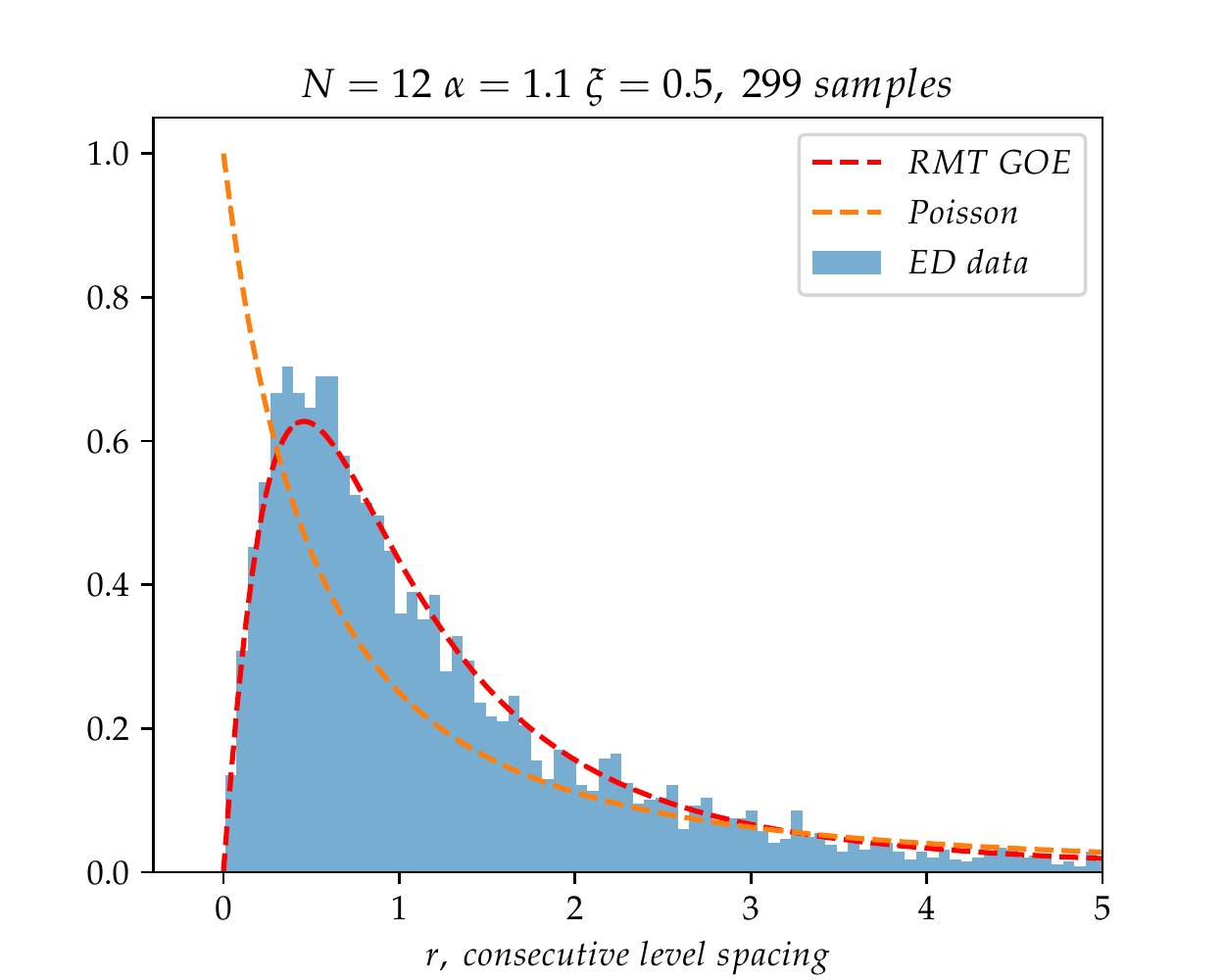}
\endminipage
\minipage{0.47\textwidth}
\includegraphics[scale=0.62]{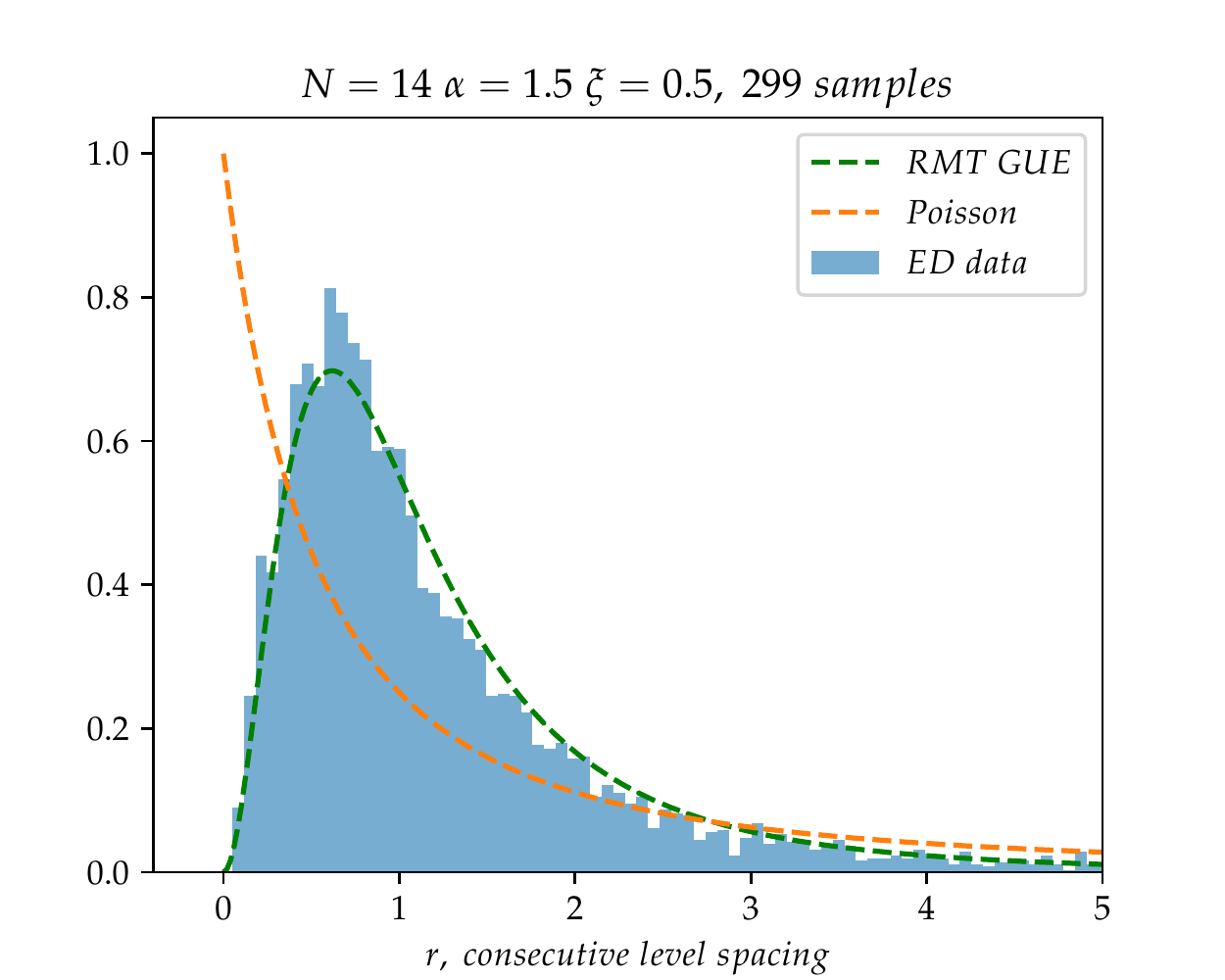}
\endminipage
\caption{Distribution of $r$ for various $\alpha,\xi,N$. Random matrix
prediction uses the surmise (\ref{eq:sur}). For comparison, we included
the exact result (\ref{eq:Poisson}) for Poisson-distributed gaps.}
\label{fig:level}
\end{figure}

The numerical results are shown in Figure \ref{fig:level}.
We used 20 lowest eigenvalues after projecting out global symmetries. 
We see a perfect agreement with the surmise (\ref{eq:sur}).
This suggests the absence of spin-glass phase at low energies.

\section{Conclusion}
\label{sec:conclusion}

In this paper we have presented a simple coupled SYK model. In the limit of large $N$ and low energies this model, 
like SYK,
has an approximate time-reparametrization
symmetry. However, unlike any previously known SYK-type model, 
the action for reparametrizations is dominated by a non-local action rather
than the (local) Schwarzian. To verify this claim studied numerically different physical quantities, 
such as thermodynamic energy, subleading correction to 2-point function and 4-point function. 
Our approach was to solve large $N$ equations numerically.
We saw that the non-local action indeed dominates everywhere. We double-checked some of our results using
finite $N$ exact diagonalization.

Also we discussed other physical features of the coupled model and the non-local action. 
It turned out that the residual entropy and (maximal) chaos exponent are exactly the same as in SYK. 
However, the heat capacity and
diffusion constant(in chain models) are very different from the models dominated by the Schwarzian.
Also certain aspects of time-ordered 4-point function are different too.
Also we presented a limited discussion of $1/N$ corrections. We computed the density of states near zero
and saw that it does not agree with 1-loop prediction of the non-local action. This shows that the partition
function is not 1-loop exact.

Let us comment on other models which can have an operator with dimension $1<h<3/2$ and thus exhibit the same physics.
First of all, in the coupled model we have studied,
apart from the operator (\ref{eq:theoperator}), there is another set of operators which may have 
the dimension $1<h<3/2$, \cite{Kim:2019upg}:
\beq
\Oc_{3,n} = \sum_i \l \psi^1_i \pr_u^{2n+1} \psi^2_i + \psi^2_i \pr_u^{2n+1} \psi^1_i \r
\eeq
Their dimensions are determined by
\beq
\frac{2(\alpha+\alpha^2)}{1+ 3 \alpha^2} g_A(h) = 1
\eeq
Therefore the dimension of $\Oc_{3,0}$ can be in the window $(1,3/2)$ for $-1<\alpha<0$.
However the operator $\Oc_{3,0}$ introduces non-diagonal(in $1,2$ indices) kinetic term. 
It can be diagonalized by a linear transformation of fermions, making it almost identical to the model
we considered. We expect the physics to be the same as in our model.
Two coupled SYK models(with 4-fermion interaction) admit another marginal interaction term:
\beq
\Lc'_{int} = \sum_{ijkl} B^1_{i;jkl} \psi^1_i \psi^2_{j} \psi^2_{k} \psi^2_l + B^2_{ijk;l} \psi^1_i \psi^1_j \psi^1_k
\psi^2_l
\eeq
Compared to eq. (\ref{eq:Lint}) it couples 3 fermions from one side to 1 fermion from the other side.
The resulting SD equations and the spectrum of conformal dimensions are very similar to the ones we studied. 
We again expect that in a certain range of parameters this model is dominated by the non-local action.

Let us conclude by a list of open questions:
\begin{itemize}
\item The most interesting question is to fully quantize the non-local theory (\ref{k:s_nonloc}). 
Is the Schwarzian piece important for this? Could it be that it starts dominating again in the
strong-coupling region $\beta J \gg N$?
\item Can we learn anything about JT gravity with matter from studying this model?
\item Is there spin-glass phase? Our results about the level statistics suggest that there is no such phase.
\item The model we described has an obvious generalization to $q$ interacting fermions.
\item Unfortunately, we could not obtain much analytic progress in the large $q$ limit. Solving the model
in this limit will give a partial analytical control over the models without the Schwarzian dominance.
\item What is tensor-model counterpart? Some tensor models are different from SYK in $1/N$ corrections and they are 
not captured by the Schwarzian.
\item What would be the physics of eternal traversable wormhole \cite{MQ}?
\item What is the physics of the spectral form factor \cite{cotler2017black} ?
\item It would be instructive to incorporate complex fermions(or global symmetries in general) and study the interplay 
between them and the non-local action \cite{KitaevRecent}. Models with complex fermions can have operators
with dimensions $1<h<3/2$ too \cite{Klebanov:2020kck}.
\item Schwarzian term gives rise to the famous linear-temperature dependence of electrical resistivity in certain models
\cite{Guo_2020}. It would be very interesting to generalize these models so that they are dominated by the non-local action.
Presumably it will lead to a tunable temperature dependence in the resistivity.
\item It would be interesting to investigate the dynamics of entanglement \cite{gu2017spread} in the chain models.
\item Finally, it is worth mentioning that in our model the point $|\alpha|=1$ seems to be special. At this value of $\alpha$
there is a field with $h=3/2$. However, because of $\cos(\pi h)$ in $m_h$, the 2-point function of reparametrizations
(\ref{eq:snonloc_n}) blows up. 
\end{itemize}

\section*{Acknowledgment}
The author is forever indebted to I.~Klebanov, G.~Tarnopolsky and W.~Zhao for many comments and discussion
throughout this project. I am grateful to  A.~Gorsky, J.~Turiaci and especially D.~Stanford and 
Z.~Yang for comments, and F.~Popov for discussions and very useful comments on the manuscript. 
I would like to thank C.~King for help with the manuscript and moral support.
This material is based upon work supported by the Air Force Office of Scientific Research under 
award number FA9550-19-1-0360. It was also supported in part by funds from the University of California. 
Use was made of computational facilities purchased with funds from the National
Science Foundation (CNS-1725797) and administered by the Center for Scientific
Computing (CSC). The CSC is supported by the California NanoSystems Institute
and the Materials Research Science and Engineering Center (MRSEC; NSF DMR
1720256) at UC Santa Barbara.

\appendix

\section{Lorentzian Schwinger--Dyson equations}
\label{sec:lorentz}
Self-energies in \textit{Lorentzian signature} are:
\begin{align}
\label{sd:lorentz}
\Sigma^>_{11} = -\frac{1}{4}J^2 \l 4 (G^>_{11})^3 + 12 \alpha^2 G^>_{11} (G^>_{22})^2 \r, \nonumber \\
\Sigma^>_{22} = -\frac{1}{4}J^2 \l 4 (G^>_{22})^3 + 12 \alpha^2 G^>_{22} (G^>_{11})^2 \r.
\end{align}
The relation between the self-energy and the retarded Green's function is
\beq
G_a^R(\omega) = \frac{1}{(1-\xi_a)\om - \Sigma^R_a}, \quad \xi_{11}=\xi, \xi_{22} = -\xi.
\eeq
To close the system we need the fluctuation--dissipation theorem to relate $G^>$ to $G^R$:
\beq
G_a^>(\omega) = 2i n_F(\omega) \Im G_a^R(\omega), \quad n_F(\omega) = \frac{1}{e^{\beta \omega}+1},
\eeq
Note that we can easily put $\beta = +\infty$.
These equations can be solved by iterations, exactly like the Euclidean case.
However one has to introduce a large interval in the time domain. So there will
be two cut-offs: the time step $dt$ and the interval length $L$.

\bibliography{refs}
\end{document}